\documentclass[
    twocolumn,
    trackchanges,
]{aastex701} 

\usepackage{graphicx}
\usepackage{dcolumn}
\usepackage{bm}
\usepackage{color}
\usepackage{xspace}
\usepackage{ulem}
\newcommand{\fermi}{\emph{Fermi}\xspace}

\newcommand{\lat}{\emph{Fermi}-LAT\xspace}
\newcommand{\fbs}{\fermi bubbles\xspace}
\newcommand{\gce}{GC excess\xspace}
\newcommand{\dd}{{\rm d}}
\newcommand{\dnde}{\dd N/\dd E}
\newcommand{\jf}{J-factor\xspace}

\newcommand{\gr}{$\gamma$-ray\xspace}
\newcommand{\grs}{$\gamma$ rays\xspace}

\renewcommand{\deg}{\ensuremath{^{\circ}}\xspace}

\newcommand\raa{RAA}

\begin{document}
\title{Observations of the \emph{Fermi} bubbles and the Galactic center excess with the DArk~Matter~Particle~Explorer}

\author[orcid=0000-0003-2021-9205]{F.~Alemanno}
\affiliation{Dipartimento di Matematica e Fisica E. De Giorgi, Universit\`a del Salento, I-73100, Lecce, Italy}
\affiliation{Istituto Nazionale di Fisica Nucleare (INFN) - Sezione di Lecce, I-73100, Lecce, Italy}
\email{francesca.alemanno@gssi.it}

\author{Q.~An}
\altaffiliation[]{Deceased}
\affiliation{State Key Laboratory of Particle Detection and Electronics, University of Science and Technology of China, Hefei 230026, China} 
\affiliation{Department of Modern Physics, University of Science and Technology of China, Hefei 230026, China}
\email{anqi@ustc.edu.cn}

\author[orcid=0000-0001-8275-5738]{P.~Azzarello}
\affiliation{Department of Nuclear and Particle Physics, University of Geneva, CH-1211, Switzerland}
\email{philipp.azzarello@unige.ch}

\author[orcid=0000-0003-0751-6731]{F.~C.~T.~Barbato}
\affiliation{Gran Sasso Science Institute (GSSI), Via Iacobucci 2, I-67100 L'Aquila, Italy}
\affiliation{Istituto Nazionale di Fisica Nucleare (INFN) - Laboratori Nazionali del Gran Sasso, I-67100 Assergi, L'Aquila, Italy}
\email{felicia.barbato@gssi.it}

\author[orcid=0000-0002-6530-3227]{P.~Bernardini}
\affiliation{Dipartimento di Matematica e Fisica E. De Giorgi, Universit\`a del Salento, I-73100, Lecce, Italy}
\affiliation{Istituto Nazionale di Fisica Nucleare (INFN) - Sezione di Lecce, I-73100, Lecce, Italy}
\email{paolo.bernardini@le.infn.it}

\author[orcid=0000-0002-5334-9754]{X.~J.~Bi}
\affiliation{University of Chinese Academy of Sciences, Beijing 100049, China}
\affiliation{Particle Astrophysics Division, Institute of High Energy Physics, Chinese Academy of Sciences, Beijing 100049, China}
\email{bixj@mail.ihep.ac.cn}

\author[orcid=0009-0004-6010-9486]{H.~V.~Boutin}
\affiliation{Department of Nuclear and Particle Physics, University of Geneva, CH-1211, Switzerland}
\email{hugo.boutin@cern.ch}

\author[orcid=0000-0001-8822-5914]{I.~Cagnoli}
\affiliation{Gran Sasso Science Institute (GSSI), Via Iacobucci 2, I-67100 L'Aquila, Italy}
\affiliation{Istituto Nazionale di Fisica Nucleare (INFN) - Laboratori Nazionali del Gran Sasso, I-67100 Assergi, L'Aquila, Italy}
\email{irene.cagnoli@gssi.it}

\author[orcid=0000-0002-9940-3146]{M.~S.~Cai}
\affiliation{Key Laboratory of Dark Matter and Space Astronomy, Purple Mountain Observatory, Chinese Academy of Sciences, Nanjing 210023, China}
\affiliation{School of Astronomy and Space Science, University of Science and Technology of China, Hefei 230026, China}
\email{caimsh@pmo.ac.cn}

\author[orcid=0009-0003-6044-3428]{E.~Casilli}
\affiliation{Gran Sasso Science Institute (GSSI), Via Iacobucci 2, I-67100 L'Aquila, Italy}
\affiliation{Istituto Nazionale di Fisica Nucleare (INFN) - Laboratori Nazionali del Gran Sasso, I-67100 Assergi, L'Aquila, Italy}
\email{elisabetta.casilli@gssi.it}

% ...

\collaboration{all}{(DAMPE Collaboration)}

\correspondingauthor{DAMPE Collaboration}
\email{dampe@pmo.ac.cn}

\received{2026 January 7}
\revised{2026 March 27}
\accepted{2026 March 27}

\submitjournal{ApJS}

\begin{abstract}
The DArk Matter Particle Explorer (DAMPE) is a space-borne high-energy particle detector that surveys the $\gamma$-ray sky above $\sim 2~\rm GeV$ with a peak acceptance of $\sim 0.2~\rm m^2\,sr$. With the 102 months of data collected by DAMPE, we show that the \emph{Fermi} bubbles are detected at a significance of $\sim 26\sigma$ and identify a GeV excess in the direction of Galactic center at $\sim 7 \sigma$ confidence. Both spectra and morphology are consistent with those observed by \emph{Fermi}-LAT and the GeV excess component can be interpreted by the dark matter annihilation with a mass of $\sim 50$ GeV and a velocity-averaged cross section of $\sim 10^{-26}~{\rm cm^{3}~s^{-1}}$ for the $\chi \chi \rightarrow b\bar{b}$ channel. Our results thus provide the first independent detection of these two intriguing diffuse gamma-ray sources besides \emph{Fermi}-LAT. 
\end{abstract}



\section{Introduction}\label{sec::intro}
The Galactic Center (GC) is the most extreme region in our Galaxy.
It harbors a supermassive black hole (SMBH), dense clouds and star clusters, a wealth of supernova remnants and pulsar wind nebulae, and a large amount of dark matter (DM).
The GC region attracts wide interest in the high-energy astrophysics community.
It is a high-energy cosmic-ray accelerator as shown with the central point source~\citep{Aharonian2004,Chernyakova2011} and the diffuse emission within the inner $\sim 1\deg$ region~\citep{HESS2016,MAGIC2020,Huang2021,Albert2024}.
It is a favorable target for DM indirect search since it is the place where the DM density peaks~\citep{Bertone2005,Bertone2018}, and an extended excess is reported in the GC~\citep{Charles2016,Murgia2020}.
It may also be responsible for the large-scale bubbles in the inner Galaxy which originated from the energetic activities of the SMBH in the past~\citep{Su2010a,Predehl2020}.
In the past decades, the \fermi Large Area Telescope (\lat) has contributed greatly to the understanding of the GC.
Two particularly interesting sources are the \gce and the \fbs.

The \gce is an unexpected extended structure on top of the regular diffuse emission from Galactic cosmic-ray sea found in \lat data~\citep{Goodenough2009,2009arXiv0912.3828V,Hooper2011,Abazajian2011,2011PhRvD..84l3005H,Abazajian2012,Hooper2013,Gordon2013,Daylan2016,Zhou2015,Calore2015b,Calore2015,Ajello2016,Huang2016,Ackermann2017}.
It is a $\sim 10\deg$ rounded excess located round the GC and has a spectrum peaked at $\sim 1-3~\rm GeV$ with a high-energy tail.
The excess is robust against systematic uncertainties in the analyses of \lat data~\citep{Zhou2015,Calore2015b,Ackermann2017,Zhong2020,Pohl2022,Cholis2022}.
The \gce is usually explained with the prompt emission from DM annihilation products~\citep[e.g.][]{Goodenough2009,Karwin2017} or the curvature emission from an unresolved millisecond pulsar (MSP) population~\citep{Abazajian2011,Mirabal2013,Brandt2015,Yuan:2014rca,Yuan2015,Gautam2022}.
Extensive morphological studies have been conducted to find out the origin of the excess~\citep{Macias2018,Macias2019,Bartels2018,Leane2019,Buschmann:2020adf,DiMauro2021,McDermott2023,Zhong2024,Song2024,Ramirez2025}.
According to recent hydrodynamical simulations, the DM distribution of the Milky Way is found elongated along the disk as well~\citep{Muru2025}, reducing the morphological difference between the two hypotheses.
The DM origin may be supported by the possible GeV anti-proton excess \citep{Cui2017,Cuoco2017,Fan2022a,Zhu2022,Duan2025b} identified in the AMS-02 data \citep{Aguilar2016,Aguilar2025}.
But some DM channels may be in tension with the observations of the Milky Way dwarf spheroidal satellite galaxies (dSphs)~\citep{2015PhRvL.115w1301A,2017ApJ...834..110A,2020PhRvD.102f1302A,McDaniel2024,2026JHEAp..5100536L}.

The \fbs, first discovered in the \lat data, consist of two large lobes, each of which is approximately $40\deg$ wide and extends to $55\deg$ above and below the GC~\citep{Su2010a}.
Their spectrum is hard from 0.1 GeV to 100 GeV and begins to soften at around 100~GeV~\citep{Su2010a,Ackermann2014}.
The two lobes share a similar spectral shape, and the interior of the bubbles is uniform~\citep{Su2010a,Hooper2013,Ackermann2014,Yang2014,Narayanan2017}.
The \fbs are believed to originate from the GC region, but their radiation mechanism is still under debate (see~\citet{Sarkar:2024mjm} for a recent review).
The bubbles could be originated from past jet activity~\citep{Guo2012,Yang2022}, accretion wind~\citep{Zubovas2012,Mou2014}, or star driven outflow~\citep{Crocker2011,Cheng2011}.

The DArk Matter Particle Explorer (DAMPE) is a space-borne high-energy particle detector launched on 17 December 2015, aiming to measure cosmic rays and $\gamma$ rays in a wide energy range~\citep{DAMPE2017,DAMPE2019b,DAMPE2021,DAMPE2022,DAMPE2025a,DAMPE2025b,DAMPE2025c,DAMPE2025d}.
It surveys the \gr sky from $\sim 2~\rm GeV$ to 10~TeV  twice a year with a peak acceptance of $\lesssim 2000~\rm cm^2\,sr$~\citep{Chang2017,Ambrosi2019}.
Thanks to the unprecedentedly high energy resolution~\citep{Xu2022}, DAMPE has set stringent constraints on the \gr line produced by the dark matter annihilation~\citep{Alemanno2022_line,Cheng:2023chi,Fan:2024rcr}.
In this work, we report the observations of the \fbs and the \gce with DAMPE.
The two sources were only detected by \lat prior to DAMPE, and now they are, for the first time, verified by an independent telescope. Furthermore, with higher energy resolution and lower cosmic ray background comparable to \lat, DAMPE is able to perform such measurements with potentially smaller systematic uncertainties. 
The paper is structured as follows:
In Section~\ref{sec::data}, the data set, the baseline model, and the fitting algorithm are described.
We firstly analyze the \fbs in Section~\ref{sec::fbs} and then proceed to the \gce in Section~\ref{sec::gce}.
For both sources, the spectrum and morphology are studied, and the systematic uncertainties are evaluated.
Finally, the conclusion is made in Section~\ref{sec::summary}.

\section{Data Analysis}\label{sec::data}

\subsection{DAMPE data}\label{sec::data:selection}
A sophisticated algorithm is developed to distinguish photons from background events with a high efficiency: the cosmic-ray contamination is $\lesssim 15\%$ that of the extragalactic diffuse \gr emission above $\sim 3~\rm GeV$ and is negligible above $20~\rm GeV$~\citep{Xu2018}.
Recently, the {\tt v6.0.3} photon data set was produced using the updated selection algorithm, which further reduces the proton and electron contamination by $\lesssim 50\%$ and $\lesssim 15\%$ respectively at the cost of merely $\sim 3\%$ loss in acceptance~\citep{Shen2023ICRC2}.
The new version of the data set also benefits from the recent calibration parameters~\citep{Ambrosi2019,Jiang2020,Cui2023}.
In this work, we use this updated data set.\footnote{
    The photon data above 3~GeV up to 2024 January 1 are publicly available in \url{https://dampe.nssdc.ac.cn}.
    The filters of SAA and solar flares have been applied in the public data set in default.
}

We select photon events collected from 1 January 2016 to 30 June 2024 (DAMPE Mission Elapsed Time $94608001-362793602$~sec).
To avoid the pile-up of the instrument, the observations collected within the South Atlantic Anomaly (SAA) or during strong solar flares are excluded.
We only choose the High-Energy-Trigger (HET) events above 2~GeV.
After the data reduction, around 359 thousand photon events remain.

\begin{table}[!bt]
    \centering
    \caption{\label{tab::data_selction}
        Summary table of the data selection for the analyses of the \fbs (second column) and the \gce (third column).
        The observed intensity maps in the regions of interest of \fbs and \gce in the plate carr\'ee (CAR) projection are shown in Figure~\ref{fig::bubble:fluxmap} and Figure~\ref{fig::gce:fluxmap}, respectively.
        The point source (PS) masks are only applied around the sources listed in the DAMPE 8.7~yr catalog.
        Details can be found in Section~\ref{sec::data:selection}.
    }
    \begin{tabular}{l|l|l}
    \hline\hline
    Selection                       & \fbs                        & \gce                      \\
    \hline
    ROI                             & $5\deg \leq|b|\leq 60\deg$, & $1\deg \leq|b|\leq 20\deg$, \\
                                    & $|\ell|\leq 60\deg$         & $|\ell|\leq 20\deg$        \\
    Spatial bins                    & HEALPix,   order 7          & CAR,  $400\times400$       \\
    Energy bins                     & 2$-$500~GeV, 24 bins        & 2$-$200~GeV, 20 bins       \\
    PS mask                         & $1\fdg5$ (${\rm TS}\geq25$) & $0\fdg3$ ($25\leq{\rm TS}<49$),\\
                                    &                             & $0\fdg7$ (${\rm TS}\geq49$)\\
    \hline\hline
    \end{tabular}
\end{table}

\begin{table*}[!hbt]
    \centering
    \caption{\label{tab::templates}
        Summary table of the template maps adopted in the data analyses of the \fbs (second column) and the \gce (third column).
        The baseline GALPROP templates are labelled with $\rm ^SL^Z10^T150$ from~\citet{1SC2016}.
        The maps of some components and the mask in the analyses of \fbs and \gce are shown in Figure~\ref{fig::appx:bubble:flux_comps} and Figure~\ref{fig::appx:gce:flux_comps}, respectively, in the Appendix.
        More descriptions on the components can be found in Section~\ref{sec::data:gr_components}.
    }
    \begin{tabular}{l|l|l}
    \hline\hline
    Component                          & Analysis of \fbs (Figure~\ref{fig::appx:bubble:flux_comps}) & Analysis of \gce (Figure~\ref{fig::appx:gce:flux_comps}) \\
    \hline
    \ion{H}{1}$+$\ion{H}{2} gas        & GALPROP, binned into three rings:   & GALPROP, all rings combined:        \\
                                       & 0$-$8~kpc (free), 8$-$10~kpc (free),& 0$-$30~kpc (free)                   \\
                                       & and 10$-$30~kpc (fixed)             &                                     \\
    $\rm H_2$ gas                      & GALPROP, all rings combined (free)  & GALPROP, all rings combined (free)  \\
    Inverse Compton                    & GALPROP, all ISRF combined (free)   & GALPROP, all ISRF combined (free)   \\
    Loop I                             & Geometric template (free)           & Geometric template (free)           \\
    Isotropic                          & Proportional to exposure map (free) & Proportional to exposure map (free) \\
    Point sources                      & DAMPE point sources (fixed)         & DAMPE point sources (fit \& fixed), \\
                                       &                                     & and weak \lat sources (free)        \\
    \fbs                               & Flat template in this work (free)   & Flat template (free with priors)    \\
    \gce                               &  $\cdots$                           & DM annihilation template (free)     \\
    \hline\hline
    \end{tabular}
\end{table*}

Instead of fitting the two targets based on the all-sky data, which can greatly suffer from systematic uncertainties, we define different regions of interest (ROIs) for them:
the central $120\deg \times 120\deg$ region excluding $10\deg$ Galactic plane for \fbs, whereas a $40\deg \times 40\deg$ one excluding $2\deg$ Galactic plane for \gce.
These ROIs can enclose the main parts of the sources while not introducing too much astrophysical background emission. 
The intensity maps for the two ROIs are presented in Figure~\ref{fig::bubble:fluxmap} and Figure~\ref{fig::gce:fluxmap}.
We further partitioned the data into spatial and energy bins to perform binned likelihood analyses.
For the \gce, the photons from 2~GeV to 200~GeV are binned into $400\times 400$ pixels in the plate carr\'ee (CAR) projection ($0\fdg10$ pixel size) and 20 logarithmically spaced energy bins.
For the \fbs, the HEALPix pixelization\footnote{\url{http://healpix.sourceforge.net}}~\citep{Healpix2005} with the order of 7 ($\approx 0\fdg46$ pixel size) is adopted to facilitate the convolution of instrumental responses.
The energy range also extends up to 500~GeV to better fit the high-energy of the bubbles, and the data are split into 24 energy bins.

To reduce the number of free parameters, the bright sources listed in the DAMPE point source catalog (DAMPE Collaboration 2026, in preparation) are masked.
In the analysis of the \fbs, circular masks with the radii of $1\fdg5$ are applied, which can remove 95\% of the point source emission at 2~GeV~\citep{Duan2025}.
On the other hand, the masks dependent on the source significance are employed for the \gce to make better use of the observed data: $0\fdg7$ ($0\fdg3$) is chosen for those with TS larger (smaller) than 49 in the catalog.
The data selection and binning are summarized in Table~\ref{tab::data_selction}.

\subsection{$\gamma$-ray emitting components}\label{sec::data:gr_components}

Several components make up the GeV $\gamma$-ray sky observed by DAMPE, including the Galactic diffuse emission (GDE), the isotropic background, and many point-like or extended/diffuse sources.
In this subsection, we will describe the baseline model, summarized in Table~\ref{tab::templates}, and leave the systematic uncertainties for the following sections.

GDE originates from the interactions of Galactic cosmic rays (CRs) with interstellar gas via inelastic nucleon-nucleon collision and electron bremsstrahlung, and with interstellar radiation field (ISRF) via inverse Compton (IC) process~\citep{Acero2016_FermiGDE}.
In the baseline model, we employ the GDE emission model calculated with the CR propagation code {\tt Galprop}\footnote{\url{https://galprop.stanford.edu/}}~\citep{Strong:1998pw,Strong:1998fr,Porter:2008ve}
using the parameter set named as {\tt Lorimer\_z10\_Ts150\_P7} (denoted as $\rm ^SL^Z10^T150$ hereafter) developed in the first \lat supernova remnant catalog~\citep[1SC,][]{1SC2016}.\footnote{\url{https://fermi.gsfc.nasa.gov/ssc/data/access/lat/1st_SNR_catalog/}}
The model is constructed under the assumptions of the CR sources distribution following that traced by pulsars~\citep{Lorimer2006}, the propagation halo with height of $z_{\rm h}=10\rm~kpc$ and radius of $R_{\rm h}=30\rm~kpc$, and the uniform \ion{H}{1} spin temperature of $T_{\rm S}=150\rm~K$.
The hadronic component and bremsstrahlung component, spatially correlated with the gas density, are combined into gas-correlated templates associated with either neutral hydrogen \ion{H}{1} or molecular hydrogen $\rm H_2$.\footnote{Ionized hydrogen and dark gas are incorporated in the \ion{H}{1} map \citep{Ackermann2012_GalpropGDE,1SC2016}. Helium and heavier element gases are assumed to be mixed with hydrogen uniformly~\citep{Casandjian2015}.}
To be more flexible in the analyses of the bubbles, the \ion{H}{1} map is further split into three annuli: an inner ring spanning the galactocentric radius $0-8~\rm kpc$, a local ring spanning $8-10~\rm kpc$, and an outer ring spanning $10-30~\rm kpc$.
The IC template is made by simply summing up the contributions of all the ISRF components.

Loop~I is a giant structure spanning $\sim 100\deg$ on the sky visible in radio~\citep{Berkhuijsen1971,Planck2016}, X-ray~\citep{Egger1995,Predehl2020} and \gr bands~\citep{Casandjian2009}.
It is originally proposed to be associated with the Sco-Cen OB association at a distance of $120-140~\rm pc$~\citep{Salter1983}, but growing evidence shows that the Loop~I may be located beyond the OB association~\citep{Planck2016} and might even be a source of GC origin~\citep{Sofue2015}.
Irrespective of its origin, we adopt the geometric model from~\citet{Wolleben2007} following the previous works~\citep{Ackermann2014,Ajello2016,Ackermann2017}.
This model is made up of two spherical shells, each of which is described with the coordinates ($\ell_{\rm s}, b_{\rm s}$) of the shell center, the distance to the center $d_{\rm s}$, and the inner and outer radius ($r_{\rm in}$, $r_{\rm out}$).
The parameter sets $(\ell_{\rm s}, b_{\rm s}, d_{\rm s}, r_{\rm in}, r_{\rm out})$ for the two shells are are $\rm (341\deg,3\deg,78~pc, 62~pc, 81~pc)$ and $\rm (332\deg,37\deg,95~pc, 58~pc, 82~pc)$.
The \gr emissivities of the shells are assumed to be the same.
We make the Loop~I spatial template by integrating the emission from the shells along the line of sight.

The isotropic diffuse background comprises the unresolved extragalactic sources, extragalactic diffuse emission, and misclassified CR particles~\citep{Ackermann2015_IGRB,Ajello2015}.
DAMPE has a lower CR contamination but a worse point source sensitivity than \lat, so the spectrum of the isotropic background is expected to be different from that of \fermi~\citep{Ackermann2015_IGRB}. 
Here, we adopt the power-law spectral shape in the global fitting.

We also consider the point sources within the ROIs.
Around 360 point sources are detected with TS values larger than 25 in 8.7 years of DAMPE photon data, given the standard \lat Galactic interstellar emission model~\citep{Acero2016_FermiGDE}.
For the analysis of \fbs, we compute an expected counts cube based on the fitted spectral parameters listed in the source catalog.
Since the masks are large enough to remove the majority of emissions from the point sources, we keep the model fixed in the fitting.
However, for the analysis of \gce, the masks are not that large and their spectral parameters may be relevant.
To ensure the parameters are compatible with the background emission, we first fit the point sources without masking them, and then adopt the fitted parameters in the analysis afterwards when the point source masks are applied.
The weak sources below the DAMPE sensitivity are also accounted for when analyzing the \gce.
We create a new template containing the sources listed in the \lat 14-yr catalog\footnote{\url{https://fermi.gsfc.nasa.gov/ssc/data/access/lat/14yr_catalog/gll_psc_v32.xml}}~\citep[4FGL-DR4;][]{4FGL2022,Ballet2023_4FGLDR4} but not in the DAMPE catalog.
We use a power law as a scale and fit the component as a whole in the analysis.

The last two components are \fbs and \gce, which are the targets in this work.
Since the two components do not have counterparts in the wavelength other than \grs, we derive them from the DAMPE data.
Firstly, we analyze the \fbs and extract their spatial templates from the residual map.
Then, based on the obtained template of the bubbles, we conduct the \gce analysis, where the spatial models of DM annihilation and millisecond pulsars are tested.
More details on the construction of the templates will be presented in the following sections.

\subsection{Binned likelihood analysis}
The total \gr intensity $I_{\rm tot}$ in the direction $(\ell, b)$ at the energy $E$ is the linear combinations of the components
\begin{equation}
    I_{\rm tot} (\ell, b, E; {\bm \Theta}) = \sum_m f_m(E; \Theta_m) I_m (\ell, b, E),
\end{equation}
where $f_m(E; \Theta_m)$ is the scaling factor parameterized with $\Theta_m$ for the template of $m$-th component $I_m (\ell, b, E)$.
In the global fitting, we use the power-law with exponential cutoff spectrum for the bubbles and the log-parabola spectrum for the \gce.
For other components, the power-law scaling factors are adopted.\footnote{See \url{https://fermi.gsfc.nasa.gov/ssc/data/analysis/scitools/source_models.html} for the definitions of the spectral models.}

\begin{figure*}
    \centering
    \includegraphics[width=0.45\textwidth]{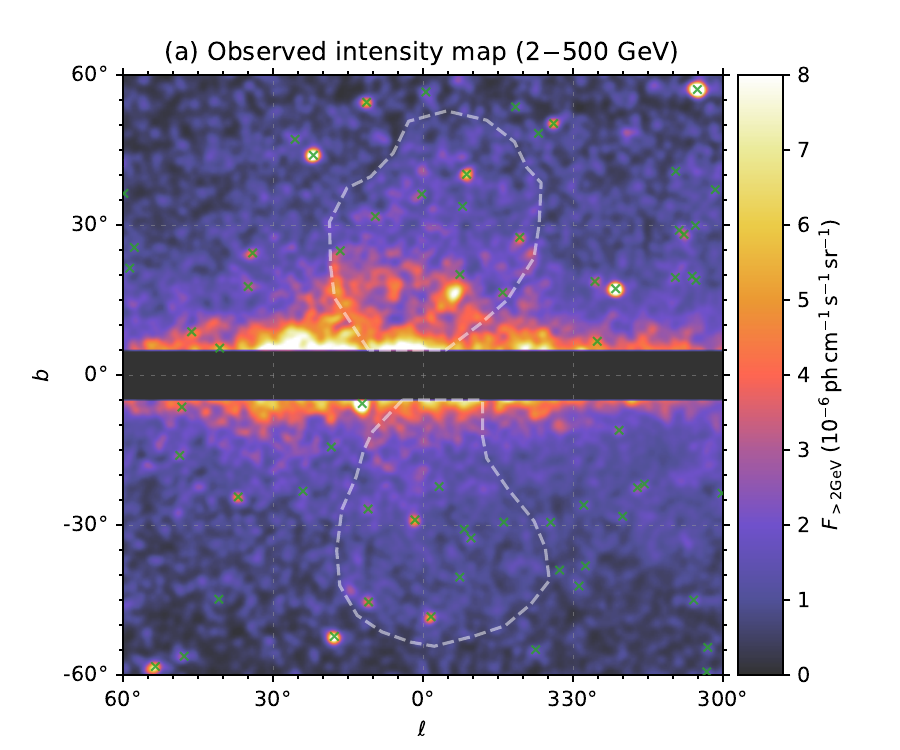}
    \includegraphics[width=0.45\textwidth]{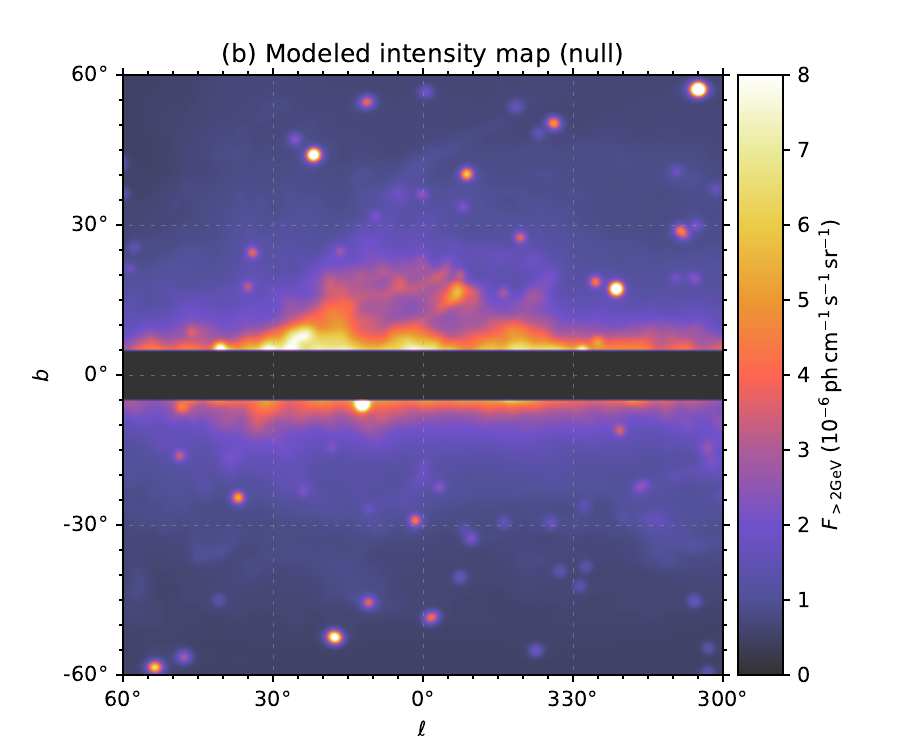}
    \includegraphics[width=0.45\textwidth]{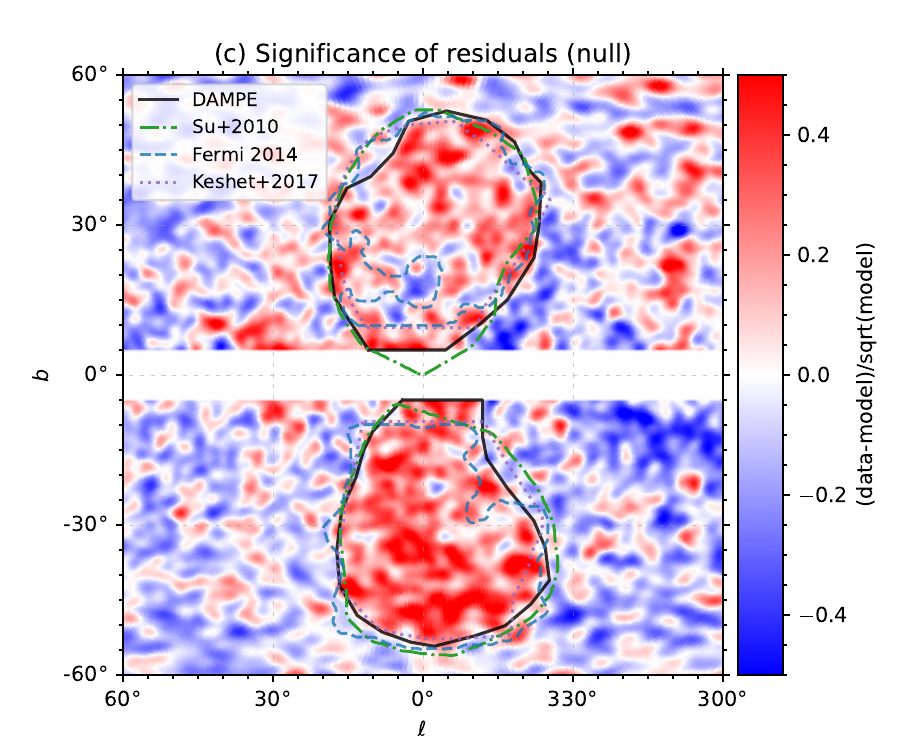}
    \includegraphics[width=0.45\textwidth]{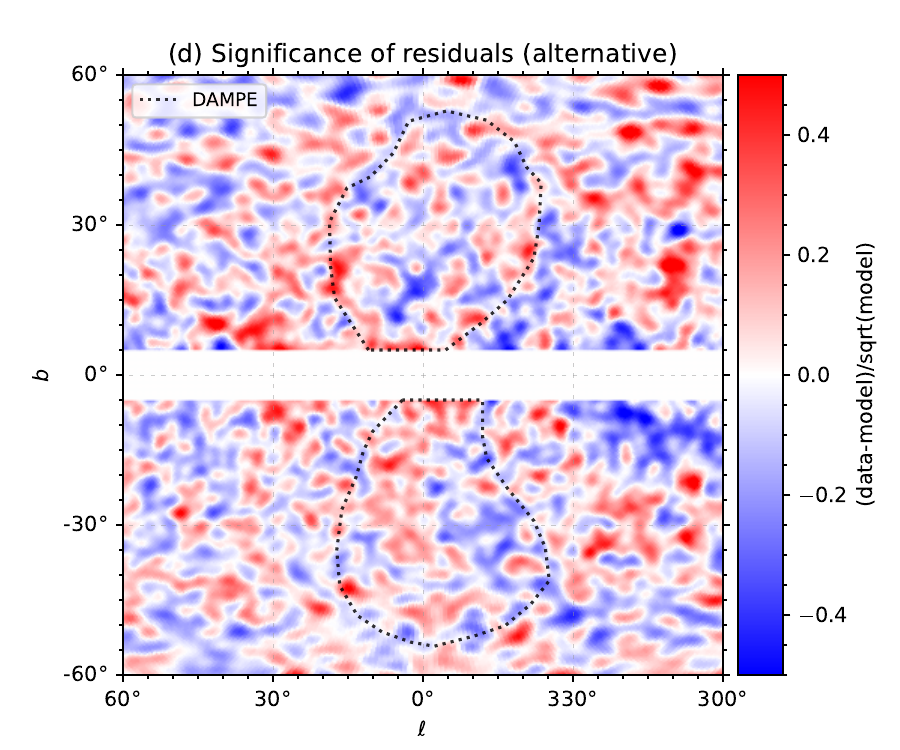}
    \caption{
    {\it Upper panels:}
    The intensity map from 2~GeV to 500~GeV in CAR projection, smoothed with a Gaussian kernel with $\sigma=0\fdg75$ for (a) the observed data and (b) best-fit model without the bubbles.
    The Galactic plane region ($|b|\leq 5\deg$) is masked.
    The green crosses mark the point sources in the DAMPE catalog, whereas the gray contour encloses the \fbs.
    {\it Lower panels:}
    The significance of the residual maps smoothed with a $1\deg$ Gaussian kernel for the models (c) without and (d) with the bubbles.
    The black curve shows the boundary derived in this work, while the other lines represent the ones from \lat.
    }
\label{fig::bubble:fluxmap}
\end{figure*}

We utilize the DAMPE \gr analysis toolkit {\tt DmpST}~\citep{Duan2019} to apply the instrumental responses (IRFs).
The recent version of this software implements the up-to-date IRFs calibration parameters~\citep{Shen2024,Duan2025}.
We calculate the exposure map $\varepsilon(\ell, b, E)$ considering the azimuth-angle dependence of the effective area.
Then the total intensity map is multiplied by the exposure map, convolved with the point-spread function (PSF), and integrated to achieve the expected counts map in an energy bin: $\tilde{\mu}_{\rm tot}=\int \dd E\, [(I_{\rm tot}\cdot \varepsilon) * {\rm PSF}] \Delta\Omega$, where $\Delta \Omega$ is the solid angle of a pixel.

Binned likelihood analysis is performed in this work.
The likelihood function $\mathcal{L}(\Theta)$ is defined as~\citep{Mattox1996,Duan2019}
\begin{equation}
    \ln \mathcal{L}({\bm \Theta}) = \sum_{ij} [N_{ij}\ln (\tilde{\mu}_{{\rm tot},ij}) - \tilde{\mu}_{{\rm tot},ij}],
\end{equation}
where $N_{ij}$ and $\tilde{\mu}_{{\rm tot},ij} \equiv \tilde{\mu}_{\rm tot}(\hat{\bf s}_i, E_j; {\bm \Theta})$ are the observed and expected photon counts in $i$-th pixel centering at the coordinate $\hat{\bf s}_i$ and $j$-th energy bin, respectively.
The likelihood is optimized with the {\tt Minuit} algorithm~\citep{MINUIT1975}.

\section{\emph{Fermi} bubbles}\label{sec::fbs}
\begin{figure*}
    \centering  
    \includegraphics[width=0.45\textwidth]{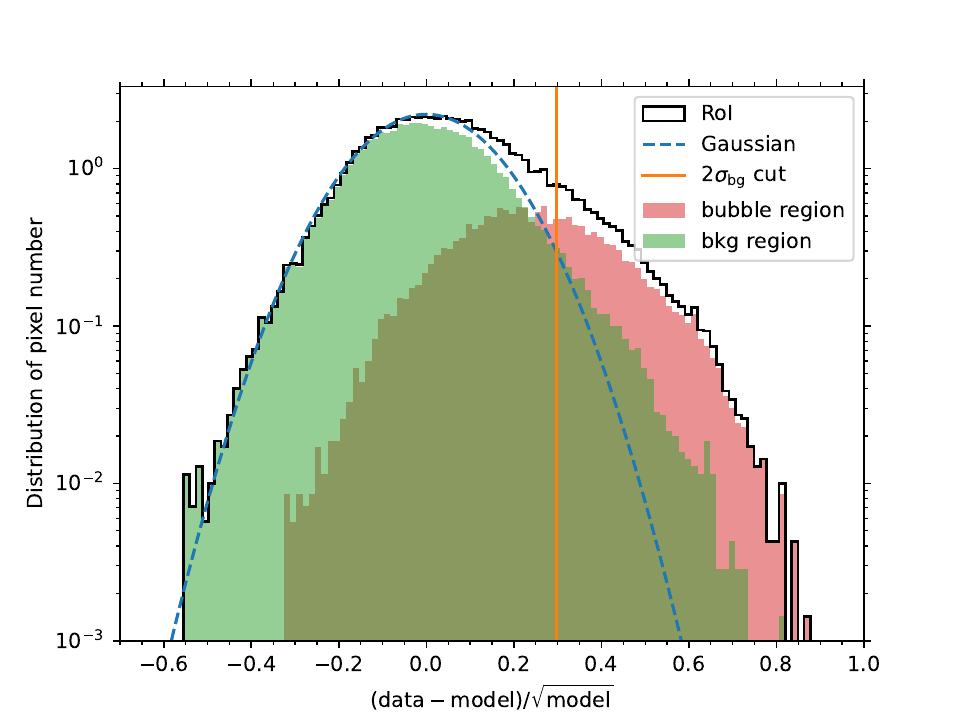}
    \includegraphics[width=0.4\textwidth]{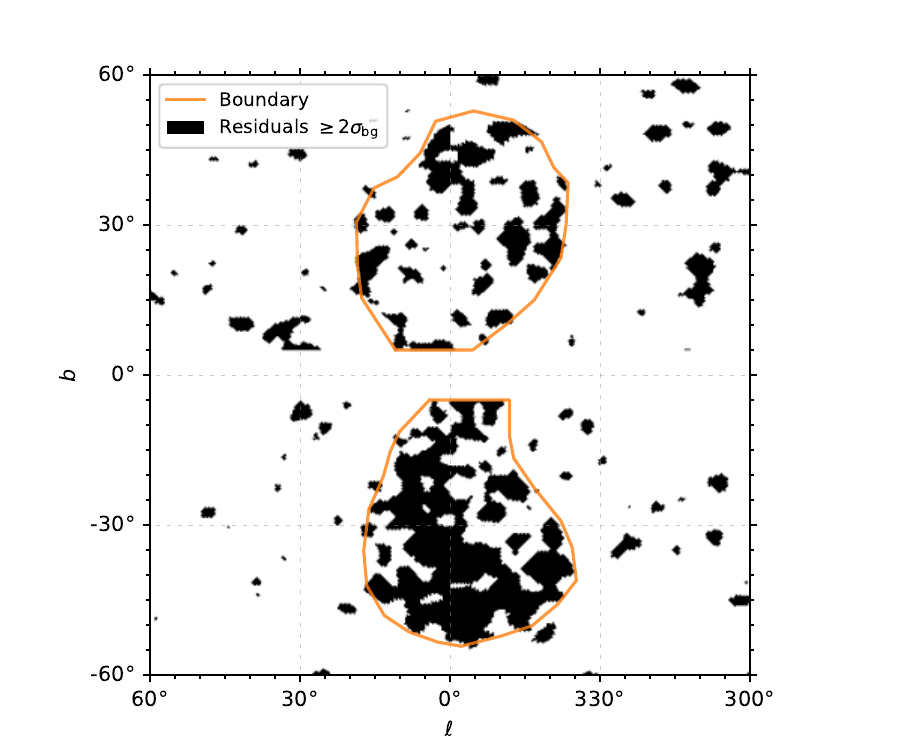}
    \caption{
        {\it Left panel:}
        Distribution of the values in the significance map.
        The map, in HEALPix projection, is smoothed with a $1\deg$ Gaussian kernel.
        The black histogram shows the distribution of pixel numbers within the ROI, whereas the red and green ones show those within and outside the bubbles.
        The blue dashed line is the Gaussian profile with a mean value of zero fitted to the background histogram.
        The orange solid line represents the $2\sigma_{\rm bg}$ cut adopted to define the bubbles' boundary.
        {\it Right panel:}
        The map with the significance larger than $2\sigma_{\rm bg}$ cut.
        The bubble template in this work consists of two polygons whose boundaries are illustrated with the solid orange lines.
    }
\label{fig::bubble:template}
\end{figure*}

\begin{figure}
    \centering  
    \includegraphics[width=0.45\textwidth]{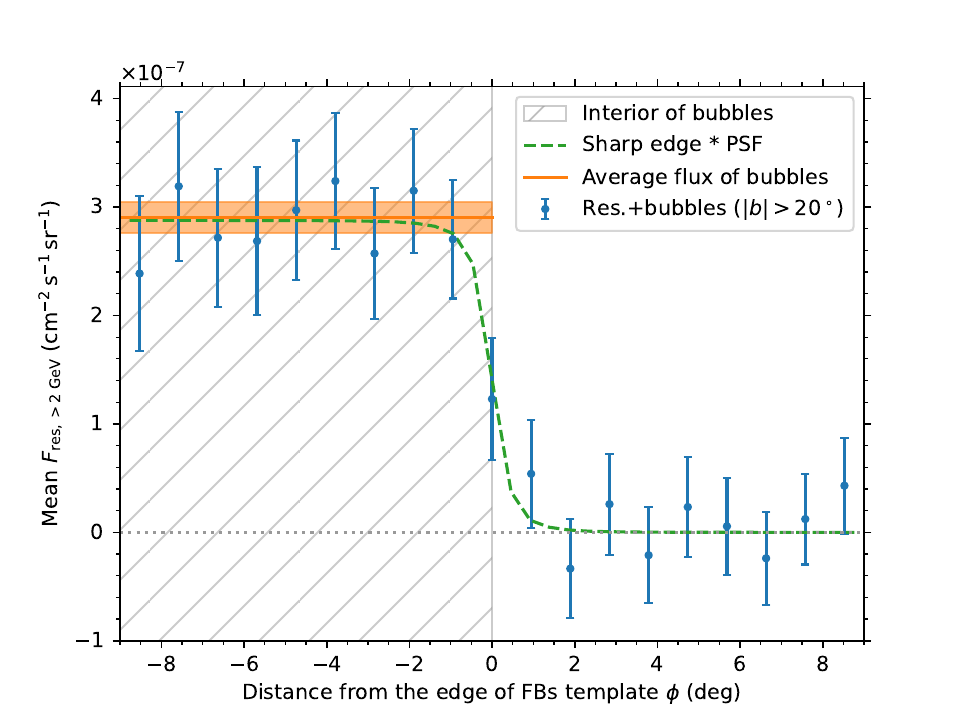}
    \caption{
    The average residual flux as a function of the angular distance to the edge in the high-latitude region with $|b|>20\deg$ (blue points).
    The residual is obtained by extracting the best-fit model except for the bubbles from the data.
    The negative (positive) $x$-axis values represent the region inside (outside) the bubbles.
    The green dashed line is the sharp edge convolved with the PSF.
    The orange band is the average flux of the bubble template as a whole.
    }
\label{fig::bubble:boundflux}
\end{figure}

\subsection{Defining a template for the bubbles}\label{sec::fbs:template}
The photon events are selected from the $120\deg \times 120\deg$ ROI defined in Section~\ref{sec::data:selection}.
In Figure~\ref{fig::bubble:fluxmap}a, we present the observed flux map integrated above 2~GeV, which is simply the counts map divided by the exposure, i.e., $F(\hat{\bm s}_i,E>{2~\rm GeV}) = \int \dd E\, F(\hat{\bm s}_i, E) \approx \sum_j N_{ij}/[\varepsilon(\hat{\bm s}_i, E_j)\,\Delta \Omega_i ]$.
To remove the background components, we set the baseline model without the bubbles as the null hypothesis and perform the binned likelihood analysis.
The fitted components are labelled in Table~\ref{tab::templates}.
Both the Galactic plane and point sources are masked.
Figure~\ref{fig::bubble:fluxmap}b shows the flux map given the optimized null model (see Figure~\ref{fig::appx:bubble:specgl} for the mean flux of the components in the global fittings).
Most of the emission is well described by the model, as we expected.
However, if the null model is subtracted from the data, some weak excess is visible in the region of \fbs.
Figure~\ref{fig::bubble:fluxmap}c shows the significance of the residual $(N_{i}-\tilde{\mu}_i)/\sqrt{\tilde{\mu}_i}$, where $N_{i}$ and $\tilde{\mu}_i$ are the observed and predicted photon counts in $i$-th pixel integrated over $2-500$~GeV.
The excess is in good agreement with the boundaries extracted from the \lat data, including the green dot-dashed line from~\citet{Su2010a}, the blue dashed line from~\citet{Ackermann2014}, and the purple dotted line from~\citet{Keshet2017}.
Besides, the normalizations of the isotropic and IC components are exaggerated in the fitting, which compensates for the absence of the bubbles and leads to a photon deficit outside the boundary.

\begin{figure*}
    \centering  
    \includegraphics[width=0.45\textwidth]{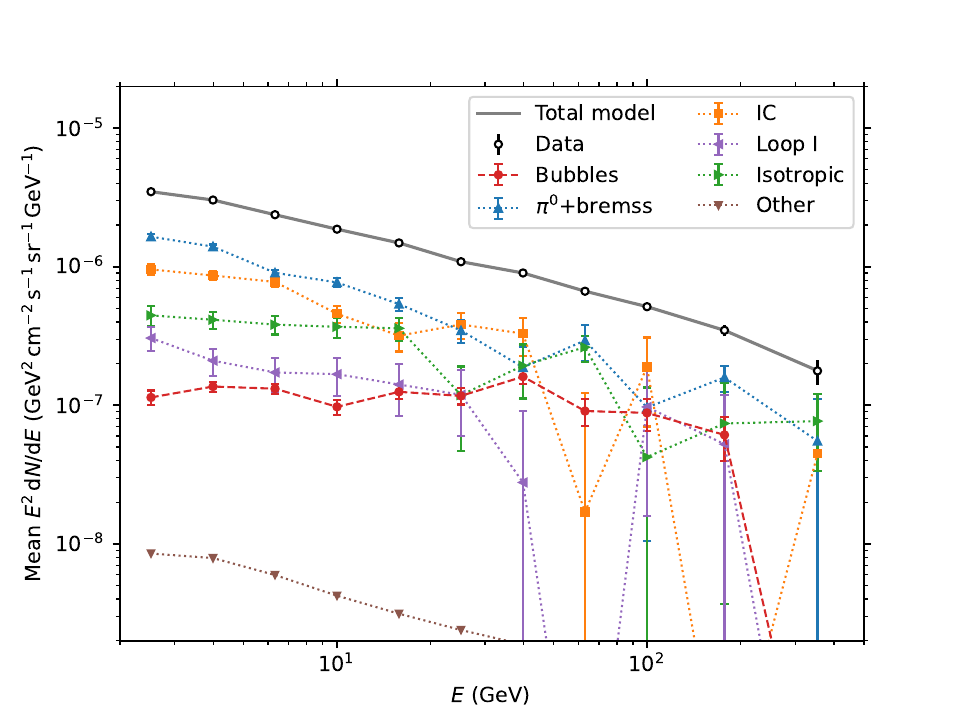}
    \includegraphics[width=0.45\textwidth]{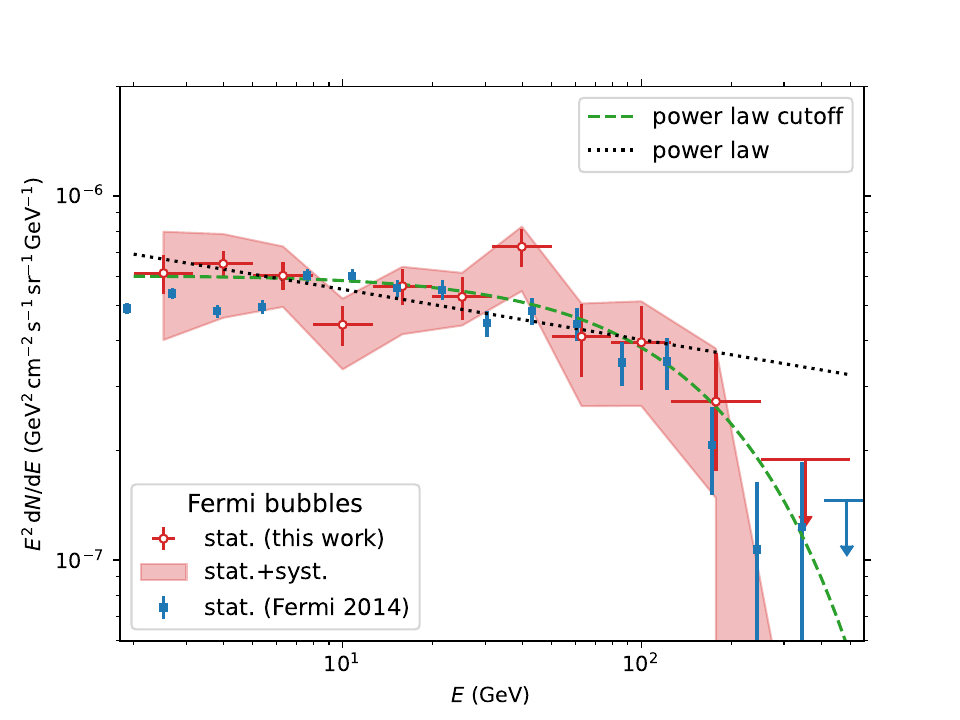}
    \caption{
    {\it Left panel:}
    The average flux of the observed data (black hollow points) and various fitted components in the baseline model (solid points connected with lines) within the ROI.
    The red points show the flux of the bubbles.
    The masked Galactic plane and point sources are excluded from the calculation.
    {\it Right panel:}
    The spectral energy distribution (SED) of the \fbs.
    The red points show the best-fit SED from DAMPE given the baseline model, whereas the blue points show those from \lat~\citep{Ackermann2014}.
    95\% upper limit is given when the TS value is below 9.
    The red band exhibits the total errors from the statistical and systematic uncertainties.
    The optimal single power law model and power law cutoff model are also presented with black dotted and green dashed lines, respectively.
    }
\label{fig::bubble:sed}
\end{figure*}

Considering the slight difference between the residual and the previous templates, we construct a new template based on the DAMPE data.
Instead of deriving the template directly from the residuals of the null model, whose backgrounds may be biased due to not including the \fbs, we adopt the one from~\citet{Su2010a} as the initial template, fit the data, and use the model subtracting the bubbles to calculate the residuals.
The residuals in significance are calculated using the observed and fitted (without bubbles) counts map integrated over energies $2-500~\rm GeV$.
We make a distribution of the values inside and outside the bubble region, denoted as the bubble region and background region, as shown in Figure~\ref{fig::bubble:template} (left).
To determine the threshold for the template of the bubbles, the background histogram is fitted using a Gaussian profile.
We only fit the negative fluctuation in the histogram and assume the mean value of the Gaussian to be zero.
The best-fit Gaussian profile, whose width is $\sigma_{\rm bg} = 0.15$, is shown with a blue dashed line.
We make a map with significance exceeding $2\sigma_{\rm bg}$ as given in Figure~\ref{fig::bubble:template} (right).
The southern bubble is clearly traced in the map, but the northern one is less obvious because of the strong background emission in the northern hemisphere.
Unlike~\citet{Ackermann2014}, which directly defines the template with the significance map above a threshold, we define the bubbles as two simply connected polygons because of the relatively low statistics of DAMPE data.
Two criteria are used to determine the boundaries:
it contains most of the significant excess in the region of the bubbles, and, at the same time, does not greatly deviate from~\citet{Su2010a}.
What we achieve is the flat template enclosed by the boundaries as illustrated with the orange solid line in Figure~\ref{fig::bubble:template} (right) and with the black line in Figure~\ref{fig::bubble:fluxmap}.
The derived lobes closely resemble the previous results (with a slight difference of $\sim 6\%$).

We further inspect the template of the \fbs.
Firstly, we check the residual map given the alternative model, including the new template.
As shown in Figure~\ref{fig::bubble:fluxmap}d, the residual map is almost consistent with the statistical fluctuation, confirming the good fit of the flat template to the data.
A mild excess can be spotted in the eastern part of the southern bubble and is likely due to the Cocoon discovered in the \lat data~\citep{Su2012}. 
Then, we derive the flux near the boundary of the bubbles.
We subtract the best-fit alternative model from the data, add the bubble model back to the residual, and calculate the average flux as a function of angular distance $\phi$ from the edge using the residual map.
The average flux is calculated using $F_{\rm res,>2GeV}=\sum_j [\sum_{i \in {\rm Ring}_k} N_{{\rm res},ij}]/[\sum_{i \in {\rm Ring}_k} \varepsilon_{ij}\Delta \Omega_i]$, where $N_{{\rm res},ij}$ is the residual counts in $i$-th pixel and $j$-th energy bin, ${\rm Ring}_k$ represents the $k$-th angular distance ring.
The residual flux in the high-latitude region with $|b|>20\deg$ is drawn with the blue points in Figure~\ref{fig::bubble:boundflux}.
The edge of the bubbles is sharp.
We find the flux evolution around the edge consistent with a template with a sharp edge convolved with the PSF (green dashed line).
Figure~\ref{fig::bubble:boundflux} also shows a flat residual flux inside the bubbles, which is well consistent with the average flux for the bubbles as a whole (orange band).
To conclude, the sharp edge and the uniform flux support the flat bubbles template.
This template is available
at ScienceDB (\dataset[doi: 10.57760/sciencedb.space.03534]{\doi{10.57760/sciencedb.space.03534}}).

\subsection{Spectral analysis}\label{sec::fbs:sed}

Once we have the new template of the bubbles, we can perform the bin-by-bin analyses to derive the spectral energy distribution (SED).
To increase the statistics, several energy bins are combined together in fittings: two (three) bins below (above) 126~GeV.
In each combined energy bin, the spectral indices of the sources are kept as those from the global fit, while the normalizations are fitted.
The average flux of the components within the ROI is given in the left panel of  Figure~\ref{fig::bubble:sed}.
The bubbles only make up $3\%-10\%$ of the \gr photons in the ROI, but they can be well distinguished from other components due to the unique morphology.
The right panel of Figure~\ref{fig::bubble:sed} presents the SED values and the statistical uncertainties of the bubbles with red points.
They are close to those from \lat~\citep{Ackermann2014} as drawn with blue points, exhibiting a hard spectrum with a cutoff at $\sim 100~\rm GeV$.
The test statistic (TS) of the source is ${\rm TS}\equiv -2 \sum_i \ln(\hat{\mathcal{L}}_{{\rm null},i}/\hat{\mathcal{L}}_{{\rm alt},i}) = 757.4$, where $\hat{\mathcal{L}}_{{\rm null},i}$ and $\hat{\mathcal{L}}_{{\rm alt},i}$ are the likelihood values for the optimal null and alternative models in the $i$-th SED bin.
Since the TS value follows the $\chi^2$ distribution with 11 degrees of freedom in the null hypothesis~\citep{Wilks1938}, the significance of the bubbles is $26.5\sigma$.

\begin{table}[!bt]
    \centering
    \caption{\label{tab::bubble:spec}
        The differential spectrum per unit angle for the \fbs, as in the right panel of Figure~\ref{fig::bubble:sed}.
        The first and second errors are statistical and systematic uncertainties, respectively.
        The third column is the TS values in the energy windows given the baseline background model.
    }
    \begin{tabular}{ccc}
    \hline\hline
    $E$ Range & $E^2 \dnde$                                   & TS value\\
      (GeV)   & $(10^{-7}~\rm GeV\,cm^{-2}\,s^{-1}\,sr^{-1})$ &  \\
    \hline
    $2.00 -	 3.17$ 	&$6.14 	 \pm 0.75 	 ^{+1.68}_{-1.98}$ & $72.6$\\
    $3.17 -	 5.02$ 	&$6.51 	 \pm 0.54 	 ^{+1.23}_{-1.80}$ & $160.4$\\
    $5.02 -	 7.95$ 	&$6.04 	 \pm 0.53 	 ^{+1.11}_{-0.94}$ & $147.5$\\
    $7.95 -	 12.6$ 	&$4.43 	 \pm 0.56 	 ^{+0.56}_{-0.94}$ & $69.9$\\
    $12.6 -	 20.0$ 	&$5.65 	 \pm 0.64 	 ^{+0.36}_{-1.34}$ & $90.2$\\
    $20.0 -	 31.6$ 	&$5.28 	 \pm 0.71 	 ^{+0.50}_{-0.52}$ & $68.6$\\
    $31.6 -	 50.1$ 	&$7.25 	 \pm 0.88 	 ^{+0.45}_{-1.54}$ & $92.6$\\
    $50.1 -	 79.4$ 	&$4.11 	 \pm 0.93 	 ^{+0.23}_{-1.13}$ & $25.2$\\
    $79.4 -	 126$ 	&$3.96 	 \pm 1.03 	 ^{+0.56}_{-0.82}$ & $19.4$\\
    $126 -	 251$ 	&$2.72 	 \pm 0.97 	 ^{+0.49}_{-0.78}$ & $11.0$\\
    $251 -	 500$ 	&$< 1.89$                              & $0.0$ \\
    \hline\hline
    \end{tabular}
\end{table}

We estimate the systematic uncertainty of the spectrum induced by inaccurate background models.
Firstly, the GDE model is tested.
The Galprop GDE model is derived assuming that the CR particles are injected following a CR source distribution, propagate throughout the Galaxy within a cylindrical propagation halo, and interact with gas and photons to produce \grs.
Two CR source distribution tracers are tested, either from supernova remnants~\citep{Case1998} or from pulsars~\citep{Lorimer2006}.
The height of the propagation halo ($z_{\rm h}=4~\rm kpc$ or $z_{\rm h}=10~\rm kpc$) and opacity of \ion{H}{1} gas (opaque with $T_{\rm S}=150~\rm K$ spin temperature or optically thin) are also verified.
Secondly, we substitute the geometric Loop~I template with the map of 408~MHz radio continuum around GC~\citep{Haslam1982,Remazeilles2015}.\footnote{\url{https://lambda.gsfc.nasa.gov/data/foregrounds/haslam_2014/haslam408_dsds_Remazeilles2014.fits}}
The map is further restricted to the region of $|\ell|<70\deg$ and $5\deg<|b|<90\deg$ where the Loop~I is prominent.
Furthermore, the bubble template is also tested.
The bubble templates from~\citet{Su2010a} and~\citet{Ackermann2014} are taken as alternatives.
We construct $8\times2\times3-1=47$ alternative models in total by changing the three factors of uncertainties one by one and repeating the bin-by-bin analyses to achieve the SEDs.
We find the TS value of the bubbles ranges from 521 to 974 in these models.
The systematic uncertainty of the spectrum is defined with the envelope of all the tested models, which is listed in Table~\ref{tab::bubble:spec}.
The red band in Figure~\ref{fig::bubble:sed} (right) shows the root sum square of the statistical and systematic errors.

\begin{figure}
    \centering  
    \includegraphics[width=0.45\textwidth]{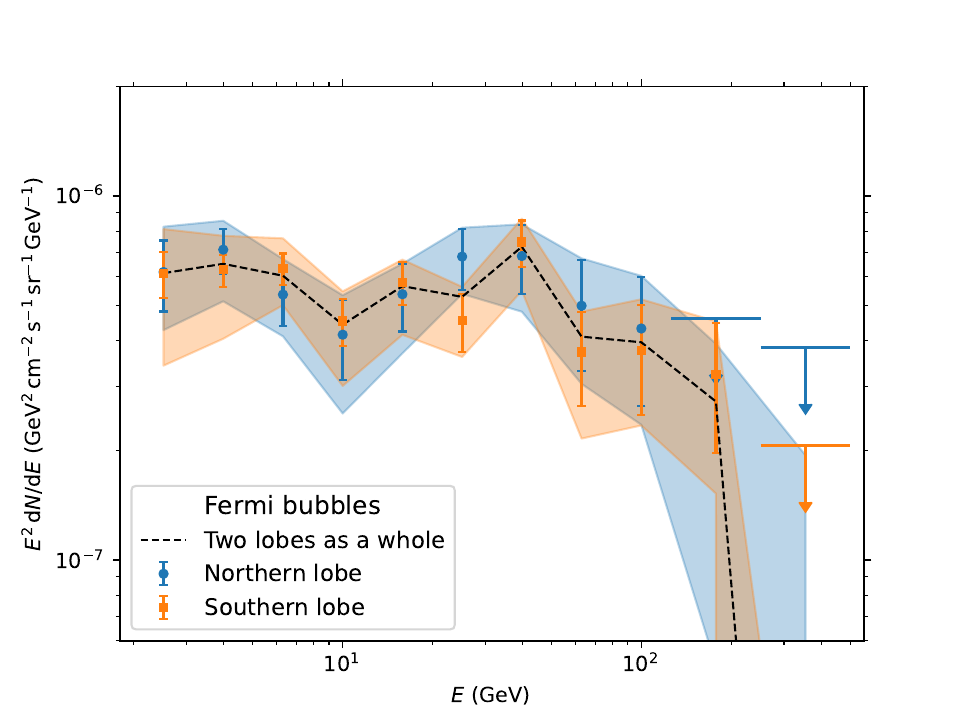}
    \caption{
        The SEDs of the northern (blue) and southern (orange) lobes.
        The points show the baseline spectrum and the statistical uncertainty, whereas the bands give the total errors.
        The black dashed line is the best-fit SED of the whole bubbles.
    }
\label{fig::bubble:sed_ns}
\end{figure}

A curved spectrum is suggested in the SED.
We fit the likelihood profiles in these energy bins with the power law with an exponential cutoff (PLEC) model: $\dnde \propto E^{\Gamma_1} \exp(-E/E_{\rm cut} )$.
The best-fit spectrum in the baseline model is shown with the green dashed line in Figure~\ref{fig::bubble:sed} (right).
Compared with the single power law (PL) model, the TS value difference is $\rm TS_{curv} \equiv TS_{plec}-TS_{pl} = 9.3$, showing the spectrum curved at $3\sigma$ significance~\citep{4FGL2022}.
Taken the systematic uncertainty into account, the optimal spectral index is $\Gamma_1=-1.99 \pm 0.06 {\rm [stat]} ^{+0.10}_{-0.09} {\rm [syst]}$, the cutoff energy is $E_{\rm cut}=204^{+120}_{-60}{\rm [stat]} ^{+63}_{-48}{\rm [syst]}~\rm GeV$, and the integrated intensity above 2~GeV is $ (2.91 \pm 0.15 {\rm [stat]} ^{+0.58}_{-0.69}{\rm [syst]})\times 10^{-7}~\rm ph\,cm^{-2}\,s^{-1}\,sr^{-1}$.
So the luminosity of the bubbles with $|b|>5\deg$ is {$\approx (3.15 \pm 0.17  {\rm [stat]} ^{+0.38}_{-0.56}  {\rm [syst]})\times 10^{37}~\rm erg\,s^{-1}$} above 2~GeV, assuming the distance to the center of each lobe ($|b|=25\deg$) is $R \approx 9.14~\rm kpc$~\citep{Ackermann2014} and the solid angle is $\Omega \approx 0.79~\rm sr$.

\begin{figure*}
    \centering
    \includegraphics[width=0.45\textwidth]{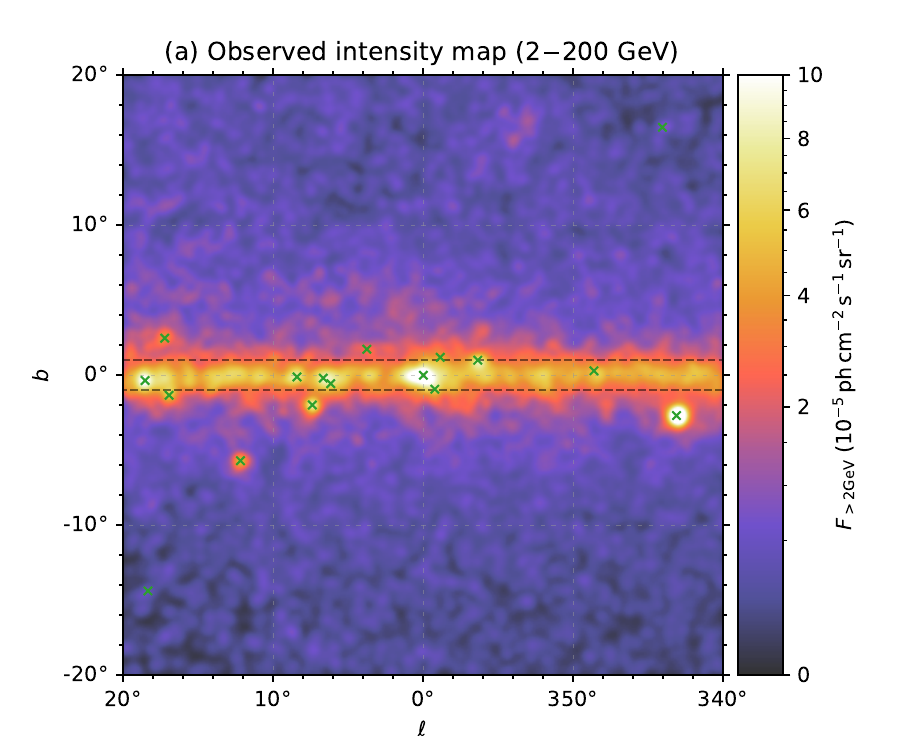}
    \includegraphics[width=0.45\textwidth]{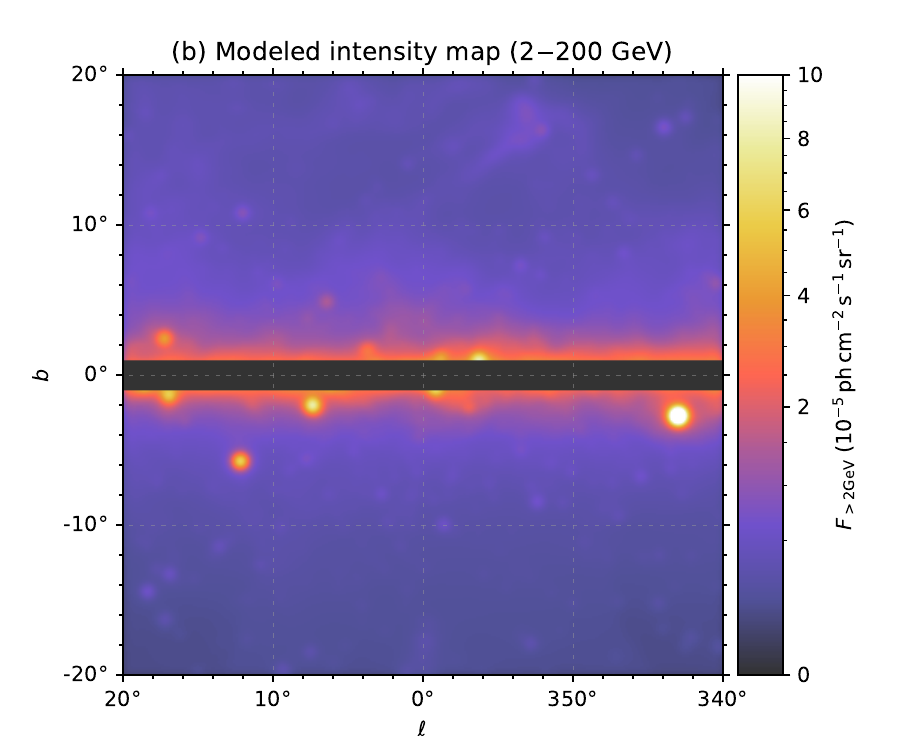}
    \includegraphics[width=0.45\textwidth]{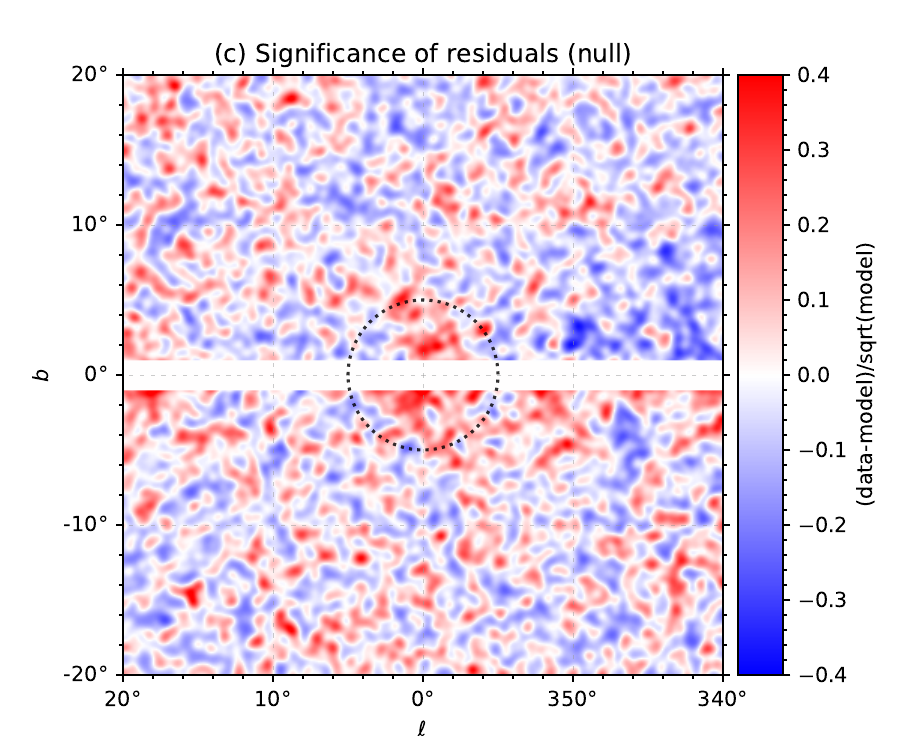}
    \includegraphics[width=0.45\textwidth]{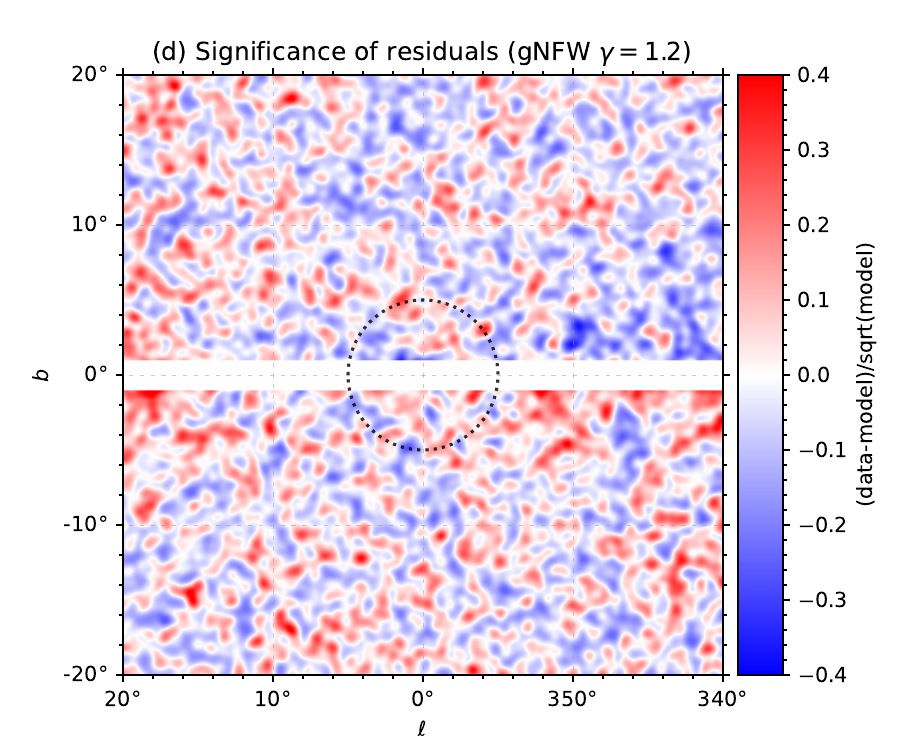}
    \caption{
        {\it Upper panels:}
        The integrated intensity map of the ROI from (a) the observation and (b) the best-fit null model.
        The green crosses represent the point sources in the DAMPE catalog.
        The black dashed line encloses the Galactic plane region ($|b|<1\deg$) that is masked in the likelihood analysis.
        {\it Lower panels:}
        The significance of the residual maps for the model (c) without and (d) with \gce model.
        The black dotted line shows a $5\deg$ circle centered at GC.
        All the maps are smoothed with a $\sigma=0\fdg3$ Gaussian kernel.
    }
\label{fig::gce:fluxmap}
\end{figure*}

We further split the bubbles into two templates and assign the free parameters for each of the lobes.
Similar likelihood analysis is performed, and the derived SEDs are shown in Figure~\ref{fig::bubble:sed_ns}.
The points are the spectra of the lobes given the baseline background model, and the color bands represent the total uncertainty.
The TS values for the northern and southern lobes are 236.8 and 533.5, respectively, corresponding to $14.0\sigma$ and $22.0\sigma$ significance.
The northern lobe is less significant than the other one because the background emission in the northern sky is stronger (Figure~\ref{fig::bubble:fluxmap}b).
We fit the SEDs with the PLEC model.
The spectrum of the northern lobe can be described with the index of $\Gamma_1=-1.95\pm0.12 {\rm [stat]}$ and $E_{\rm cut} = 174^{+220}_{-70} {\rm [stat]}~\rm GeV$,
whereas the southern lobe has the spectrum of $\Gamma_1=-2.01\pm0.08 {\rm [stat]}$ and $E_{\rm cut} = 221^{+190}_{-80} {\rm [stat]}~\rm GeV$.
The spectra of the two lobes are in excellent agreement, supporting the identical physical origin of the two lobes.

\section{Galactic center excess}\label{sec::gce}

\subsection{\gce in the baseline model}
To search for the emission from \gce, we use a smaller ROI, $40\deg$ wide as defined in Section~\ref{sec::data:selection}, than that of the \fbs.
Figure~\ref{fig::gce:fluxmap}a shows the flux map, which is corrected for the exposure and integrated above 2~GeV.
The point sources in the catalog are marked as green crosses.
We firstly fit the data globally with the baseline null model, which contains the gas templates, IC template, Loop I, the bubbles, point sources, weak point sources, and the isotropic emission.
Notably, the baseline \fbs template only has the high latitude part, but we will later incorporate the lower latitude part to evaluate the impact on \gce.
The description of the baseline model can be found in Section~\ref{sec::data:gr_components}.
In the fitting, the mask for the Galactic plane ($|b|\geq 1\deg$) and sources is adopted to reduce the systematic uncertainties on the diffuse emission and point sources.
To reduce correlations between parameters, we use the values and errors of the \fbs in Section~\ref{sec::fbs} as the priors.
The normalizations of the \ion{H}{1} and $\rm H_2$ are assumed to differ by one single free factor throughout the energy range to account for the CO-to-$\rm H_2$ conversion factor $X_{\rm CO}$.
The fitted components are labelled in Table~\ref{tab::templates}.
Figure~\ref{fig::gce:fluxmap}b shows the integrated flux map of the best-fit null model (see Figure~\ref{fig::appx:gce:specgl} for the mean component flux in the global fittings).
Figure~\ref{fig::gce:fluxmap}c presents the significance of the residual given the null model.
An extended excess is visible within the innermost $5\deg$ region.

To test the alternative hypothesis, we model the excess by DM annihilation with a template proportional to the \jf map
\begin{equation}
   J(\ell,b)\equiv \int_{\rm l.o.s.} \rho_{\rm dm}^2(r(\ell,b,s))\,{\rm d}s,
\end{equation}
where $s$ is the line-of-sight (l.o.s.) distance and $r$ is the Galactocentric distance.
The \jf map traces the prompt emission emitted by annihilating DM particles.
We assume the DM follows the generalized Navarro–Frenk–White (gNFW) density profile~\citep{Navarro1996}
\begin{equation}
    \rho_{\rm dm}(r)=\frac{\rho_{0}}{(r/r_{\rm s})^\gamma (1+r/r_{\rm s})^{3-\gamma}}.
\end{equation}
To be compatible with the previous works, we set the scale radius of $r_{\rm s}=20~\rm kpc$ and the inner density slope of $\gamma=1.2$.
The normalization $\rho_0$ is determined by $\rho_{\rm dm}(R_0)=0.4~\rm GeV\,cm^{-3}$~\citep{Catena2010} and $R_0=8.5~\rm kpc$~\citep{Ghez2008}.
The \jf template is shown in Figure~\ref{fig::appx:gce:nfw_bulge}a.
There are indeed updated measurements on local DM density~\citep[e.g.][]{McMillan2017,deSalas2021} and distance to GC~\citep{Keck2019,GRAVITY2024}.
If we take the parameters $(R_0, r_{\rm s}, \rho_{\rm dm}(R_0))=({8.28~\rm kpc}, {25~\rm kpc}, {0.38~\rm GeV\,cm^{-3}})$~\citep{McMillan2017,GRAVITY2024}, the \jf would be decreased by $\sim 22\%$ on average.
But the change can be mostly absorbed into the normalization of the \gce in the fittings.

\begin{figure}
    \centering  
    \includegraphics[width=0.45\textwidth]{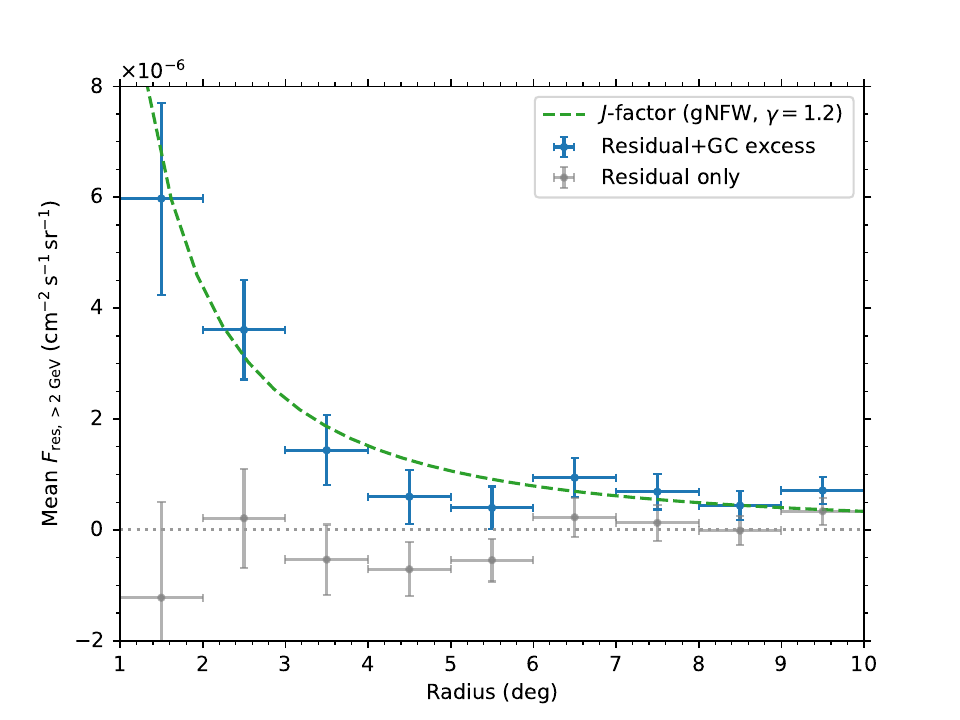}
    \caption{
        The average residual flux in annuli centered at GC.
        The gray points show the residual flux given the alternative model, whereas the blue points show the flux that includes both the residual and \gce.
        The green dashed line represents the flux of the best-fit \jf model.
    }
\label{fig::gce:flux_radius}
\end{figure}

We add the \gce to the model and fit the model to the observed data.
Figure~\ref{fig::gce:fluxmap}d shows the residual map using the alternative model.
The residual in the central $5\deg$ region is well consistent with the statistical fluctuation.
To further visualize the spatial distribution of the excess, we split the residual map according to the angular distance to GC and calculate the average flux within several annuli of width $1\deg$.
In Figure~\ref{fig::gce:flux_radius}, the gray points represent the average flux of the residual for the model including the \gce, which confirms that the residuals are mostly flat.
The weak negative residual in the inner region suggests the intensity of the excess is slightly overstated due to the complex background.
We also re-add the \gce model to the residual map and show the average flux with blue points.
The excess is indeed peaked at GC, and the radial distribution is consistent with the baseline \jf map.

\begin{figure*}
    \centering  
    \includegraphics[width=0.45\textwidth]{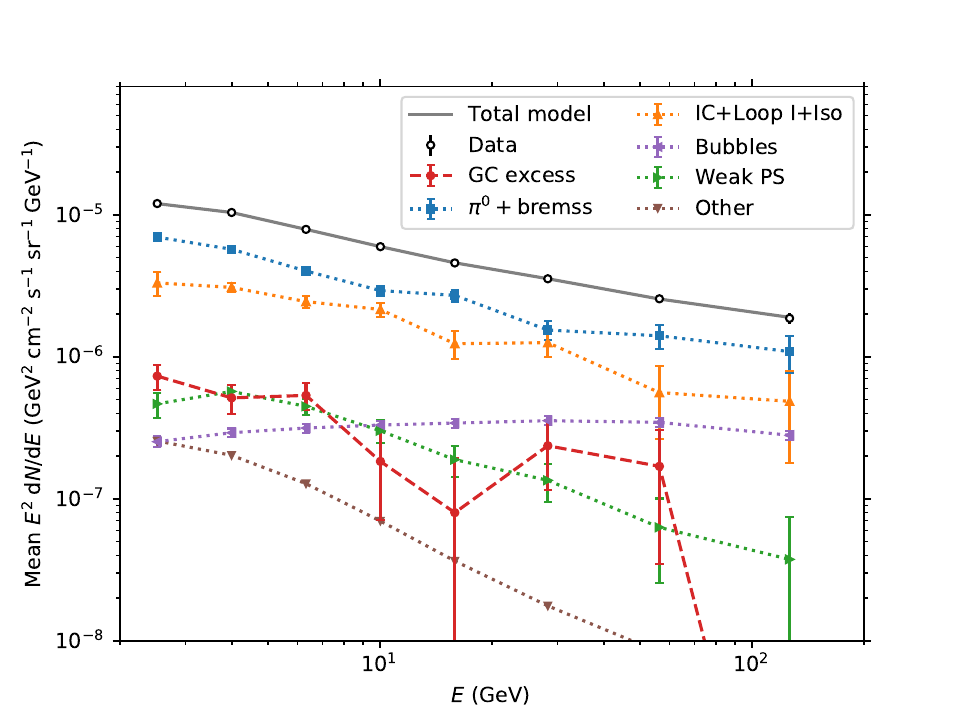}
    \includegraphics[width=0.45\textwidth]{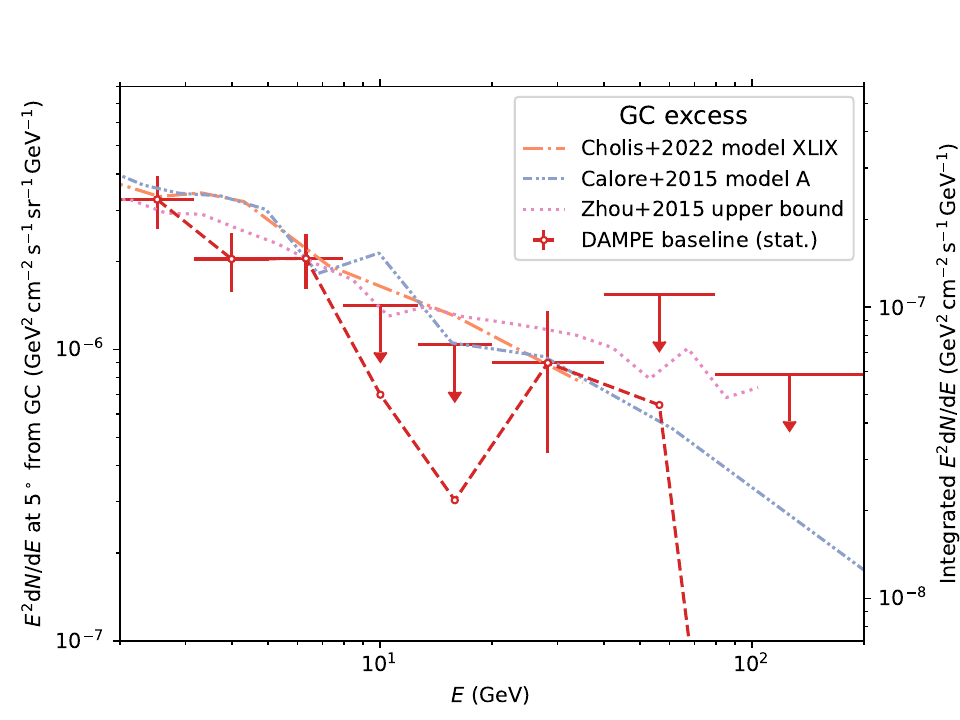}
    \caption{
        {\it Left panel:}
        The mean flux of the observed data and the components in the baseline model within the ROI.
        The red points show the flux of the \gce.
        The masked regions are excluded from the calculation.
        {\it Right panel:}
        The baseline SED of the \gce observed by DAMPE (red points).
        The left $y$-axis represents the SED at an angle of $5\deg$ from GC, whereas the right $y$-axis is the flux integrated over the circle $<10\deg$ from GC excluding $2\deg$ Galactic plane.
        95\% upper limits are shown when the TS values are below 4.
        The red dashed line connects the best-fit flux points.
        The \gce follows the \jf template given the gNFW profile with $\gamma=1.2$.
        Some of the \lat observations are also shown with lines.
    }
\label{fig::gce:sed_gce}
\end{figure*}

Bin-by-bin analysis is conducted to derive the SED of \gce.
Several energy bins are combined together to increase statistics: two bins below 20~GeV, three bins between 20~GeV and 80~GeV, and four bins above 80~GeV.
Based on the global-fit model, we fix the spectral indices, tie the normalizations of \ion{H}{1} and $\rm H_2$ together while keeping the ratio, and fit the parameters in each combined energy bin.
The left panel of Figure~\ref{fig::gce:sed_gce} shows the mean flux of the \gr components in the ROI.
On average, the \gce contributes more photons than \fbs and point sources below $\sim \rm 8~GeV$.
But it is still much weaker than the Galactic diffuse emission and therefore is prone to the systematics of the background model.
The SED of the \gce at $5\deg$ from GC, given the baseline background model, is shown in the right panel of Figure~\ref{fig::gce:sed_gce} and listed in Table~\ref{tab::gce:spec}.
The DAMPE results are in good agreement with the spectra extracted from \lat data~\citep{Zhou2015,Calore2015b,Cholis2022}.\footnote{
    \citet{Calore2015b} and~\citet{Cholis2022} present the spectra of \gce averaged over the ROI of $|\ell|\leq 20\deg$ and $2\deg\leq|b|\leq20\deg$, so $4.41$ is multiplied to their SEDs to convert to the ones measured at $5^\circ$ from GC.
    \citet{Zhou2015} shows the mean SED within $10\deg$ of GC, hence a conversion factor of $0.766$ is adopted.}
A slight tension might present in the energy range of $8-20$~GeV since the upper limits are below the observed flux from \lat.
The TS value for the \gce in the whole energy range is 80.1, showing $7.5\sigma$ significance given eight degrees of freedom~\citep{Wilks1938}.
A weak excess with a TS value of 6.3 is presented between 20~GeV and 80~GeV, which might suggest a high-energy tail or is merely a statistical fluctuation.
If we only account for the first five energy bins ($2-20~\rm GeV$), which covers the main spectral component of \gce, the TS value decreases to 73.8, but the significance increases to $7.7\sigma$.

\begin{table}[!bt]
    \centering
    \caption{\label{tab::gce:spec}
        The SED of the \gce at an angle of $5\deg$ from GC given the baseline background model, as in the right panel of Figure~\ref{fig::gce:sed_gce}.
        The error bars only show the statistical uncertainties.
        95\% upper limits are given when the TS values are below 4.
        The third column is the TS values in the energy windows.
    }
    \begin{tabular}{ccc}
    \hline\hline
    $E$ Range & $E^2 \dnde$                                   & TS value\\
      (GeV)   & $(10^{-6}~\rm GeV\,cm^{-2}\,s^{-1}\,sr^{-1})$ &  \\
    \hline
    $2.00 -	 3.17$ 	&$3.27 	 \pm 0.67$ & $25.3$\\
    $3.17 -	 5.02$ 	&$2.05 	 \pm 0.47$ & $21.0$\\
    $5.02 -	 7.96$ 	&$2.05 	 \pm 0.44$ & $24.2$\\
    $7.96 -	 12.6$ 	&$< 1.42$ 	       & $2.9$ \\
    $12.6 -	 20.0$ 	&$< 1.04$ 	       & $0.5$ \\
    $20.0 -	 39.9$ 	&$0.90 	 \pm 0.46$ & $4.5$\\
    $39.9 -	 79.6$ 	&$< 1.55$ 	       & $1.8$ \\
    $79.6 -	 200$ 	&$< 0.82$ 	       & $0.0$ \\
    \hline\hline
    \end{tabular}
\end{table}

\subsection{Morphological study}
\begin{figure}[!hbt]
    \centering
    \includegraphics[width=0.45\textwidth]{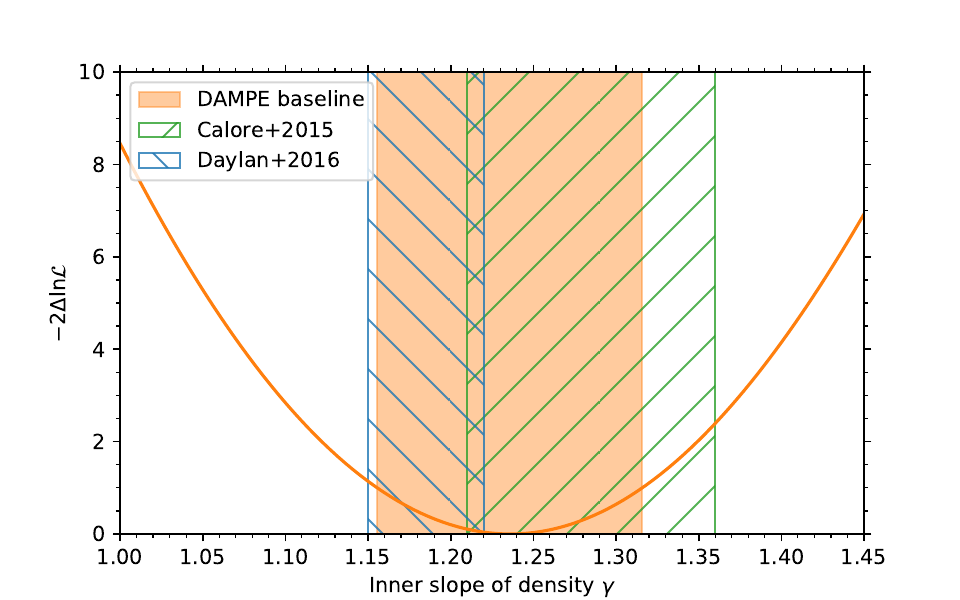} 
    \includegraphics[width=0.45\textwidth]{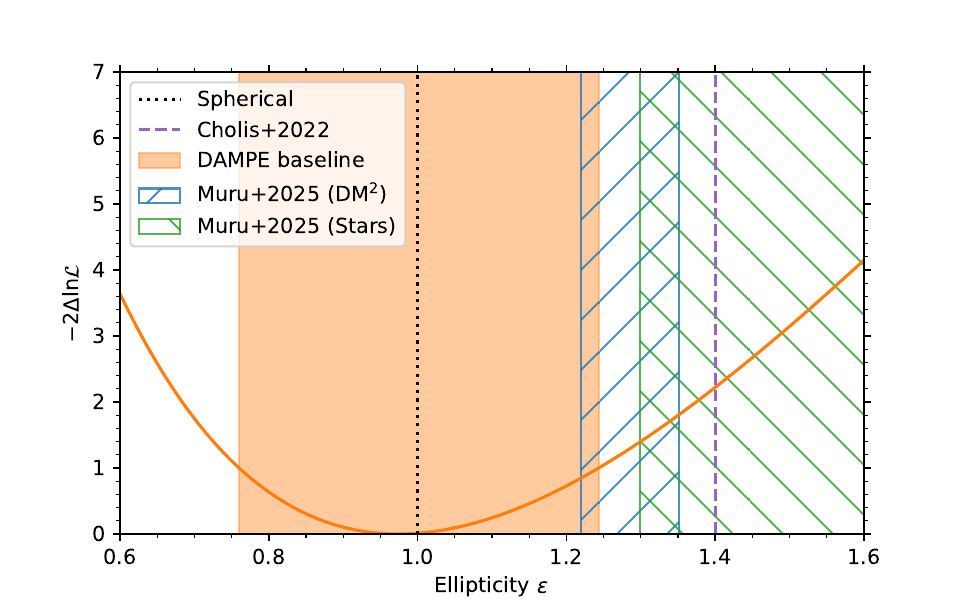} 
    \includegraphics[width=0.45\textwidth]{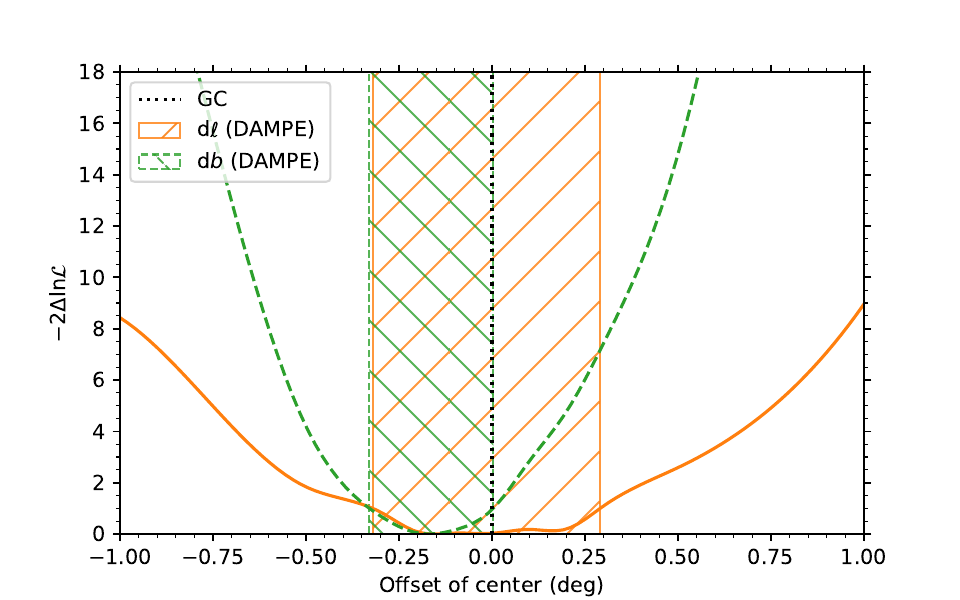} 
    \caption{
        The change of log-likelihood value as a function of the steepness (upper), the ellipticity (middle), and the center position (lower) of the DM density profile derived with the $2-20~\rm GeV$ data assuming the baseline background model.
        In the lower panel, the orange solid and green dashed lines show the likelihood variation versus the Galactic longitude and latitude, respectively.
    }
\label{fig::gce:gnfw_morph}
\end{figure}

\begin{figure*}[!hbt]
    \centering  
    \includegraphics[width=0.32\textwidth]{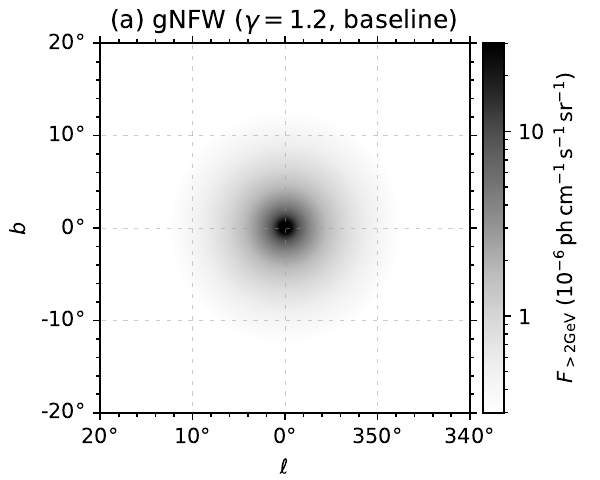}
    \includegraphics[width=0.32\textwidth]{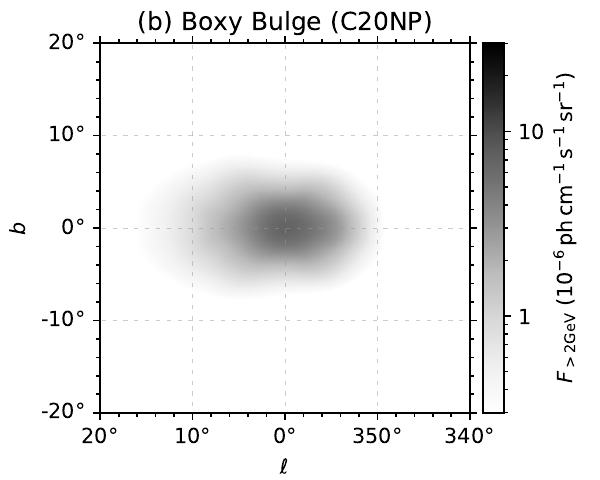}
    \includegraphics[width=0.32\textwidth]{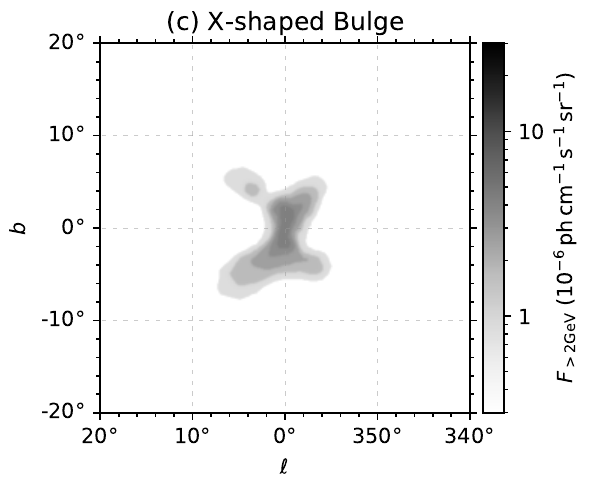}
    \caption{
        Several spatial models for the \gce: (a) the \jf map given the gNFW profile with inner slope of $\gamma=1.2$, (b) the non-parameteric bulge model~\citep{Coleman2020} and (c) the X-shaped bulge map~\citep{Macias2018,McDermott2023}.
        The intensity of the each component is based on the best-fit model given the \gce template.
        We do not convolve the maps with PSF.
    }
\label{fig::appx:gce:nfw_bulge}
\end{figure*}

The morphology of the \gce can help unveil its nature, which is extensively studied in \lat observations~\citep{Macias2018,Macias2019,Bartels2018,DiMauro2021,McDermott2023,Zhong2024,Song2024,Ramirez2025}.
If the excess originated from DM annihilation, its spatial shape would be brighter in the center and more spherical than the MSP origin.
In this subsection, we will compare the excess with the gNFW profile and then test various bulge templates.
We adopt the same backgrounds in the baseline model in the analysis since it is among the best-fitted models to the data.

We change the inner density slope of the gNFW profile $\gamma$, calculate the \jf template, and repeat the bin-by-bin analysis in the energy range from 2~GeV to 20~GeV.
The upper panel of Figure~\ref{fig::gce:gnfw_morph} demonstrates the difference of the likelihood value $-2\Delta \ln (\mathcal{L})\equiv -2\ln[\mathcal{L}(\gamma)/\mathcal{L}(\gamma_{\rm best})]$ versus the inner slope $\gamma$, where $\mathcal{L}$ is the product of the likelihood values for the lowest five energy bins, and $\gamma_{\rm best}$ is the optimal density slope.
The orange band in the figure shows the $1\sigma$ statistical uncertainty of $\gamma$.
Given the baseline background model, the slope is found to be $\gamma=1.23\pm0.08$ with DAMPE data.
Such a density steeper than the regular NFW profile ($\gamma=1.0$) can be reasonable for MW due to the contraction of DM halo alongside the accretion of baryons at GC~\citep{Cantun2020}.
Our result also agrees with the \lat observations very well.
Shown with the hatched bands are the $1\sigma$ errors from~\citet{Daylan2016} and~\citet{Calore2015b}.
Recent analyses~\citep{DiMauro2021,Cholis2022} confirm the preference for the contracted NFW profile with $\gamma\approx1.2-1.3$.

The ellipticity of the \gce is also analyzed.
Following~\citet{Cholis2022}, we deform the opening angle from GC $\psi$ by the ellipticity parameter $\varepsilon$ such that $\cos(\psi)=\cos(b)\cos(\ell/\varepsilon)$. 
The spherical symmetric gNFW model corresponds to $\varepsilon=1$.
The template is elongated along the Galactic plane for $\varepsilon>1$, whereas perpendicular to the plane for $\varepsilon<1$.
The inner density slope is kept to $\gamma=1.2$.
The middle panel of Figure~\ref{fig::gce:gnfw_morph} demonstrates the likelihood versus the ellipticity.
Our analysis slightly favors the spherical shape of the excess given the best-fit ellipticity of $\varepsilon=1.0^{+0.3}_{-0.2}$, which is consistent with the baseline model of~\citet{Cholis2022} (purple dashed line) within $2\sigma$ uncertainty.
A recent simulation shows the DM density profile of MW-like galaxies is flattened along the Galactic plane~\citep{Muru2025}.
In their simulations, the minor-to-major axis ratios for \jf range from 0.74 to 0.82 in the inner $\sim 5\deg$ region, which corresponds to $\varepsilon=1.22-1.35$ as shown with the blue hatch in the figure.
On the other hand, the morphology is slightly more elongated with the axis ratios of $0.54-0.77$ ($\varepsilon=1.30-1.85$).
Our baseline result is compatible with the ellipticity of \jf in the simulation within $1\sigma$ uncertainty.

\begin{table}[!bt]
    \centering
    \caption{\label{tab::gce:excess_morph}
        Comparison of the \gce models.
        The second (third) column is the TS value of the excess for the data from 2~GeV to 200~GeV (20~GeV).
        The baseline background model described in Section~\ref{sec::data:gr_components} is adopted.
    }
    \begin{tabular}{lcc}
    \hline\hline
    Excess model        & $\rm TS_{2-200}$ & $\rm TS_{2-20}$ \\
    \hline
    gNFW ($\gamma=1.2$) & $80.1$ & $73.8$\\
    X-shaped bulge      & $66.4$ & $62.8$\\
    Boxy bulge (F98S)   & $71.0$ & $64.7$\\
    Boxy bulge (C20NP)  & $77.8$ & $70.8$\\
    \hline\hline
    \end{tabular}
\end{table}

The central position of the DM template is also tested.
We shift the center of the \jf template along the Galactic plane and perpendicular to the plane with an angle, conduct the same bin-by-bin analysis between 2~GeV and 20~GeV, and calculate the change of likelihood value with the angle.
In this analysis, we assume the spherical gNFW density profile with $\gamma=1.2$.
The result is demonstrated in the lower panel of Figure~\ref{fig::gce:gnfw_morph}, where the solid orange and dashed green lines are for the models with the longitude and latitude of the center changed, respectively.
The optimal longitude of the center is $\dd \ell=-0\fdg1^{+0\fdg4}_{-0\fdg2}$ given the latitude of $\dd b=0\deg$; while the optimal latitude is $\dd b=-0\fdg2 \pm 0\fdg2$ given $\dd \ell=0\deg$.
The center of the excess is slightly offset from the dynamic center of GC, located at its southwest.
Similar offset with an angle of $\sim 0\fdg1$ was reported in \lat observations as well~\citep{DiMauro2021,Ackermann2017}.
Nevertheless, the offset is not significant in the DAMPE data.

Finally, we substitute the \jf templates with the bulge models, which are considered the tracer of MSPs.
We adopt the bulge/bar model (model S) developed in~\citet{Freudenreich1998} (F98S), the X-shaped bulge~\citep{Macias2018} from~\citet{McDermott2023},\footnote{\url{https://github.com/samueldmcdermott/gcepy}} and the non-parameterized bulge model from~\citet{Coleman2020} (C20NP).\footnote{\url{https://github.com/chrisgordon1/galactic_bulge_templates}}
Some of the bulge models are shown in Figure~\ref{fig::appx:gce:nfw_bulge}.
We use these bulge models as the template of the excess and repeat the bin-by-bin analyses in the energy range of $2-200~\rm GeV$.
The resultant TS values are given in Table~\ref{tab::gce:excess_morph}.
Since they share the same null model, the template with a larger TS value also has a larger likelihood value and thereby fits the data better.
The C20NP bulge model is the best among the three bulge models, showing a TS value of $77.8$, but it is worse than the baseline \jf map produced from the gNFW profile.
The preference is still valid if we restrict the energy range to $2-20~\rm GeV$.
The result is not surprising since the spherical shape of the \gce is slightly preferred as shown previously.

\subsection{Systematic uncertainty}\label{sec::gce:sys}
\begin{figure}[!hbt]
    \centering
    \includegraphics[width=0.45\textwidth]{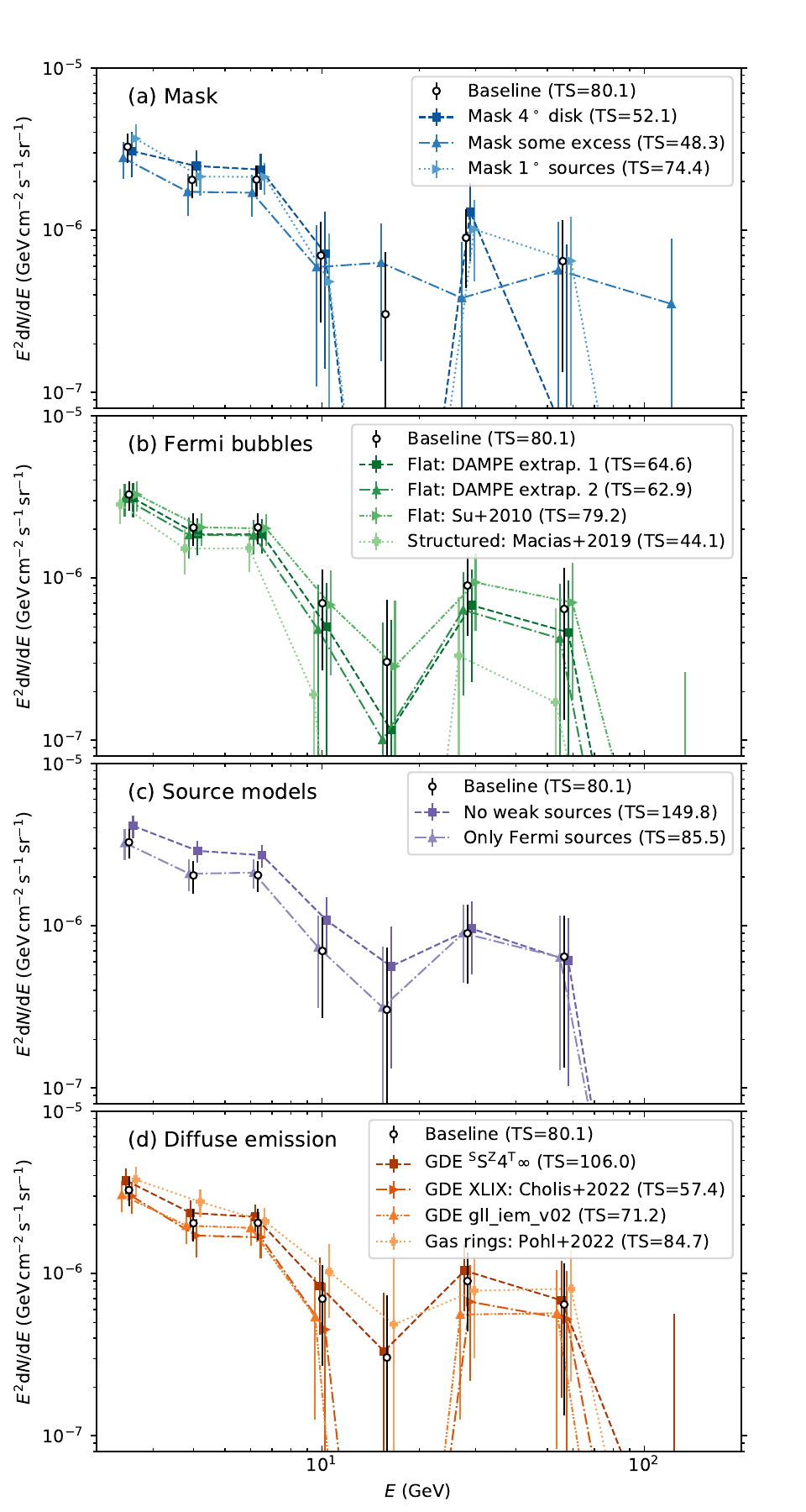}
    \caption{
        The systematic uncertainty of the \gce caused by the analysis procedures and the background models, including (a) the mask, (b) the \fbs template, (c) the point/extended sources, and (d) the GDE models.
        The points show the SEDs and the statistical uncertainties measured at $5\deg$ from GC.
        The TS values in the labels are for the $2-200~\rm GeV$ data.
        The central energies are slightly offset to avoid overlapping.
    }
\label{fig::gce:spec_sys1}
\end{figure}

Though the presence of \gce is quite robust, its flux and spectrum are significantly influenced by systematic uncertainties as revealed in the data analysis of \lat ~\citep[e.g.][]{Zhou2015,Calore2015b,Ackermann2017,Cholis2022}.
In this subsection, we will discuss the impact of the mask, the \fbs template, the source model, and the GDE templates on the observation of the \gce with DAMPE.
We still adopt \jf map with gNFW density slope of $\gamma=1.2$ as the template of the \gce.
The main results are given in Figure~\ref{fig::gce:spec_sys1} and Table~\ref{tab::gce:excess_sys}.

\begin{figure*}[!hbt]
    \centering
    \includegraphics[width=0.95\textwidth]{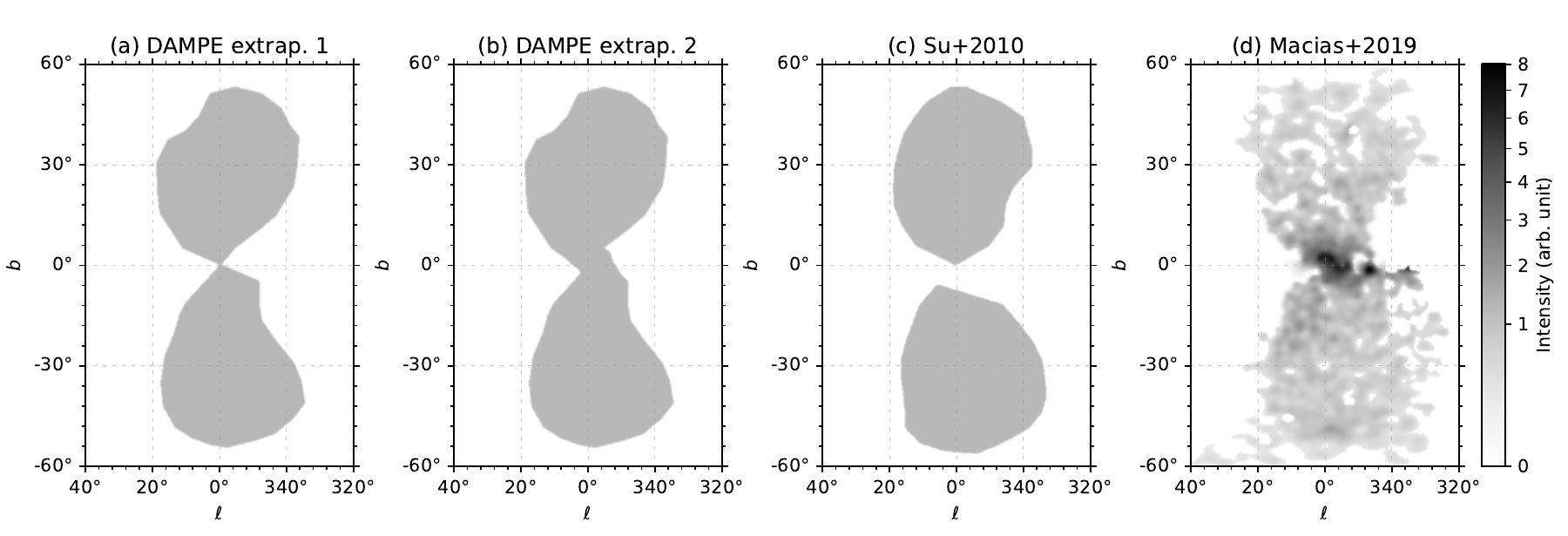}
    \caption{
        Alternative bubbles templates with low-latitude component for evaluating the systematic uncertainty of the \gce:
        (a--b) the extrapolated maps from the baseline template and (c--d) the bubbles derived from \lat observations.
    }
\label{fig::gce:fbs_templs}
\end{figure*}

Firstly, we test three types of masks in order to understand the impact of the Galactic plane, the excess in the residual map, and the events from point sources spilled out of the mask.
In the first case, we extend the Galactic plane mask to $|b|<2\deg$.
In the second case, we exclude the region with $340\deg<\ell<355\deg$ and $|b|<10\deg$ where some excess and deficit exist in the residual map (see Figure~\ref{fig::gce:fluxmap}d).
Such structures are also presented in \lat~\citep[e.g.][]{Calore2015b}.
In the last case, we adopt a larger point source mask with a radius of $1\deg$.
The resultant spectra of the \gce are shown in Figure~\ref{fig::gce:spec_sys1}a along with the TS values listed in the label.
In all three cases, the significance decreases a little.
For cases 1 and 3, the decrease is likely caused by the removal of the signal region since their SEDs are close to or even higher than the baseline.
However, for case 2, the spectrum is lower than the baseline, so the low TS value of $\sim 48.3$ is induced by the absorption of the \gce by the background.
Interestingly, the SED point at $\sim 16~\rm GeV$ grows a bit in this case, so the weak disagreement with the \lat can be systematic origin.

\begin{table}[!bt]
    \centering
    \caption{\label{tab::gce:excess_sys}
        Comparison of the \gce in various background models.
        The second column is the log-likelihood values with the gNFW ($\gamma=1.2$) model of \gce using the $2-200~\rm GeV$ data.
        The third and fourth columns are the TS values of the excess given the gNFW and boxy bulge (C20NP) templates, respectively.
        The baseline mask is adopted throughout the analyses.
    }
    \begin{tabular}{l|ccc}
    \hline\hline
    Background & $\ln(\mathcal{L}_{\rm gNFW})$ & $\rm TS_{gNFW}$ & $\rm TS_{C20NP}$\\
    \hline
    Baseline                & $-129068.8$ & $80.1$  & $77.8$ \\
    \hline
    Bubbles {\tt extrap.1}  & $-129066.3$ & $64.6$  & $61.0$ \\
    Bubbles {\tt extrap.2}  & $-129065.1$ & $62.9$  & $57.6$ \\
    Bubbles Su+2010         & $-129085.8$ & $79.2$  & $69.2$ \\
    Bubbles Macias+2019     & $-129070.0$ & $44.1$  & $37.3$ \\
    \hline
    No weak sources         & $-129248.1$ & $149.8$ & $153.5$\\
    Only Fermi sources      & $-129083.6$ & $85.5$  & $78.8$ \\
    \hline
    GDE $\rm ^SS^Z4^T\infty$& $-129069.5$ & $106.0$ & $101.1$\\
    GDE XLIX                & $-129130.3$ & $57.4$  & $69.4$ \\
    GDE {\tt gll\_iem\_v02} & $-129225.3$ & $71.2$  & $103.6$\\
    GDE Pohl+2022           & $-129155.6$ & $84.7$  & $120.1$ \\
    \hline\hline
    \end{tabular}
\end{table}

The low-latitude \fbs also affects the spectrum of the \gce.
However, due to the low statistics and complex background, it is hard for DAMPE to make such a template.
We test four bubbles templates with low-latitude structure for the systematic uncertainty (Figure~\ref{fig::gce:fbs_templs}).
Two templates are based on the \lat observation:
one is the flat template extracted from the residual map in~\citet{Su2010a} (Figure~\ref{fig::gce:fbs_templs}c),
and the other one is the structured map~\citep{Macias2019} after applying the Laplace inpainting algorithm to the hard component in the residual~\citep{Ackermann2017} (Figure~\ref{fig::gce:fbs_templs}d).
Two other templates are extrapolated from the baseline \fbs templates extracted in Section~\ref{sec::fbs:template}.
In the first case (denoted as {\tt extrap.1}), we
connect the base of bubbles with two lines to make a intersection point, and define the low-latitude bubbles with the northern and southern parts among the four regions separated by the lines (see Figure~\ref{fig::gce:fbs_templs}a).
In the second case ({\tt extrap.2}), the low-latitude template is extracted based on the shape of the inpainted \fbs template.
But different from the latter one, we exclude some excess along the disk and assume the template is uniform (Figure~\ref{fig::gce:fbs_templs}b).
The baseline \fbs template is substituted, and the same fittings are performed.
The results are shown in Figure~\ref{fig::gce:spec_sys1}b.
The extrapolated bubbles templates fit the data slightly better than the baseline template.
For the {\tt extrap.1} bubbles template, the log-likelihood increase is 2.4 and the TS value for \gce is 64.6 ($6.5\sigma$);
whereas for {\tt extrap.2} template, the log-likelihood increase is 3.6 and the TS value is 62.9 ($6.4\sigma$).
The spectra of the \gce are also very similar in the two cases.
On the other hand, the two bubbles templates from the \lat perform worse than the baseline.
The log-likelihood decreases are 3.3 and 18.1 for the structured map~\citep{Macias2019} and flat map~\citep{Su2010a}, respectively.
The structured bubbles template is also peaked at the GC, so it may absorb the emission from the \gce, causing a smaller TS value of 44.1 ($5.0\sigma$).

Furthermore, the effect of point and extended sources is checked.
We either remove the weak source template from the model or replace all the point/extended sources with the ones defined in the \lat 14-yr catalog~\citep{Ballet2023_4FGLDR4} and then fit them as a whole.
As shown in Table~\ref{tab::gce:excess_sys}, the fitting deteriorates if the weak sources are not accounted for.
However, using the \lat source catalog is statistically slightly better than the baseline model according to the Akaike information criterion~\citep{Akaike1974}, considering the baseline model has $15\times 2 = 30$ more free spectral parameters for the sources.
The corresponding \gce spectra are presented in Figure~\ref{fig::gce:spec_sys1}c.
The low-energy part of the spectrum can be affected by the weak sources, as illustrated in the first case.
It is because the weak sources are concentrated at the center and can be absorbed by the \gce if this component is not accounted for.
However, the sources that are even unresolved by \lat can not be verified here.
We also find that altering the source catalog makes little impact on the SED.

Finally, the Galactic diffuse emission models are inspected.
We show four models as representatives.
The first one is from \lat 1SC~\citep{1SC2016}, labelled as $\rm ^SS^Z4^T\infty$, which assumes the CR source distribution traced by supernova remnants, the height of the propagation halo of 4~kpc, and the optically thin \ion{H}{1} gas.
The second one is the {\tt Galprop}-based GDE model XLIX developed in~\citet{Cholis2022}.
They fitted the injection and propagation parameters to the AMS-02 CR data~\citep[e.g.][]{Aguilar2018} and generated 80 alternative CR parameter sets and \gr templates.
They also accounted for different tracers of the CR electron and proton sources.
The model XLIX was reported as the best fit to the \lat data among all the alternatives.
We take the parameter set~\citep{Cholis2022_params}, calculate the Galactic diffuse emission templates using {\tt Galprop v54.1.984}~\citep{Strong:1998pw}, and combine the rings together to make the three components the same as the baseline model.
The third one is a data-driven model {\tt gll\_iem\_v02} from \lat,\footnote{\url{https://fermi.gsfc.nasa.gov/ssc/data/access/lat/BackgroundModels.html}}
which is often adopted in the analysis of the \gce~\citep[e.g.][]{Daylan2016}.
It consists of the hadronic and bremsstrahlung emission traced by gas column densities in six Galactocentric rings and the IC emission from {\tt Galprop}~\citep{gll_iem_v02}, so we replace the three Galactic diffuse emission components with this model.
The final model is taken from~\citet{Pohl2022,Pohl2022_data}.
They provide new gas column density maps split into four rings and the IC templates considering both the CR sources along the spiral arms~\citep{Johannesson2018} and the 3D distribution of ISRF~\citep{Porter2017}.
We keep the four gas rings for \ion{H}{1} and $\rm H_2$, assign power-law scales to them, and tie the spectral indices of the neutral and molecular gas templates in each ring together.
For the IC components, we add all six rings to make a single IC template.
The positive dust residuals are also included in the fitting as an independent component.
All the fitted spectra are shown in Figure~\ref{fig::gce:spec_sys1}d.
The latter three GDE models present worse fitting than the baseline model, with the log-likelihood differences larger than 60,
while the first model fits the data similarly well to the baseline model.
We check the residual maps of the three models and find slightly more positive residuals along the southern Galactic plane.
In all these cases, the \gce shows significant emission with the lowest TS value of 57.4 for the GDE model XLIX, corresponding to the significance of $6.0\sigma$.
The low-energy component below $\sim 10~\rm GeV$ and high-energy tail around $40~\rm GeV$ persist, while the dip at $\sim 16~\rm GeV$ is prone to the choice of the GDE model.

We also repeat the systematic analysis assuming the C20NP bulge template of the \gce (Figure~\ref{fig::appx:gce:nfw_bulge}b).
The TS values are presented in the fourth column of Table~\ref{tab::gce:excess_sys}.
The Galactic diffuse emission is a major systematic factor for the morphological study of the excess.
If the baseline GDE model ($\rm ^SL^Z10^T150$) is adopted, the bulge template is generally less favorable than the gNFW template.
The preference persists given the GDE model $\rm ^SS^Z4^T\infty$, whose log-likelihood ranks second among the five GDE models.
But the C20NP bulge model is preferred in the remaining three models.
For the last two GDE models in the table, the bulge are much better than the gNFW.
It is because more positive residue exists along the Galactic plane compared to the others, which is absorbed by the box bulge template.
We further test the X-shape bulge (Figure~\ref{fig::appx:gce:nfw_bulge}c).
The TS values of the bulge are $78.8$, $56.1$, $64.3$, and $69.3$ for the four GDE models, respectively.
The X-shaped bulge is worse than the gNFW template in these cases.

\subsection{Implications on DM parameters}

If the \gce originates from the DM annihilation, we can set constraints on the DM parameters.
The spectrum of the prompt emission emitted by annihilating DM particles can be calculated with
\begin{equation}\label{eqn::gce:dmspec}
    \frac{{\rm d}N}{{\rm d}E}= \frac{1}{4\pi}\frac{\left< \sigma v\right>_f}{2 m_\chi^2} \frac{{\rm d}N_f}{{\rm d}E} \times J,
\end{equation}
where $m_\chi$ is the mass of a DM particle, $\left< \sigma v\right>_f$ is the velocity-averaged annihilation cross section in the channel $f$.
They are the two free parameters in the fittings.
${{\rm d}N_f}/{{\rm d}E}$ is the DM prompt \gr spectrum per annihilation, which is interpolated from the tabulated data in PPPC4~\citep{PPPC2011}.
$J$ is the \jf.
Since the spectrum is normalized at $5\deg$ from GC, the \jf should also be at $\psi=5\deg$.
In this work, we focus on two annihilation channels: $\chi\chi \to b\bar{b}$ and $\chi\chi \to \tau^+\tau^-$.
We fit the DM spectrum to the SED of the \gce.
The likelihood profiles in all eight energy bins are adopted, similar to the Section~\ref{sec::fbs:sed}.

\begin{figure}[!bt]
    \centering
    \includegraphics[width=0.45\textwidth]{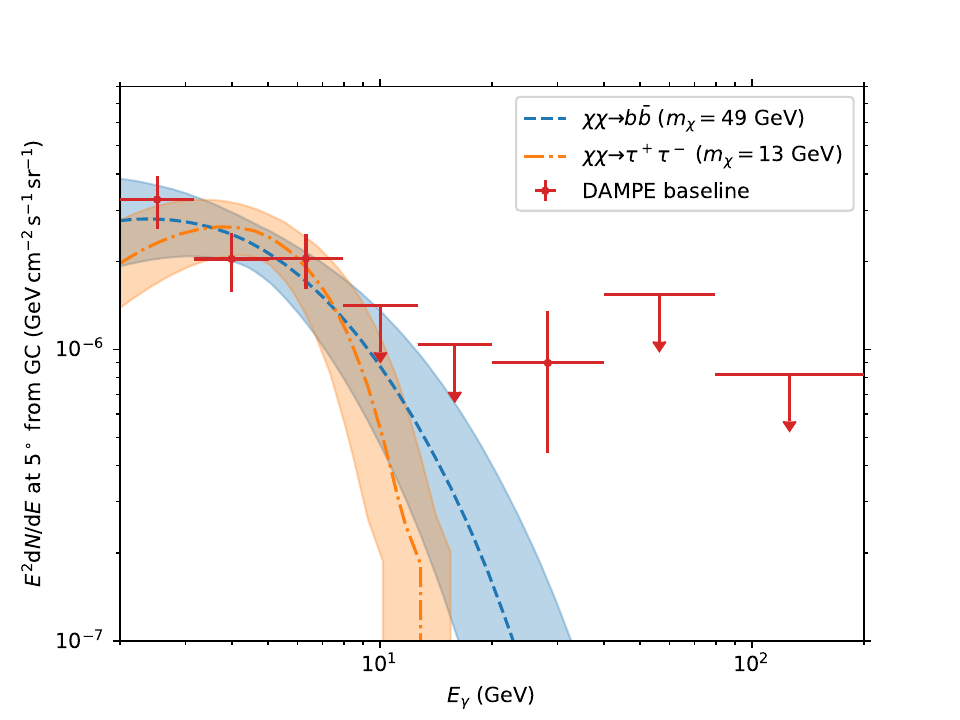}
    \caption{
        The SED of the \gce (red points) for the baseline background and the best-fit DM annihilation models.
        The blue dashed and orange dot-dashed lines correspond to the DM annihilation channel of $\chi\chi \to b\bar{b}$ and $\chi\chi \to \tau^+\tau^-$, respectively.
        The colored bands show the $1\sigma$ statistical uncertainties of the DM models.
    }
\label{fig::gce:dmfit}
\end{figure}

\begin{figure}[!bt]
    \centering
    \includegraphics[width=0.45\textwidth]{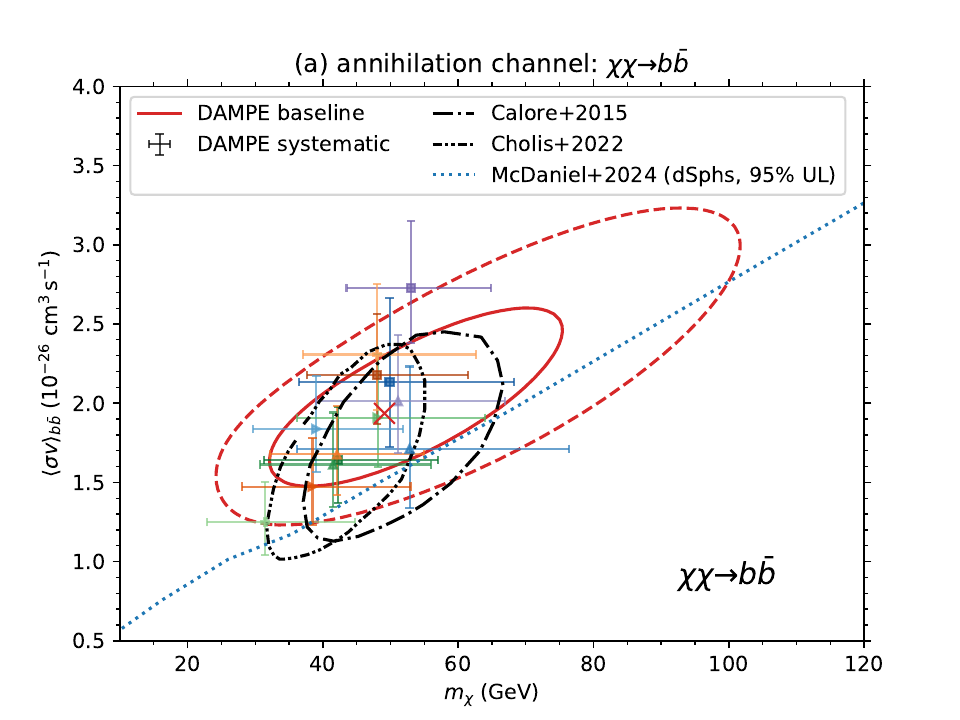}
    \includegraphics[width=0.45\textwidth]{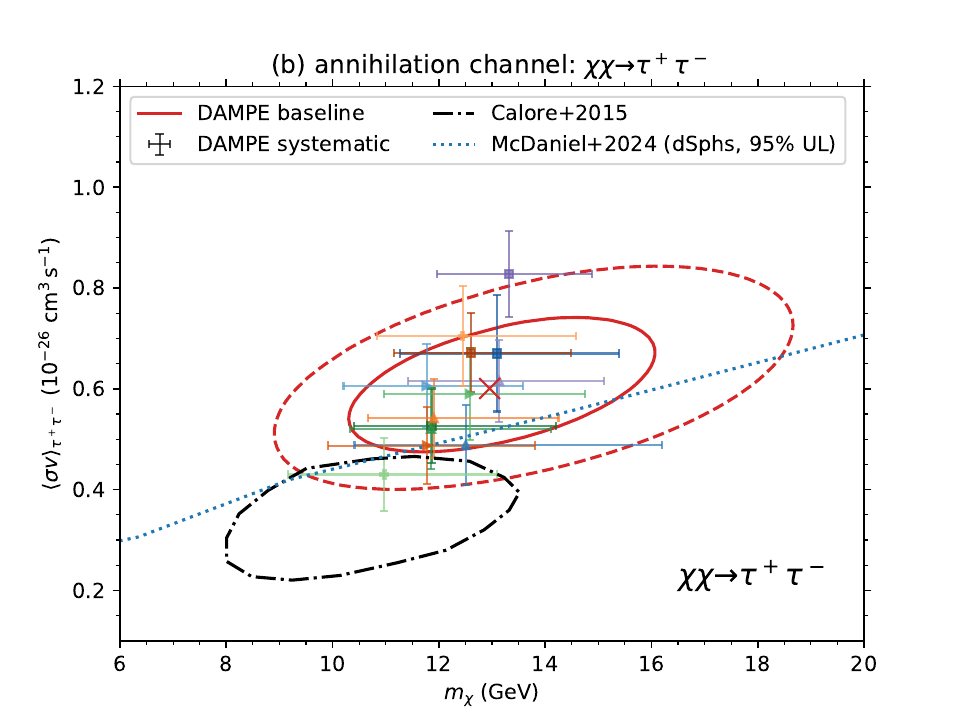}
    \caption{
        The preferred DM parameters for the annihilation channels of (a) $\chi\chi \to b\bar{b}$ and (b) $\chi\chi \to \tau^+\tau^-$.
        The red crosses are the best-fit points.
       The solid and dashed red contours are the $1\sigma$ and $2\sigma$ regions.
        The error bars show the DM parameters and the statistical uncertainties, given various background models with the same colors as those in Figure~\ref{fig::gce:spec_sys1}.
        The dot-dashed and dotted lines are the $2\sigma$ countours from the \lat observations of the \gce.
        The blue dotted lines are the 95\% confidence level upper limits from dSphs observed by \lat.
    }
\label{fig::gce:contours}
\end{figure}

Figure~\ref{fig::gce:dmfit} exhibits the DM annihilation models best fitted to the baseline spectrum along with the $1\sigma$ statistical uncertainty bands.
Both DM models can well describe the emission in the low-energy range.
For the annihilation channel to $b$ quarks, the DM mass needs to be $m_\chi = 49^{+16}_{-12}~\rm GeV$ and the cross section is $\left< \sigma v \right> = 1.9^{+0.4}_{-0.3}\times 10^{-26}~\rm cm^3\,s^{-1}$.
For the channel to the leptons $\tau^+\tau^-$, the optimal DM mass is $m_\chi = 13.0^{+1.9}_{-1.8}~\rm GeV$ and annihilation cross section is $\left< \sigma v \right> = (6.0\pm0.9)\times 10^{-27}~\rm cm^3\,s^{-1}$.
The $1\sigma$ and $2\sigma$ contours, defined with the log-likelihood differences relative to the optimal model ($-2\Delta \ln(\mathcal{L})$) of 2.30 and 6.18, are presented with the solid and dashed red curves in Figure~\ref{fig::gce:contours}, respectively.
We also show the systematic uncertainties of the DM parameters for various background models (Section~\ref{sec::gce:sys}) with the error bars, which are as large as the statistical uncertainties.
The $2\sigma$ regions from the \lat observations of \gce conducted by~\citet{Calore2015b} (dot-dashed line) and~\citet{Cholis2022} (dot-dot-dashed line) are given in the figure.
In both channels, our parameter spaces are consistent with the previous works based on the \lat data.
Drawn with the blue dotted line in the figures are the 95\% confidence level upper limits of the cross sections from dSphs observed by \lat~\citep{McDaniel2024}.
Some of the DM parameter spaces for the two channels are excluded by the dwarf galaxies.

\section{Conclusion}\label{sec::summary}
The Galactic Center region, which hosts a supermassive black hole and a large amount of dark matter as well as cosmic rays, is the most extreme region in the Milky Way and has attracted wide interest in the high-energy astrophysics community. Among the diffuse \gr sources detected in the Galactic center region, the large-scale bubbles and the Galactic center excess in the GeV band detected firstly by \lat have attracted wide attention. In particular, a dark matter origin of the Galactic center excess emerging in the GeV band has been widely speculated in the literature and the multi-messenger ``counterpart" may have been identified in the AMS-02 antiproton data.

Besides \lat, DAMPE is the other GeV \gr detector currently in performance in space that has collected a large amount of data.  With the photon events above 2 GeV observed from January 2016 to June 2024,
the \fbs and the Galactic Center excess are detected with significances of $\sim 26\sigma$ and $\sim 7 \sigma$, respectively. 
Both sources are robust against systematic uncertainties.
Their morphology and spectra are well consistent with those measured with \lat. In particular, the GeV excess component can be interpreted by the dark matter annihilation with a mass of $\sim 50$ GeV and a velocity-averaged cross section of $\sim 10^{-26}~{\rm cm^{3}~s^{-1}}$ for the $\chi \chi \rightarrow b\bar{b}$ channel, also in agreement with the previous findings made with the \lat data. 

A new space mission, the Very Large Area gamma-ray Space Telescope (VLAST), has been proposed. The VLAST mission \citep{Fan2022b,WangQ2023,PanX2024,YangZ2026,ZhangY2025} is distinguished by its extremely wide energy range (from $\sim 1$ MeV to at least 10 TeV) and the very large acceptance (with a peak of $\sim 12~{\rm m^{2}~sr}$).  If successfully launched, the nature of the \fbs as well as the GeV excess in the Galactic center will be further revealed.

\begin{acknowledgments}
    The DAMPE mission is funded by the strategic priority science and technology projects in space science of Chinese Academy of Sciences.
    In China, the data analysis is supported in part by
    the National Key Research and Development Program of China (No. 2022YFF0503301),
    the National Natural Science Foundation of China (Nos. 12588101, 12220101003, 12003074),
    the Strategic Priority Program on Space Science of Chinese Academy of Sciences (No. E02212A02S),
    the Project for Young Scientists in Basic Research of the Chinese Academy of Sciences (Nos. YSBR-092, YSBR-061),
    the Youth Innovation Promotion Association CAS,
    the New Cornerstone Science Foundation through the XPLORER PRIZE,
    and the Entrepreneurship and Innovation Program of Jiangsu Province.
    In Europe, the activities and the data analysis are supported by
    the Swiss National Science Foundation (SNSF), Switzerland,
    the National Institute for Nuclear Physics (INFN), Italy,
    and the European Research Council (ERC) under the European Union's Horizon 2020 research and innovation program.
\end{acknowledgments}
    
\begin{contribution}
    This work is the result of the contributions and efforts of all the participating institutes.
    All authors have reviewed, discussed, and commented on the results and on the manuscript.
    In line with the collaboration policy, the authors are listed alphabetically.
\end{contribution}

\facility{DAMPE}
\software{
        NumPy~\citep{numpy2020},
        SciPy~\citep{scipy2020},
        Matplotlib~\citep{matplotlib2007},
        Astropy~\citep{astropy2022},
        healpy~\citep{Healpix2005,healpy2019},
        iminuit~\citep{iminuit,MINUIT1975},
        Galprop~\citep{Strong:1998pw,Strong:1998fr},
        DmpST~\citep{Duan2019}.
}

\appendix

\section{Maps of the $\gamma$-ray components in the baseline model}\label{appx::comp_maps}
\begin{figure}[!hbt]
    \centering  
    \includegraphics[width=0.32\textwidth]{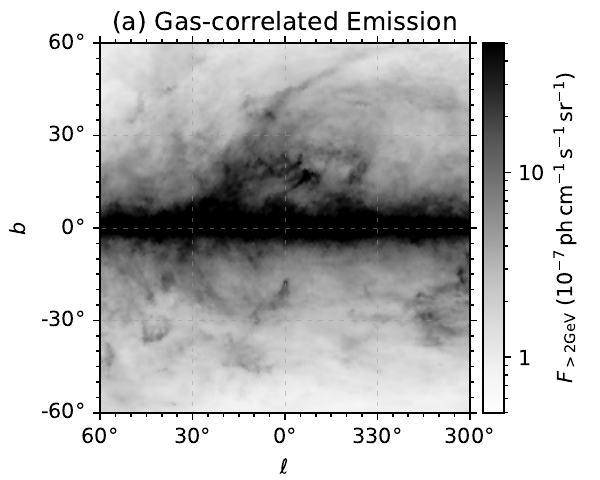}
    \includegraphics[width=0.32\textwidth]{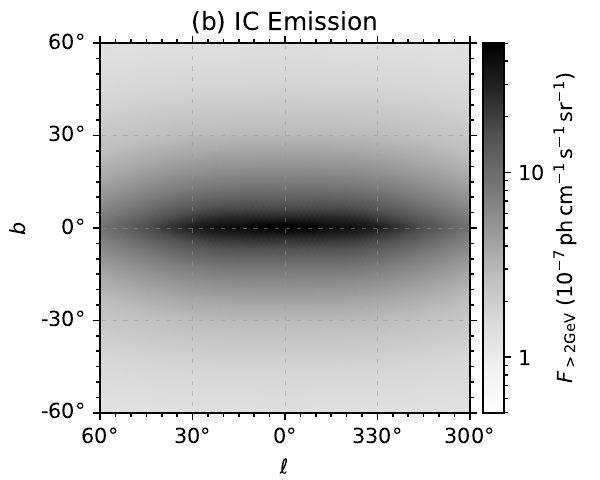}
    \includegraphics[width=0.32\textwidth]{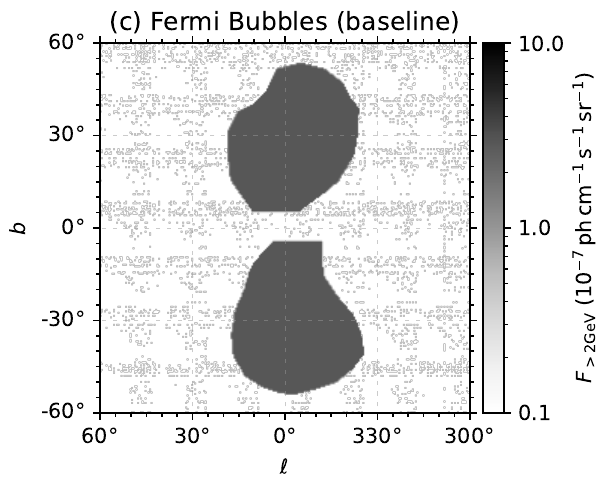}
    \includegraphics[width=0.32\textwidth]{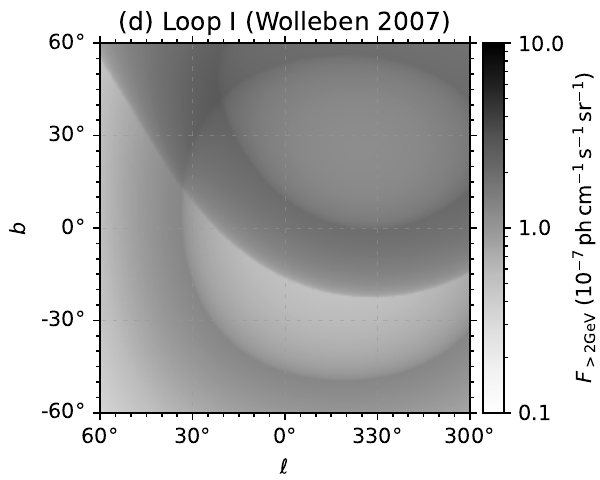}
    \includegraphics[width=0.32\textwidth]{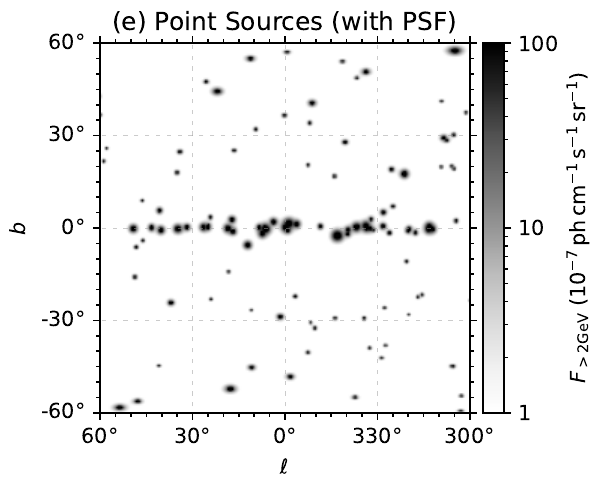}
    \includegraphics[width=0.32\textwidth]{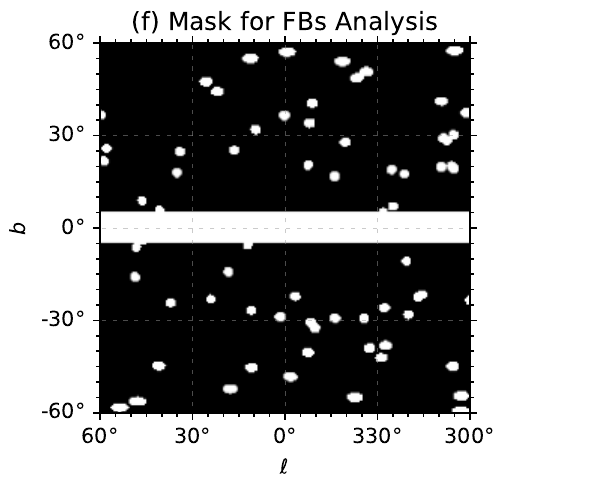}
    \caption{
        (a--e) The maps of the components and (f) mask in the analyses of the \fbs.
        The maps include (a) the hadronic and bremsstrahlung emission associated with gas, (b) the inverse Compton emission, (c) the flat template of \fbs, (d) the geometric template of the Loop~I, and (e) the bright point sources.
        We do not convolve the first four maps with the PSF.
        The white regions in the mask are excluded from the data analysis.
        Both the maps and mask are converted from the HEALPix projection.
    }
\label{fig::appx:bubble:flux_comps}
\end{figure}

\begin{figure}[!hbt]
    \centering  
    \includegraphics[width=0.32\textwidth]{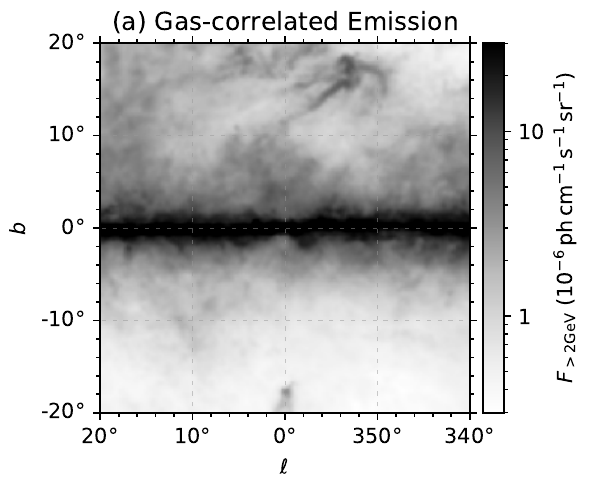}
    \includegraphics[width=0.32\textwidth]{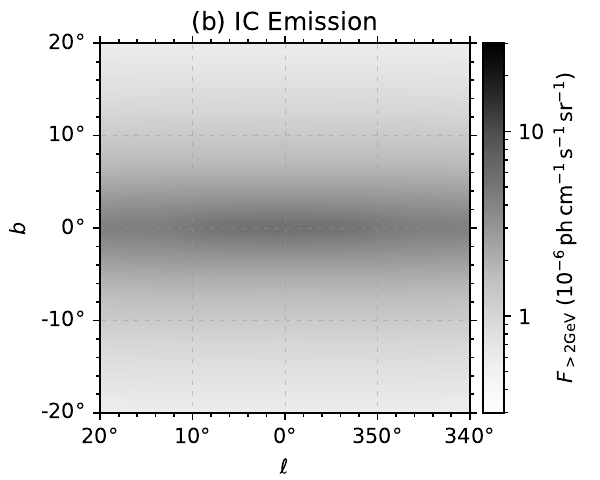}
    \includegraphics[width=0.32\textwidth]{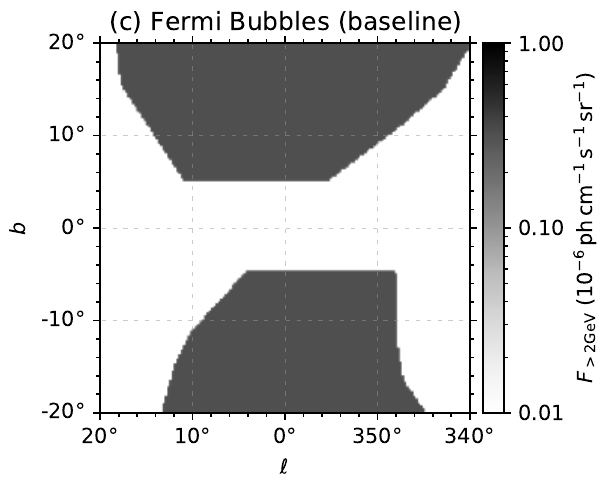}
    \includegraphics[width=0.32\textwidth]{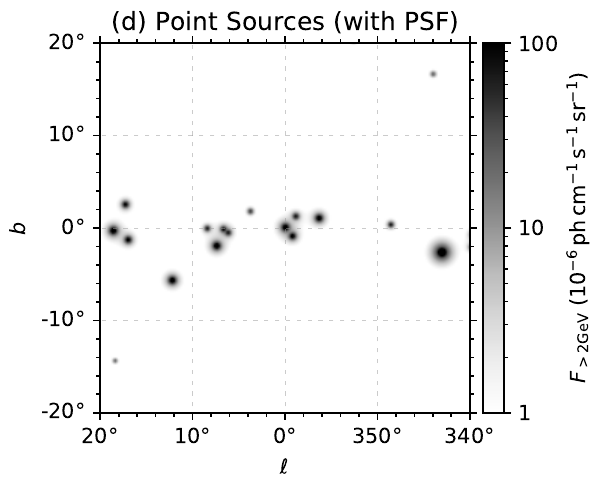}
    \includegraphics[width=0.32\textwidth]{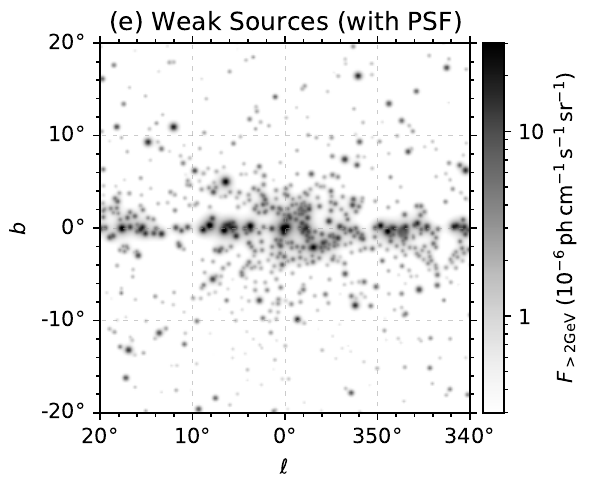}
    \includegraphics[width=0.32\textwidth]{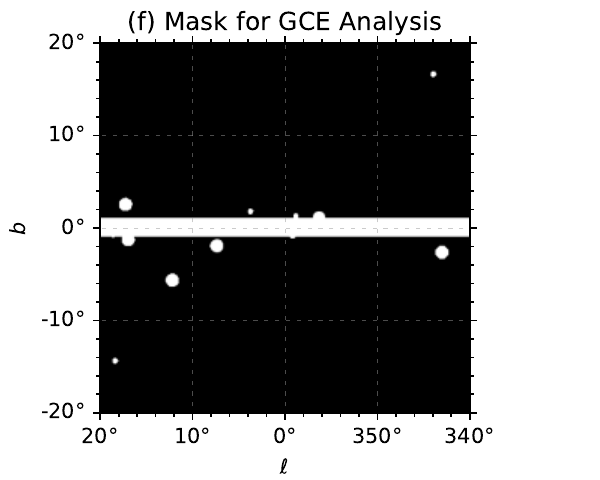}
    \caption{
        (a--e) The maps of the components and (f) mask in the analyses of the \gce.
        The first four components are the same as those in the analyses of the bubbles.
        The color bars are adjusted for better visualization.
        The fifth map is for the weak sources detected by \fermi but not by DAMPE.
    }
\label{fig::appx:gce:flux_comps}
\end{figure}

In Section~\ref{sec::data:gr_components} and Table~\ref{tab::templates}, we introduce the \gr emitting components of the baseline model.
To better understand the structures of the components, we show their intensity maps within the ROIs of the two targets in this section.

Figure~\ref{fig::appx:bubble:flux_comps} presents the intensity maps and mask adopted in the analyses of the \fbs.
The first two sub-figures show (a) the map of hadronic and bremsstrahlung emission and (b) the map of inverse Compton emission from the \lat supernova catalog~\citep{1SC2016}.
The third plot shows the template of \fbs derived with DAMPE data, while the fourth presents the geometric template of Loop~I~\citep{Wolleben2007}.
To better visualize the structures in these components, the maps above are not convolved with the PSF.
The final intensity map is for the bright point sources, whose parameters are from the 8.7-yr DAMPE source catalog (DAMPE Collaboration 2026, in preparation).
The intensities of all the components are calculated based on the optimal spectral parameters of the alternative model, given the baseline background emission model.

Figure~\ref{fig::appx:gce:flux_comps} presents the intensity maps and mask adopted in the analyses of the \gce.
The first four maps are for (a) the hadronic and bremsstrahlung emission, (b) the inverse Compton emission, (c) the \fbs, and (d) the bright point sources, respectively.
The fifth map shows the weak sources that are included in the \lat 4FGL-DR4 source catalog but not significantly detected by DAMPE.
The intensities are based on the best-fit model containing the generalized NFW template with slope of $\gamma=1.2$.

\section{Global spectra of the $\gamma$-ray components in the baseline model}\label{appx::comp_glspec}
\begin{figure*}[!hbt]
    \centering  
    \includegraphics[width=0.45\textwidth]{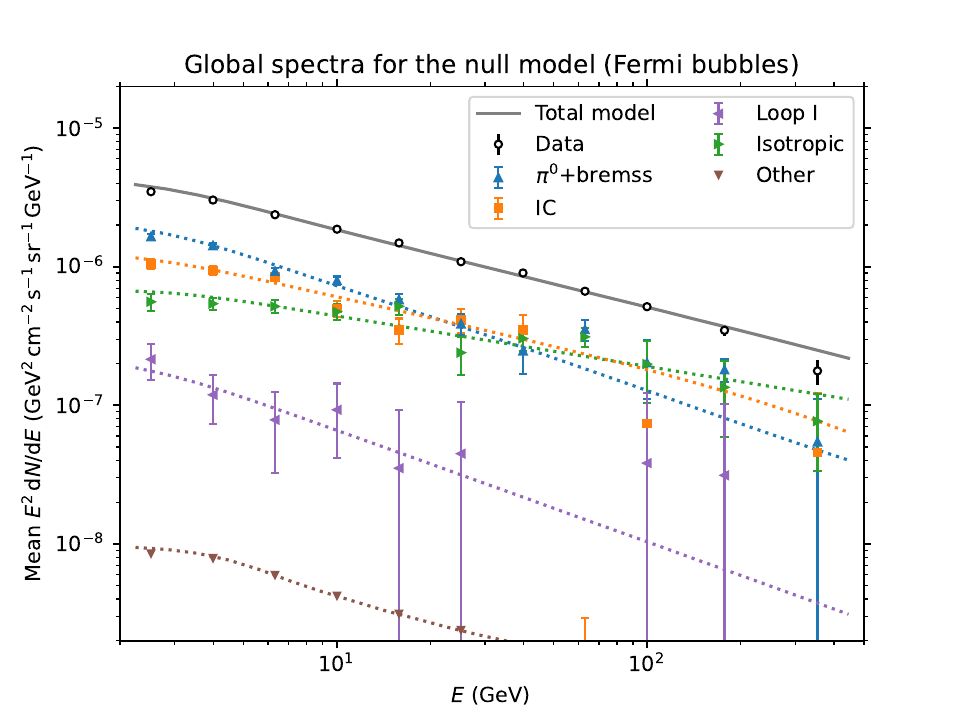}
    \includegraphics[width=0.45\textwidth]{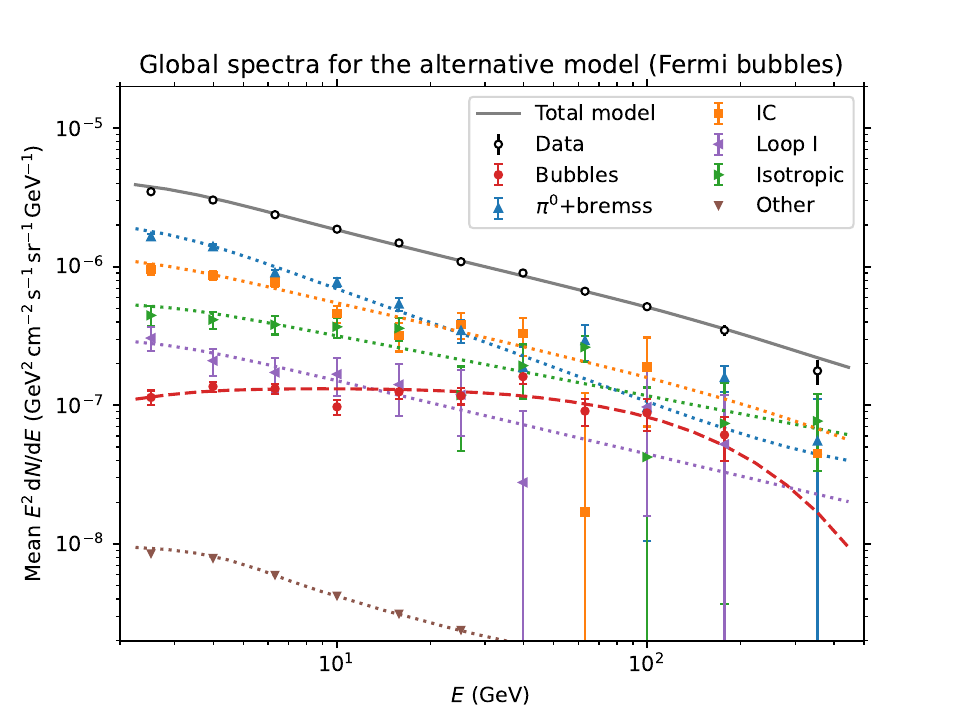}
    \caption{
        The mean flux of the components within the ROI for the best-fit global null (left) and alternative (right) models of the \fbs analysis.
        The data points show the best-fit SEDs from the bin-by-bin analysis.
    }
\label{fig::appx:bubble:specgl}
\end{figure*}

\begin{figure*}[!hbt]
    \centering  
    \includegraphics[width=0.45\textwidth]{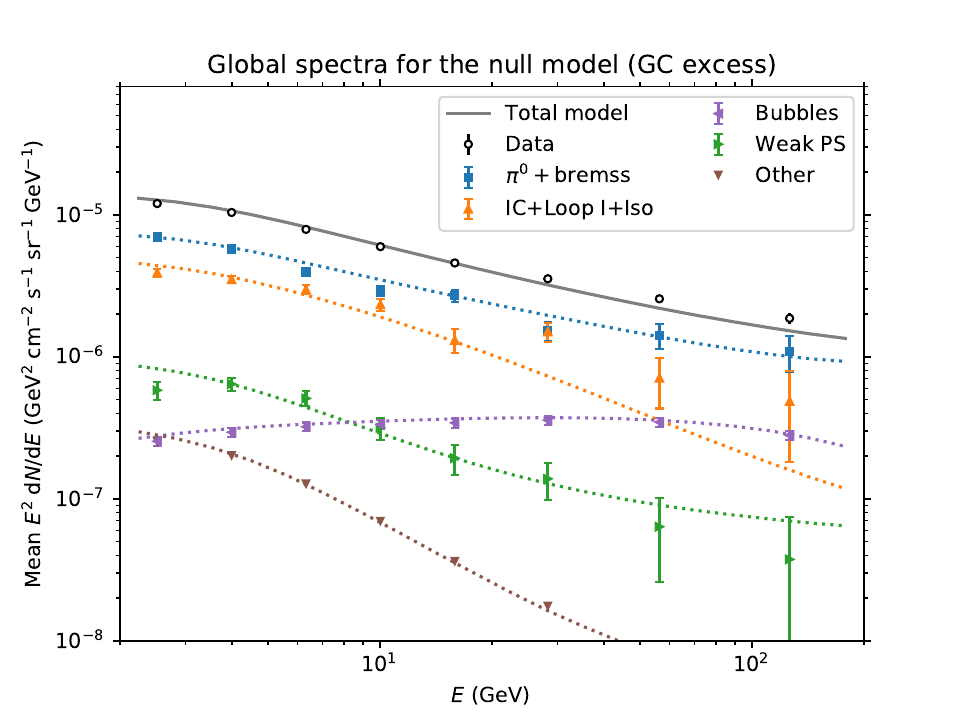}
    \includegraphics[width=0.45\textwidth]{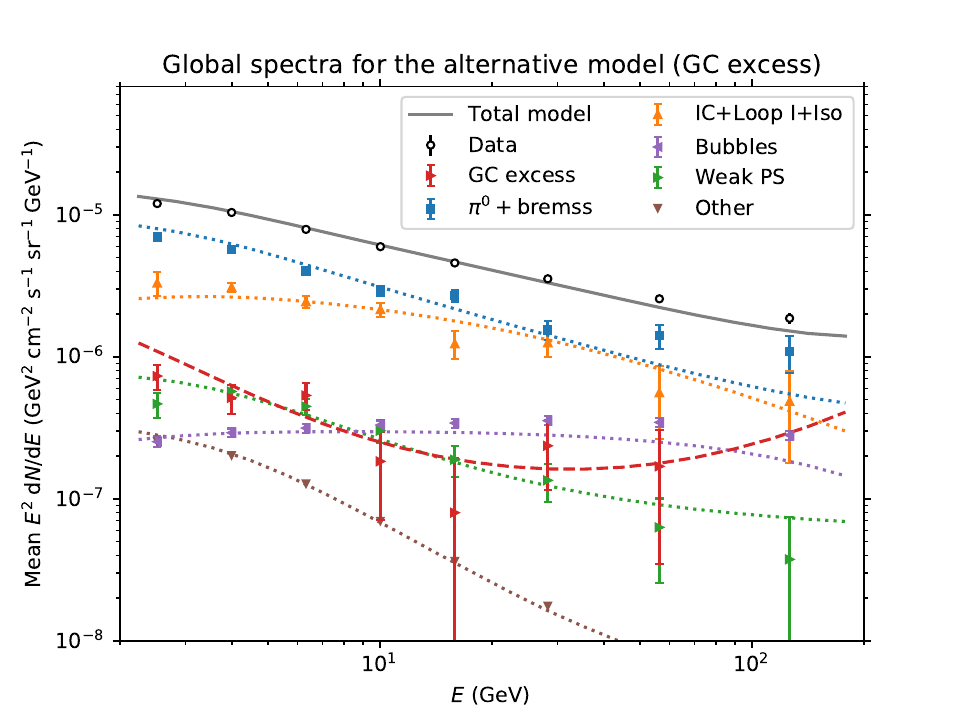}
    \caption{
        The mean flux of the components within the ROI for the best-fit global null (left) and alternative (right) models of the \gce analysis.
    }
\label{fig::appx:gce:specgl}
\end{figure*}

We use the best-fit global models to calculate the residual maps in the analyses of the \fbs (Section~\ref{sec::fbs}) and \gce (Section~\ref{sec::gce}).
In this section, we present the mean flux of the components within the ROI for the best-fit global models.
The left and right panels of Figure~\ref{fig::appx:bubble:specgl} show the component flux from the best-fit global null and alternative models for the \fbs, whereas the Figure~\ref{fig::appx:gce:specgl} is for the \gce.
The data points correspond to the SEDs from the bin-by-bin analyses.


\begin{thebibliography}{}
\expandafter\ifx\csname natexlab\endcsname\relax\def\natexlab#1{#1}\fi
\providecommand{\url}[1]{\href{#1}{#1}}
\providecommand{\dodoi}[1]{doi:~\href{http://doi.org/#1}{\nolinkurl{#1}}}
\providecommand{\doeprint}[1]{\href{http://ascl.net/#1}{\nolinkurl{http://ascl.net/#1}}}
\providecommand{\doarXiv}[1]{\href{https://arxiv.org/abs/#1}{\nolinkurl{https://arxiv.org/abs/#1}}}

\bibitem[{K.~N. {Abazajian}(2011){Abazajian}}]{Abazajian2011}
{Abazajian}, K.~N. 2011, \bibinfo{title}{{The consistency of Fermi-LAT observations of the galactic center with a millisecond pulsar population in the central stellar cluster},} \jcap, 2011, 010, \dodoi{10.1088/1475-7516/2011/03/010}

\bibitem[{K.~N. {Abazajian} \& M. {Kaplinghat}(2012){Abazajian} \& {Kaplinghat}}]{Abazajian2012}
{Abazajian}, K.~N., \& {Kaplinghat}, M. 2012, \bibinfo{title}{{Detection of a gamma-ray source in the Galactic Center consistent with extended emission from dark matter annihilation and concentrated astrophysical emission},} \prd, 86, 083511, \dodoi{10.1103/PhysRevD.86.083511}

\bibitem[{K. {Abd El Dayem} {et~al.}(2024){Abd El Dayem}, {Abuter}, {Aimar}, {Amaro Seoane}, {et~al.}}]{GRAVITY2024}
{Abd El Dayem}, K., {Abuter}, R., {Aimar}, N., {Amaro Seoane}, P., {et~al.} 2024, \bibinfo{title}{{Improving constraints on the extended mass distribution in the Galactic center with stellar orbits},} \aap, 692, A242, \dodoi{10.1051/0004-6361/202452274}

\bibitem[{S. {Abdollahi} {et~al.}(2022){Abdollahi}, {Acero}, {Baldini}, {et~al.}}]{4FGL2022}
{Abdollahi}, S., {Acero}, F., {Baldini}, L., {et~al.} 2022, \bibinfo{title}{{Incremental Fermi Large Area Telescope Fourth Source Catalog},} \apjs, 260, 53, \dodoi{10.3847/1538-4365/ac6751}

\bibitem[{A. {Abramowski} {et~al.}(2016){Abramowski}, {Aharonian}, {Benkhali}, {Akhperjanian}, {Ang{\"u}ner}, {Backes}, {Balzer}, {Becherini}, {Tjus}, {Berge}, {Bernhard}, {Bernl{\"o}hr}, {Birsin}, {Blackwell}, {B{\"o}ttcher}, {Boisson}, {Bolmont}, {Bordas}, {Bregeon}, {Brun}, {Brun}, {Bryan}, {Bulik}, {Carr}, {Casanova}, {Chakraborty}, {Chalme-Calvet}, {Chaves}, {Chen}, {Chr{\'e}tien}, {Colafrancesco}, {Cologna}, {Conrad}, {Couturier}, {Cui}, {Davids}, {Degrange}, {Deil}, {Dewilt}, {Djannati-Ata{\"\i}}, {Domainko}, {Donath}, {Drury}, {Dubus}, {Dutson}, {Dyks}, {Dyrda}, {Edwards}, {Egberts}, {Eger}, {Ernenwein}, {Espigat}, {Farnier}, {Fegan}, {Feinstein}, {Fernandes}, {Fernandez}, {Fiasson}, {Fontaine}, {F{\"o}rster}, {F{\"u}{\ss}ling}, {Gabici}, {Gajdus}, {Gallant}, {Garrigoux}, {Giavitto}, {Giebels}, {Glicenstein}, {Gottschall}, {Goyal}, {Grondin}, {Grudzi{\'n}ska}, {Hadasch}, {H{\"a}ffner}, {Hahn}, {Hawkes}, {Heinzelmann}, {Henri}, {Hermann}, {Hervet}, {Hillert}, {Hinton}, {Hofmann}, {Hofverberg}, {Hoischen}, {Holler}, {Horns}, {Ivascenko}, {Jacholkowska}, {Jamrozy}, {Janiak}, {Jankowsky}, {Jung-Richardt}, {Kastendieck}, {Katarzy{\'n}ski}, {Katz}, {Kerszberg}, {Kh{\'e}lifi}, {Kieffer}, {Klepser}, {Klochkov}, {Klu{\'z}niak}, {Kolitzus}, {Komin}, {Kosack}, {Krakau}, {Krayzel}, {Kr{\"u}ger}, {Laffon}, {Lamanna}, {Lau}, {Lefaucheur}, {Lefranc}, {Lemi{\'e}re}, {Lemoine-Goumard}, {Lenain}, {Lohse}, {Lopatin}, {Lu}, {Lui}, {Marandon}, {Marcowith}, {Mariaud}, {Marx}, {Maurin}, {Maxted}, {Mayer}, {Meintjes}, {Menzler}, {Meyer}, {Mitchell}, {Moderski}, {Mohamed}, {Mor{\r{a}}}, {Moulin}, {Murach}, {de Naurois}, {Niemiec}, {Oakes}, {Odaka}, {{\"O}ttl}, {Ohm}, {Opitz}, {Ostrowski}, {Oya}, {Panter}, {Parsons}, {Arribas}, {Pekeur}, {Pelletier}, {Petrucci}, {Peyaud}, {Pita}, {Poon}, {Prokoph}, {P{\"u}hlhofer}, {Punch}, {Quirrenbach}, {Raab}, {Reichardt}, {Reimer}, {Reimer}, {Renaud}, {de Los Reyes}, {Rieger}, {Romoli}, {Rosier-Lees}, {Rowell}, {Rudak}, {Rulten}, {Sahakian}, {Salek}, {Sanchez}, {Santangelo}, {Sasaki}, {Schlickeiser}, {Sch{\"u}ssler}, {Schulz}, {Schwanke}, {Schwemmer}, {Seyffert}, {Simoni}, {Sol}, {Spanier}, {Spengler}, {Spies}, {Stawarz}, {Steenkamp}, {Stegmann}, {Stinzing}, {Stycz}, {Sushch}, {Tavernet}, {Tavernier}, {Taylor}, {Terrier}, {Tluczykont}, {Trichard}, \& {Tuffs}}]{HESS2016}
{Abramowski}, A., {Aharonian}, F., {Benkhali}, F.~A., {et~al.} 2016, \bibinfo{title}{{Acceleration of petaelectronvolt protons in the Galactic Centre},} \nat, 531, 476, \dodoi{10.1038/nature17147}

\bibitem[{V.~A. {Acciari} {et~al.}(2020){Acciari}, {Ansoldi}, {Antonelli}, {Arbet Engels}, {Baack}, {Babi{\'c}}, {Banerjee}, {Barres de Almeida}, {Barrio}, {Becerra Gonz{\'a}lez}, {Bednarek}, {Bellizzi}, {Bernardini}, {Berti}, {Besenrieder}, {Bhattacharyya}, {Bigongiari}, {Biland}, {Blanch}, {Bonnoli}, {Bo{\v{s}}njak}, {Busetto}, {Carosi}, {Ceribella}, {Chai}, {Chilingaryan}, {Cikota}, {Colak}, {Colin}, {Colombo}, {Contreras}, {Cortina}, {Covino}, {D'Elia}, {da Vela}, {Dazzi}, {de Angelis}, {de Lotto}, {Delfino}, {Delgado}, {Depaoli}, {di Pierro}, {di Venere}, {Do Souto Espi{\~n}eira}, {Dominis Prester}, {Donini}, {Dorner}, {Doro}, {Elsaesser}, {Fallah Ramazani}, {Fattorini}, {Fern{\'a}ndez-Barral}, {Ferrara}, {Fidalgo}, {Foffano}, {Fonseca}, {Font}, {Fruck}, {Fukami}, {Garc{\'\i}a L{\'o}pez}, {Garczarczyk}, {Gasparyan}, {Gaug}, {Giglietto}, {Giordano}, {Godinovi{\'c}}, {Green}, {Guberman}, {Hadasch}, {Hahn}, {Herrera}, {Hoang}, {Hrupec}, {H{\"u}tten}, {Inada}, {Inoue}, {Ishio}, {Iwamura}, {Jouvin}, {Kerszberg}, {Kubo}, {Kushida}, {Lamastra}, {Lelas}, {Leone}, {Lindfors}, {Lombardi}, {Longo}, {L{\'o}pez}, {L{\'o}pez-Coto}, {L{\'o}pez-Oramas}, {Loporchio}, {Machado de Oliveira Fraga}, {Maggio}, {Majumdar}, {Makariev}, {Mallamaci}, {Maneva}, {Manganaro}, {Mannheim}, {Maraschi}, {Mariotti}, {Mart{\'\i}nez}, {Masuda}, {Mazin}, {Mi{\'c}anovi{\'c}}, {Miceli}, {Minev}, {Miranda}, {Mirzoyan}, {Molina}, {Moralejo}, {Morcuende}, {Moreno}, {Moretti}, {Munar-Adrover}, {Neustroev}, {Nigro}, {Nilsson}, {Ninci}, {Nishijima}, {Noda}, {Nogu{\'e}s}, {N{\"o}the}, {Nozaki}, {Paiano}, {Palacio}, {Palatiello}, {Paneque}, {Paoletti}, {Paredes}, {Pe{\~n}il}, {Peresano}, {Persic}, {Prada Moroni}, {Prandini}, {Puljak}, {Rhode}, {Rib{\'o}}, {Rico}, {Righi}, {Rugliancich}, {Saha}, {Sahakyan}, {Saito}, {Sakurai}, {Satalecka}, {Schmidt}, {Schweizer}, {Sitarek}, {{\v{S}}nidari{\'c}}, {Sobczynska}, {Somero}, {Stamerra}, {Strom}, {Strzys}, {Suda}, {Suri{\'c}}, {Takahashi}, {Tavecchio}, {Temnikov}, {Terzi{\'c}}, {Teshima}, {Torres-Alb{\`a}}, {Tosti}, {Tsujimoto}, {Vagelli}, {van Scherpenberg}, {Vanzo}, {Vazquez Acosta}, {Vigorito}, {Vitale}, {Vovk}, {Will}, \& {Zari{\'c}}}]{MAGIC2020}
{Acciari}, V.~A., {Ansoldi}, S., {Antonelli}, L.~A., {et~al.} 2020, \bibinfo{title}{{MAGIC observations of the diffuse {\ensuremath{\gamma}}-ray emission in the vicinity of the Galactic center},} \aap, 642, A190, \dodoi{10.1051/0004-6361/201936896}

\bibitem[{F. {Acero} {et~al.}(2016{\natexlab{a}}){Acero}, {Ackermann}, {Ajello}, {et~al.}}]{1SC2016}
{Acero}, F., {Ackermann}, M., {Ajello}, M., {et~al.} 2016{\natexlab{a}}, \bibinfo{title}{{The First Fermi LAT Supernova Remnant Catalog},} \apjs, 224, 8, \dodoi{10.3847/0067-0049/224/1/8}

\bibitem[{F. {Acero} {et~al.}(2016{\natexlab{b}}){Acero}, {Ackermann}, {Ajello}, {et~al.}}]{Acero2016_FermiGDE}
{Acero}, F., {Ackermann}, M., {Ajello}, M., {et~al.} 2016{\natexlab{b}}, \bibinfo{title}{{Development of the Model of Galactic Interstellar Emission for Standard Point-source Analysis of Fermi Large Area Telescope Data},} \apjs, 223, 26, \dodoi{10.3847/0067-0049/223/2/26}

\bibitem[{M. {Ackermann} {et~al.}(2015{\natexlab{a}}){Ackermann}, {Ajello}, {Albert}, {et~al.}}]{Ackermann2015_IGRB}
{Ackermann}, M., {Ajello}, M., {Albert}, A., {et~al.} 2015{\natexlab{a}}, \bibinfo{title}{{The Spectrum of Isotropic Diffuse Gamma-Ray Emission between 100 MeV and 820 GeV},} \apj, 799, 86, \dodoi{10.1088/0004-637X/799/1/86}

\bibitem[{M. {Ackermann} {et~al.}(2017){Ackermann}, {Ajello}, {Albert}, {et~al.}}]{Ackermann2017}
{Ackermann}, M., {Ajello}, M., {Albert}, A., {et~al.} 2017, \bibinfo{title}{{The Fermi Galactic Center GeV Excess and Implications for Dark Matter},} \apj, 840, 43, \dodoi{10.3847/1538-4357/aa6cab}

\bibitem[{M. {Ackermann} {et~al.}(2012){Ackermann}, {Ajello}, {et~al.}}]{Ackermann2012_GalpropGDE}
{Ackermann}, M., {Ajello}, M., {et~al.} 2012, \bibinfo{title}{{Fermi-LAT Observations of the Diffuse {\ensuremath{\gamma}}-Ray Emission: Implications for Cosmic Rays and the Interstellar Medium},} \apj, 750, 3, \dodoi{10.1088/0004-637X/750/1/3}

\bibitem[{M. {Ackermann} {et~al.}(2014){Ackermann}, {Albert}, {Atwood}, {et~al.}}]{Ackermann2014}
{Ackermann}, M., {Albert}, A., {Atwood}, W.~B., {et~al.} 2014, \bibinfo{title}{{The Spectrum and Morphology of the Fermi Bubbles},} \apj, 793, 64, \dodoi{10.1088/0004-637X/793/1/64}

\bibitem[{M. {Ackermann} {et~al.}(2015{\natexlab{b}}){Ackermann}, {Albert}, {Anderson}, {Atwood}, {Baldini}, {Barbiellini}, {Bastieri}, {Bechtol}, {et~al.}}]{2015PhRvL.115w1301A}
{Ackermann}, M., {Albert}, A., {Anderson}, B., {et~al.} 2015{\natexlab{b}}, \bibinfo{title}{{Searching for Dark Matter Annihilation from Milky Way Dwarf Spheroidal Galaxies with Six Years of Fermi Large Area Telescope Data},} \prl, 115, 231301, \dodoi{10.1103/PhysRevLett.115.231301}

\bibitem[{P.~A.~R. {Ade} {et~al.}(2016){Ade}, {Aghanim}, {et~al.}}]{Planck2016}
{Ade}, P.~A.~R., {Aghanim}, N., {et~al.} 2016, \bibinfo{title}{{Planck 2015 results. XXV. Diffuse low-frequency Galactic foregrounds},} \aap, 594, A25, \dodoi{10.1051/0004-6361/201526803}

\bibitem[{M. {Aguilar} {et~al.}(2016){Aguilar}, {Ali Cavasonza}, {Alpat}, {Ambrosi}, {Arruda}, {Attig}, {Aupetit}, {Azzarello}, {Bachlechner}, {Barao}, {Barrau}, {Barrin}, {Bartoloni}, {Basara}, {Ba{\textcommabelow s}e{\c{C}}{\textsection}mez-du Pree}, {Battarbee}, {Battiston}, {Bazo}, {Becker}, {Behlmann}, {Beischer}, {Berdugo}, {Bertucci}, {Bindi}, {Boella}, {de Boer}, {Bollweg}, {Bonnivard}, {Borgia}, {Boschini}, {Bourquin}, {Bueno}, {Burger}, {Cadoux}, {Cai}, {Capell}, {Caroff}, {Casaus}, {Castellini}, {Cernuda}, {Cervelli}, {Chae}, {Chang}, {Chen}, {Chen}, {Chen}, {Cheng}, {Chou}, {Choumilov}, {Choutko}, {Chung}, {Clark}, {Clavero}, {Coignet}, {Consolandi}, {Contin}, {Corti}, {Coste}, {Creus}, {Crispoltoni}, {Cui}, {Dai}, {Delgado}, {Della Torre}, {Demirk{\"o}z}, {Derome}, {Di Falco}, {Dimiccoli}, {D{\'\i}az}, {von Doetinchem}, {Dong}, {Donnini}, {Duranti}, {D'Urso}, {Egorov}, {Eline}, {Eronen}, {Feng}, {Fiandrini}, {Finch}, {Fisher}, {Formato}, {Galaktionov}, {Gallucci}, {Garc{\'\i}a}, {Garc{\'\i}a-L{\'o}pez}, {Gargiulo}, {Gast}, {Gebauer}, {Gervasi}, {Ghelfi}, {Giovacchini}, {Goglov}, {G{\'o}mez-Coral}, {Gong}, {Goy}, {Grabski}, {Grandi}, {Graziani}, {Guerri}, {Guo}, {Habiby}, {Haino}, {Han}, {He}, {Heil}, {Hoffman}, {Hsieh}, {Huang}, {Huang}, {Huh}, {Incagli}, {Ionica}, {Jang}, {Jinchi}, {Kang}, {Kanishev}, {Kim}, {Kim}, {Kirn}, {Konak}, {Kounina}, {Kounine}, {Koutsenko}, {Krafczyk}, {La Vacca}, {Laudi}, {Laurenti}, {Lazzizzera}, {Lebedev}, {Lee}, {Lee}, {Leluc}, {Li}, {Li}, {Li}, {Li}, {Li}, {Li}, {Li}, {Li}, {Lim}, {Lin}, {Lipari}, {Lippert}, {Liu}, {Liu}, {Lu}, {Lu}, {Luebelsmeyer}, {Luo}, {Luo}, {Lv}, {Majka}, {Ma{\~n}{\'a}}, {Mar{\'\i}n}, {Martin}, {Mart{\'\i}nez}, {Masi}, {Maurin}, {Menchaca-Rocha}, {Meng}, {Mo}, {Morescalchi}, {Mott}, {Nelson}, {Ni}, {Nikonov}, {Nozzoli}, {Nunes}, {Oliva}, {Orcinha}, {Palmonari}, {Palomares}, {Paniccia}, {Pauluzzi}, {Pensotti}, {Pereira}, {Picot-Clemente}, {Pilo}, {Pizzolotto}, {Plyaskin}, {Pohl}, {Poireau}, {Putze}, {Quadrani}, {Qi}, {Qin}, {Qu}, {R{\"a}ih{\"a}}, {Rancoita}, {Rapin}, {Ricol}, {Rodr{\'\i}guez}, {Rosier-Lees}, {Rozhkov}, {Rozza}, {Sagdeev}, {Sandweiss}, \& {Saouter}}]{Aguilar2016}
{Aguilar}, M., {Ali Cavasonza}, L., {Alpat}, B., {et~al.} 2016, \bibinfo{title}{{Antiproton Flux, Antiproton-to-Proton Flux Ratio, and Properties of Elementary Particle Fluxes in Primary Cosmic Rays Measured with the Alpha Magnetic Spectrometer on the International Space Station},} \prl, 117, 091103, \dodoi{10.1103/PhysRevLett.117.091103}

\bibitem[{M. {Aguilar} {et~al.}(2018){Aguilar}, {Ali Cavasonza}, {Alpat}, {Ambrosi}, {Arruda}, {Attig}, {Aupetit}, {Azzarello}, {Bachlechner}, {Barao}, {Barrau}, {Barrin}, {Bartoloni}, {Basara}, {Ba{\c{s}}e{\v{g}}mez-du Pree}, {Battarbee}, {Battiston}, {Becker}, {Behlmann}, {Beischer}, {Berdugo}, {Bertucci}, {Bindel}, {Bindi}, {de Boer}, {Bollweg}, {Bonnivard}, {Borgia}, {Boschini}, {Bourquin}, {Bueno}, {Burger}, {Cadoux}, {Cai}, {Capell}, {Caroff}, {Casaus}, {Castellini}, {Cervelli}, {Chae}, {Chang}, {Chen}, {Chen}, {Chen}, {Chen}, {Cheng}, {Chou}, {Choumilov}, {Choutko}, {Chung}, {Clark}, {Clavero}, {Coignet}, {Consolandi}, {Contin}, {Corti}, {Creus}, {Crispoltoni}, {Cui}, {Dadzie}, {Dai}, {Datta}, {Delgado}, {Della Torre}, {Demirk{\"o}z}, {Derome}, {Di Falco}, {Dimiccoli}, {D{\'\i}az}, {von Doetinchem}, {Dong}, {Donnini}, {Duranti}, {D'Urso}, {Egorov}, {Eline}, {Eronen}, {Feng}, {Fiandrini}, {Fisher}, {Formato}, {Galaktionov}, {Gallucci}, {Garc{\'\i}a-L{\'o}pez}, {Gargiulo}, {Gast}, {Gebauer}, {Gervasi}, {Ghelfi}, {Giovacchini}, {G{\'o}mez-Coral}, {Gong}, {Goy}, {Grabski}, {Grandi}, {Graziani}, {Guo}, {Haino}, {Han}, {He}, {Heil}, {Hoffman}, {Hsieh}, {Huang}, {Huang}, {Huh}, {Incagli}, {Ionica}, {Jang}, {Jia}, {Jinchi}, {Kang}, {Kanishev}, {Khiali}, {Kim}, {Kim}, {Kirn}, {Konak}, {Kounina}, {Kounine}, {Koutsenko}, {Kulemzin}, {La Vacca}, {Laudi}, {Laurenti}, {Lazzizzera}, {Lebedev}, {Lee}, {Lee}, {Leluc}, {Li}, {Li}, {Li}, {Li}, {Li}, {Li}, {Light}, {Lim}, {Lin}, {Lipari}, {Lippert}, {Liu}, {Liu}, {Lordello}, {Lu}, {Lu}, {Luebelsmeyer}, {Luo}, {Luo}, {Luo}, {Lyu}, {Machate}, {Ma{\~n}{\'a}}, {Mar{\'\i}n}, {Martin}, {Mart{\'\i}nez}, {Masi}, {Maurin}, {Menchaca-Rocha}, {Meng}, {Mikuni}, {Mo}, {Mott}, {Nelson}, {Ni}, {Nikonov}, {Nozzoli}, {Oliva}, {Orcinha}, {Palermo}, {Palmonari}, {Palomares}, {Paniccia}, {Pauluzzi}, {Pensotti}, {Perrina}, {Phan}, {Picot-Clemente}, {Pilo}, {Pizzolotto}, {Plyaskin}, {Pohl}, {Poireau}, {Popkow}, {Quadrani}, {Qi}, {Qin}, {Qu}, {R{\"a}ih{\"a}}, {Rancoita}, {Rapin}, {Ricol}, {Rosier-Lees}, {Rozhkov}, {Rozza}, {Sagdeev}, {Schael}, {Schmidt}, {Schulz von Dratzig}, \& {Schwering}}]{Aguilar2018}
{Aguilar}, M., {Ali Cavasonza}, L., {Alpat}, B., {et~al.} 2018, \bibinfo{title}{{Observation of Fine Time Structures in the Cosmic Proton and Helium Fluxes with the Alpha Magnetic Spectrometer on the International Space Station},} \prl, 121, 051101, \dodoi{10.1103/PhysRevLett.121.051101}

\bibitem[{M. {Aguilar} {et~al.}(2025){Aguilar}, {Ambrosi}, {Anderson}, {Arruda}, {Attig}, {Bagwell}, {Barao}, {Barbanera}, {Barrin}, {Bartoloni}, {Battiston}, {Bayyari}, {Belyaev}, {Bertucci}, {Bindi}, {Bollweg}, {Bolster}, {Borchiellini}, {Borgia}, {Boschini}, {Bourquin}, {Brugnoni}, {Burger}, {Burger}, {Cai}, {Capell}, {Casaus}, {Castellini}, {Cervelli}, {Chang}, {Chen}, {Chen}, {Chen}, {Chen}, {Chen}, {Cheng}, {Chou}, {Chouridou}, {Choutko}, {Chung}, {Clark}, {Coignet}, {Consolandi}, {Contin}, {Corti}, {Cui}, {Dadzie}, {D'Angelo}, {Dass}, {Delgado}, {Della Torre}, {Demirk{\"o}z}, {Derome}, {Di Falco}, {Di Felice}, {D{\'\i}az}, {Dimiccoli}, {von Doetinchem}, {Dong}, {Duranti}, {Egorov}, {Eline}, {Faldi}, {Fehr}, {Feng}, {Fiandrini}, {Fisher}, {Formato}, {Garc{\'\i}a-L{\'o}pez}, {Gargiulo}, {Gast}, {Gervasi}, {Giovacchini}, {G{\'o}mez-Coral}, {Gong}, {Grandi}, {Graziani}, {Haino}, {Han}, {Hashmani}, {He}, {Heber}, {Hern{\'a}ndez-Nicol{\'a}s}, {Hsieh}, {Hu}, {Huang}, {Ionica}, {Incagli}, {Jia}, {Jinchi}, {Karag{\"o}z}, {Kirn}, {Klipfel}, {Kounina}, {Kounine}, {Koutsenko}, {Krasnopevtsev}, {Kuhlman}, {Kulemzin}, {La Vacca}, {Laudi}, {Laurenti}, {LaVecchia}, {Lazzizzera}, {Lee}, {Lee}, {Li}, {Li}, {Li}, {Li}, {Li}, {Li}, {Li}, {Li}, {Li}, {Liang}, {Liao}, {Lin}, {Lippert}, {Liu}, {Liu}, {Lu}, {Lu}, {Luo}, {Luo}, {Luo}, {Luo}, {Ma{\~n}{\'a}}, {Mar{\'\i}n}, {Marquardt}, {Mart{\'\i}nez}, {Masi}, {Maurin}, {Medvedeva}, {Menchaca-Rocha}, {Meng}, {Mikhailov}, {Molero}, {Mott}, {Mussolin}, {Jozani}, {Nicolaidis}, {Nikonov}, {Nozzoli}, {Ocampo-Peleteiro}, {Oliva}, {Orcinha}, {Palmonari}, {Paniccia}, {Pashnin}, {Pauluzzi}, {Pelosi}, {Pensotti}, {Pietzcker}, {Plyaskin}, {Poluianov}, {Prid{\"o}hl}, {Qu}, {Quadrani}, {Rancoita}, {Rapin}, {Conde}, {Robyn}, {Rodr{\'\i}guez-Garc{\'\i}a}, {Romaneehsen}, {Rossi}, {Rozhkov}, {Rozza}, {Sagdeev}, {Schael}, {von Dratzig}, {Schwering}, {Seo}, {Shan}, {Shukla}, {Siedenburg}, {Silvestre}, {Song}, {Song}, {Sonnabend}, {Strigari}, {Su}, {Sun}, {Sun}, {Tabarroni}, {Tacconi}, {Tang}, {Tian}, {Tian}, {Ting}, {Ting}, {Tomassetti}, {Torsti}, {Ubaldi}, {Urban}, {Usoskin}, {Vagelli}, {Vainio}, {V{\"a}is{\"a}nen}, \& {Valencia-Otero}}]{Aguilar2025}
{Aguilar}, M., {Ambrosi}, G., {Anderson}, H., {et~al.} 2025, \bibinfo{title}{{Antiprotons and Elementary Particles over a Solar Cycle: Results from the Alpha Magnetic Spectrometer},} \prl, 134, 051002, \dodoi{10.1103/PhysRevLett.134.051002}

\bibitem[{F. {Aharonian} {et~al.}(2004){Aharonian}, {Akhperjanian}, {Aye}, {et~al.}}]{Aharonian2004}
{Aharonian}, F., {Akhperjanian}, A.~G., {Aye}, K.-M., {et~al.} 2004, \bibinfo{title}{{Very high energy gamma rays from the direction of Sagittarius A$^{*}$},} \aap, 425, L13, \dodoi{10.1051/0004-6361:200400055}

\bibitem[{M. {Ajello} {et~al.}(2016){Ajello}, {Albert}, {Atwood}, {et~al.}}]{Ajello2016}
{Ajello}, M., {Albert}, A., {Atwood}, W.~B., {et~al.} 2016, \bibinfo{title}{{Fermi-LAT Observations of High-Energy Gamma-Ray Emission toward the Galactic Center},} \apj, 819, 44, \dodoi{10.3847/0004-637X/819/1/44}

\bibitem[{M. {Ajello} {et~al.}(2015){Ajello}, {Gasparrini}, {S{\'a}nchez-Conde}, {Zaharijas}, {Gustafsson}, {Cohen-Tanugi}, {Dermer}, {Inoue}, {Hartmann}, {Ackermann}, {Bechtol}, {Franckowiak}, {Reimer}, {Romani}, \& {Strong}}]{Ajello2015}
{Ajello}, M., {Gasparrini}, D., {S{\'a}nchez-Conde}, M., {et~al.} 2015, \bibinfo{title}{{The Origin of the Extragalactic Gamma-Ray Background and Implications for Dark Matter Annihilation},} \apjl, 800, L27, \dodoi{10.1088/2041-8205/800/2/L27}

\bibitem[{H. {Akaike}(1974){Akaike}}]{Akaike1974}
{Akaike}, H. 1974, \bibinfo{title}{{A New Look at the Statistical Model Identification},} IEEE Transactions on Automatic Control, 19, 716

\bibitem[{A. {Albert} {et~al.}(2017){Albert}, {Anderson}, {Bechtol}, {Drlica-Wagner}, {Meyer}, {S{\'a}nchez-Conde}, {Strigari}, {Wood}, {et~al.}}]{2017ApJ...834..110A}
{Albert}, A., {Anderson}, B., {Bechtol}, K., {et~al.} 2017, \bibinfo{title}{{Searching for Dark Matter Annihilation in Recently Discovered Milky Way Satellites with Fermi-Lat},} \apj, 834, 110, \dodoi{10.3847/1538-4357/834/2/110}

\bibitem[{A. {Albert} {et~al.}(2024){Albert}, {Alfaro}, {Alvarez}, {Andr{\'e}s}, {Arteaga-Vel{\'a}zquez}, {Avila Rojas}, {Ayala Solares}, {Babu}, {Belmont-Moreno}, {Bernal}, {Caballero-Mora}, {Capistr{\'a}n}, {Carrami{\~n}ana}, {Casanova}, {Cotti}, {Cotzomi}, {Couti{\~n}o de Le{\'o}n}, {De la Fuente}, {de Le{\'o}n}, {Depaoli}, {Di Lalla}, {Diaz Hernandez}, {Dingus}, {DuVernois}, {D{\'\i}az-V{\'e}lez}, {Engel}, {Ergin}, {Espinoza}, {Fan}, {Fang}, {Fraija}, {Fraija}, {Garc{\'\i}a-Gonz{\'a}lez}, {Garfias}, {Goksu}, {Gonz{\'a}lez}, {Goodman}, {Groetsch}, {Harding}, {Hern{\'a}ndez-Cadena}, {Herzog}, {Hinton}, {Huang}, {Hueyotl-Zahuantitla}, {Humensky}, {H{\"u}ntemeyer}, {Iriarte}, {Kaufmann}, {Kieda}, {Lara}, {Lee}, {Lee}, {Vargas}, {Linnemann}, {Longinotti}, {Luis-Raya}, {Malone}, {Martinez}, {Mart{\'\i}nez-Castro}, {Matthews}, {Miranda-Romagnoli}, {Montes}, {Morales-Soto}, {Moreno}, {Mostaf{\'a}}, {Najafi}, {Nellen}, {Newbold}, {Nisa}, {Noriega-Papaqui}, {Olivera-Nieto}, {Omodei}, {Osorio-Archila}, {Araujo}, {P{\'e}rez-P{\'e}rez}, {Rho}, {Rosa-Gonz{\'a}lez}, {Ruiz-Velasco}, {Salazar}, {Salazar-Gallegos}, {Sandoval}, {Schneider}, {Schwefer}, {Serna-Franco}, {Smith}, {Son}, {Springer}, {Tibolla}, {Tollefson}, {Torres}, {Torres-Escobedo}, {Turner}, {Ure{\~n}a-Mena}, {Varela}, {Wang}, {Wang}, {Watson}, {Willox}, {Wu}, {Yu}, {Yun-C{\'a}rcamo}, \& {Zhou}}]{Albert2024}
{Albert}, A., {Alfaro}, R., {Alvarez}, C., {et~al.} 2024, \bibinfo{title}{{Observation of the Galactic Center PeVatron beyond 100 TeV with HAWC},} \apjl, 973, L34, \dodoi{10.3847/2041-8213/ad772e}

\bibitem[{F. {Alemanno} {et~al.}(2022{\natexlab{a}}){Alemanno}, {Altomare}, {An}, {Azzarello}, {et~al.}}]{DAMPE2022}
{Alemanno}, F., {Altomare}, C., {An}, Q., {Azzarello}, P., {et~al.} 2022{\natexlab{a}}, \bibinfo{title}{{Detection of spectral hardenings in cosmic-ray boron-to-carbon and boron-to-oxygen flux ratios with DAMPE},} Sci. Bull., 67, 2162, \dodoi{10.1016/j.scib.2022.10.002}

\bibitem[{F. {Alemanno} {et~al.}(2022{\natexlab{b}}){Alemanno}, {An}, {Azzarello}, {et~al.}}]{Alemanno2022_line}
{Alemanno}, F., {An}, Q., {Azzarello}, P., {et~al.} 2022{\natexlab{b}}, \bibinfo{title}{{Search for gamma-ray spectral lines with the DArk Matter Particle Explorer},} Sci. Bull., 67, 679, \dodoi{10.1016/j.scib.2021.12.015}

\bibitem[{F. {Alemanno} {et~al.}(2021){Alemanno}, {An}, {Azzarello}, {Barbato}, {Bernardini}, {Bi}, {Cai}, {Catanzani}, {Chang}, {Chen}, {Chen}, {Chen}, {Cui}, {Cui}, {Cui}, {Dai}, {D'Amone}, {de Benedittis}, {de Mitri}, {de Palma}, {Deliyergiyev}, {di Santo}, {Dong}, {Dong}, {Donvito}, {Droz}, {Duan}, {Duan}, {D'Urso}, {Fan}, {Fan}, {Fang}, {Fang}, {Feng}, {Feng}, {Fusco}, {Gao}, {Gargano}, {Gong}, {Gong}, {Guo}, {Guo}, {Guo}, {Han}, {Hu}, {Huang}, {Huang}, {Huang}, {Ionica}, {Jiang}, {Kong}, {Kotenko}, {Kyratzis}, {Lei}, {Li}, {Li}, {Li}, {Li}, {Liang}, {Liu}, {Liu}, {Liu}, {Liu}, {Liu}, {Liu}, {Loparco}, {Luo}, {Ma}, {Ma}, {Ma}, {Ma}, {Marsella}, {Mazziotta}, {Mo}, {Niu}, {Pan}, {Parenti}, {Peng}, {Peng}, {Perrina}, {Qiao}, {Rao}, {Ruina}, {Salinas}, {Shang}, {Shen}, {Shen}, {Shen}, {Silveri}, {Song}, {Stolpovskiy}, {Su}, {Su}, {Sun}, {Surdo}, {Teng}, {Tykhonov}, {Wang}, {Wang}, {Wang}, {Wang}, {Wang}, {Wang}, {Wang}, {Wang}, {Wang}, {Wei}, {Wei}, {Wei}, {Wen}, {Wu}, {Wu}, {Wu}, {Wu}, {Wu}, {Xia}, {Xu}, {Xu}, {Xu}, {Xu}, {Xue}, {Yang}, {Yang}, {Yang}, {Yao}, {Yu}, {Yuan}, {Yuan}, {Yue}, {Zang}, {Zhang}, {Zhang}, {Zhang}, {Zhang}, {Zhang}, {Zhang}, {Zhang}, {Zhang}, {Zhang}, {Zhang}, {Zhao}, {Zhao}, {Zhao}, {Zhou}, {Zhu}, \& {Dampe Collaboration}}]{DAMPE2021}
{Alemanno}, F., {An}, Q., {Azzarello}, P., {et~al.} 2021, \bibinfo{title}{{Measurement of the Cosmic Ray Helium Energy Spectrum from 70 GeV to 80 TeV with the DAMPE Space Mission},} \prl, 126, 201102, \dodoi{10.1103/PhysRevLett.126.201102}

\bibitem[{F. {Alemanno} {et~al.}(2025{\natexlab{a}}){Alemanno}, {Altomare}, {An}, {Azzarello}, {Barbato}, {Bernardini}, {Bi}, {Boutin}, {Cagnoli}, {Cai}, {Casilli}, {Catanzani}, {Chang}, {Chen}, {Chen}, {Chen}, {Chen}, {Coppin}, {Cui}, {Cui}, {Cui}, {de Mitri}, {de Palma}, {di Giovanni}, {Dong}, {Dong}, {Donvito}, {Droz}, {Duan}, {Duan}, {Fan}, {Fan}, {Fang}, {Fang}, {Feng}, {Feng}, {Frieden}, {Fusco}, {Gao}, {Gargano}, {Ghose}, {Gong}, {Gong}, {Guo}, {Guo}, {Han}, {Hu}, {Huang}, {Huang}, {Huang}, {Ionica}, {Jiang}, {Jiang}, {Jiang}, {Kong}, {Kotenko}, {Kyratzis}, {Lei}, {Li}, {Li}, {Li}, {Li}, {Liang}, {Liu}, {Liu}, {Liu}, {Liu}, {Liu}, {Loparco}, {Ma}, {Ma}, {Ma}, {Ma}, {Marsella}, {Mazziotta}, {Mo}, {Niu}, {Parenti}, {Peng}, {Peng}, {Perrina}, {Putti-Garcia}, {Qiao}, {Rao}, {Sarkar}, {Savina}, {Serpolla}, {Shangguan}, {Shen}, {Shen}, {Shen}, {Silveri}, {Song}, {Stolpovskiy}, {Su}, {Su}, {Sun}, {Sun}, {Surdo}, {Teng}, {Tykhonov}, {Wang}, {Wang}, {Wang}, {Wang}, {Wang}, {Wang}, {Wei}, {Wei}, {Wei}, {Wu}, {Wu}, {Wu}, {Wu}, {Xia}, {Xu}, {Xu}, {Xu}, {Xu}, {Xu}, {Xu}, {Xue}, {Yang}, {Yang}, {Yang}, {Yao}, {Yu}, {Yuan}, {Yue}, {Zang}, {Zhang}, {Zhang}, {Zhang}, {Zhang}, {Zhang}, {Zhang}, {Zhang}, {Zhang}, {Zhang}, {Zhang}, {Zhao}, {Zhao}, {Zhao}, {Zhou}, {Zhu}, \& {Dampe Collaboration}}]{DAMPE2025a}
{Alemanno}, F., {Altomare}, C., {An}, Q., {et~al.} 2025{\natexlab{a}}, \bibinfo{title}{{Observation of a Spectral Hardening in Cosmic Ray Boron Spectrum with the DAMPE Space Mission},} \prl, 134, 191001, \dodoi{10.1103/PhysRevLett.134.191001}

\bibitem[{F. {Alemanno} {et~al.}(2025{\natexlab{b}}){Alemanno}, {An}, {Azzarello}, {Barbato}, {Bernardini}, {Bi}, {Boutin}, {Cagnoli}, {Cai}, {Casilli}, {Catanzani}, {Chang}, {Chen}, {Chen}, {Chen}, {Chen}, {Coppin}, {Cui}, {Cui}, {Cui}, {de Mitri}, {de Palma}, {di Giovanni}, {Dong}, {Dong}, {Donvito}, {Duan}, {Duan}, {Fan}, {Fan}, {Fang}, {Fang}, {Feng}, {Feng}, {Frieden}, {Fusco}, {Gao}, {Gargano}, {Ghose}, {Gong}, {Gong}, {Guo}, {Guo}, {Han}, {Hu}, {Huang}, {Huang}, {Huang}, {Ionica}, {Jiang}, {Jiang}, {Jiang}, {Kong}, {Kotenko}, {Kyratzis}, {Lei}, {Li}, {Li}, {Li}, {Li}, {Li}, {Liang}, {Liu}, {Liu}, {Liu}, {Liu}, {Liu}, {Loparco}, {Luo}, {Ma}, {Ma}, {Ma}, {Ma}, {Marsella}, {Mazziotta}, {Mo}, {Nie}, {Niu}, {Parenti}, {Peng}, {Peng}, {Perrina}, {Putti-Garcia}, {Qiao}, {Rao}, {Rong}, {Sarkar}, {Savina}, {Serpolla}, {Shangguan}, {Shen}, {Shen}, {Shen}, {Silveri}, {Song}, {Su}, {Su}, {Sun}, {Sun}, {Surdo}, {Teng}, {Tykhonov}, {Wang}, {Wang}, {Wang}, {Wang}, {Wang}, {Wang}, {Wang}, {Wei}, {Wei}, {Wei}, {Wu}, {Wu}, {Wu}, {Wu}, {Xia}, {Xiong}, {Xu}, {Xu}, {Xu}, {Xu}, {Xu}, {Xu}, {Xue}, {Yang}, {Yang}, {Yang}, {Yao}, {Yan}, {Yu}, {Yuan}, {Yue}, {Zang}, {Zhang}, {Zhang}, {Zhang}, {Zhang}, {Zhang}, {Zhang}, {Zhang}, {Zhang}, {Zhang}, {Zhao}, {Zhao}, {Zhao}, {Zhou}, {Zhu}, \& {Dampe Collaboration}}]{DAMPE2025b}
{Alemanno}, F., {An}, Q., {Azzarello}, P., {et~al.} 2025{\natexlab{b}}, \bibinfo{title}{{Measurement of separate electron and positron spectra from 10 to 20 GeV with the geomagnetic field on DAMPE},} Chinese Physics C, 49, 115001, \dodoi{10.1088/1674-1137/adfa04}

\bibitem[{F. {Alemanno} {et~al.}(2025{\natexlab{c}}){Alemanno}, {An}, {Azzarello}, {Barbato}, {Bernardini}, {Bi}, {Valentin Boutin}, {Cagnoli}, {Cai}, {Casilli}, {Chang}, {Chen}, {Chen}, {Chen}, {Chen}, {Coppin}, {Cui}, {Cui}, {De Mitri}, {de Palma}, {Di Giovanni}, {Dong}, {Dong}, {Donvito}, {Duan}, {Duan}, {Fan}, {Fan}, {Fang}, {Fang}, {Feng}, {Feng}, {Fogliacco}, {Frieden}, {Fusco}, {Gao}, {Gargano}, {Ghose}, {Gong}, {Gong}, {Guo}, {Guo}, {Han}, {Hu}, {Huang}, {Huang}, {Huang}, {Ionica}, {Jiang}, {Jiang}, {Jiang}, {Kong}, {Kotenko}, {Kyratzis}, {Lei}, {Li}, {Li}, {Li}, {Li}, {Li}, {Li}, {Liang}, {Liu}, {Liu}, {Liu}, {Liu}, {Liu}, {Loparco}, {Ma}, {Ma}, {Ma}, {Ma}, {Marsella}, {Mazziotta}, {Mo}, {Nie}, {Niu}, {Parenti}, {Peng}, {Peng}, {Perrina}, {Putti-Garcia}, {Qiao}, {Rao}, {Rong}, {Serpolla}, {Sarkar}, {Savina}, {Shangguan}, {Shen}, {Shen}, {Shen}, {Silveri}, {Song}, {Su}, {Su}, {Sun}, {Sun}, {Surdo}, {Teng}, {Tykhonov}, {Wang}, {Wang}, {Wang}, {Wang}, {Wang}, {Wang}, {Wei}, {Wei}, {Wei}, {Wu}, {Wu}, {Wu}, {Wu}, {Xia}, {Xiong}, {Xu}, {Xu}, {Xu}, {Xu}, {Xu}, {Xu}, {Xue}, {Yan}, {Yang}, {Yang}, {Yang}, {Yao}, {Yu}, {Yuan}, {Yue}, {Zang}, {Zhang}, {Zhang}, {Zhang}, {Zhang}, {Zhang}, {Zhang}, {Zhang}, {Zhang}, {Zhang}, {Zhang}, {Zhao}, {Zhao}, {Zhao}, {Zhou}, {Zhu}, \& {Zhu}}]{DAMPE2025c}
{Alemanno}, F., {An}, Q., {Azzarello}, P., {et~al.} 2025{\natexlab{c}}, \bibinfo{title}{{Charge-dependent spectral softenings of primary cosmic-rays from proton to iron below the knee},} arXiv e-prints, arXiv:2511.05409, \dodoi{10.48550/arXiv.2511.05409}

\bibitem[{F. {Alemanno} {et~al.}(2025{\natexlab{d}}){Alemanno}, {An}, {Azzarello}, {Barbato}, {Bernardini}, {Bi}, {Boutin}, {Cagnoli}, {Cai}, {Casilli}, {Chang}, {Chen}, {Chen}, {Chen}, {Chen}, {Coppin}, {Cui}, {Cui}, {De Mitri}, {de Palma}, {Di Giovanni}, {Dong}, {Dong}, {Donvito}, {Duan}, {Duan}, {Fan}, {Fan}, {Fang}, {Fang}, {Feng}, {Feng}, {Fogliacco}, {Frieden}, {Fusco}, {Gao}, {Gargano}, {Ghose}, {Gong}, {Gong}, {Guo}, {Guo}, {Han}, {Hu}, {Huang}, {Huang}, {Huang}, {Ionica}, {Jiang}, {Jiang}, {Jiang}, {Kong}, {Kotenko}, {Kyratzis}, {Lei}, {Li}, {Li}, {Li}, {Li}, {Li}, {Li}, {Liang}, {Liu}, {Liu}, {Liu}, {Liu}, {Liu}, {Loparco}, {Ma}, {Ma}, {Ma}, {Ma}, {Marsella}, {Mazziotta}, {Mo}, {Nie}, {Niu}, {Parenti}, {Peng}, {Peng}, {Perrina}, {Putti Garcia}, {Qiao}, {Rao}, {Rong}, {Serpolla}, {Sarkar}, {Savina}, {Shangguan}, {Shen}, {Shen}, {Shen}, {Silveri}, {Song}, {Su}, {Su}, {Sun}, {Sun}, {Surdo}, {Teng}, {Tykhonov}, {Wang}, {Wang}, {Wang}, {Wang}, {Wang}, {Wang}, {Wei}, {Wei}, {Wei}, {Wu}, {Wu}, {Wu}, {Wu}, {Xia}, {Xiong}, {Xu}, {Xu}, {Xu}, {Xu}, {Xu}, {Xu}, {Xue}, {Yan}, {Yang}, {Yang}, {Yang}, {Yao}, {Yu}, {Yuan}, {Yue}, {Zang}, {Zhang}, {Zhang}, {Zhang}, {Zhang}, {Zhang}, {Zhang}, {Zhang}, {Zhang}, {Zhang}, {Zhang}, {Zhao}, {Zhao}, {Zhao}, {Zhou}, {Zhu}, \& {Zhu}}]{DAMPE2025d}
{Alemanno}, F., {An}, Q., {Azzarello}, P., {et~al.} 2025{\natexlab{d}}, \bibinfo{title}{{Measurement of the cosmic ray nickel energy spectrum from 10 GeV/n to 2 TeV/n with the DAMPE},} arXiv e-prints, arXiv:2512.11425, \dodoi{10.48550/arXiv.2512.11425}

\bibitem[{G. {Ambrosi} {et~al.}(2019){Ambrosi}, {An}, {Asfandiyarov}, {et~al.}}]{Ambrosi2019}
{Ambrosi}, G., {An}, Q., {Asfandiyarov}, R., {et~al.} 2019, \bibinfo{title}{{The on-orbit calibration of DArk Matter Particle Explorer},} Astropart. Phys., 106, 18, \dodoi{10.1016/j.astropartphys.2018.10.006}

\bibitem[{G. {Ambrosi} {et~al.}(2017){Ambrosi}, {An}, {Asfandiyarov}, {Azzarello}, {Bernardini}, {Bertucci}, {Cai}, {Chang}, {Chen}, {Chen}, {Chen}, {Chen}, {Cui}, {Cui}, {D'Amone}, {de Benedittis}, {De Mitri}, {di Santo}, {Dong}, {Dong}, {Dong}, {Dong}, {Donvito}, {Droz}, {Duan}, {Duan}, {Duranti}, {D'Urso}, {Fan}, {Fan}, {Fang}, {Feng}, {Feng}, {Fusco}, {Gallo}, {Gan}, {Gao}, {Gao}, {Gargano}, {Garrappa}, {Gong}, {Gong}, {Guo}, {Guo}, {Hu}, {Huang}, {Huang}, {Ionica}, {Jiang}, {Jiang}, {Jin}, {Kong}, {Lei}, {Li}, {Li}, {Li}, {Li}, {Liang}, {Liang}, {Liao}, {Liu}, {Liu}, {Liu}, {Liu}, {Liu}, {Loparco}, {Ma}, {Ma}, {Ma}, {Ma}, {Ma}, {Ma}, {Marsella}, {Mazziotta}, {Mo}, {Niu}, {Peng}, {Peng}, {Qiao}, {Rao}, {Salinas}, {Shang}, {H. Shen}, {Shen}, {Shen}, {Song}, {Su}, {Su}, {Sun}, {Surdo}, {Teng}, {Tian}, {Tykhonov}, {Vagelli}, {Vitillo}, {Wang}, {Wang}, {Wang}, {Wang}, {Wang}, {Wang}, {Wang}, {Wang}, {Wang}, {Wang}, {Wang}, {Wang}, {Wen}, {Wang}, {Wei}, {Wei}, {Wei}, {Wu}, {Wu}, {Wu}, {Wu}, {Wu}, {Xi}, {Xia}, {Xin}, {Xu}, {Xu}, {Xu}, {Xue}, {Yang}, {Yang}, {Yang}, {Yang}, {Yao}, {Yu}, {Yuan}, {Yue}, {Zang}, {Zhang}, {Zhang}, {Zhang}, {Zhang}, {Zhang}, {Zhang}, {Zhang}, {Zhang}, {Zhang}, {Zhang}, {Zhang}, {Zhang}, {Zhang}, {Zhang}, {Zhang}, {Zhang}, {Zhang}, {Zhao}, {Zhao}, {Zhao}, {Zhou}, {Zhou}, {Zhu}, {Zhu}, \& {Zimmer}}]{DAMPE2017}
{Ambrosi}, G., {An}, Q., {Asfandiyarov}, R., {et~al.} 2017, \bibinfo{title}{{Direct detection of a break in the teraelectronvolt cosmic-ray spectrum of electrons and positrons},} \nat, 552, 63, \dodoi{10.1038/nature24475}

\bibitem[{Q. {An} {et~al.}(2019){An}, {Asfandiyarov}, {Azzarello}, {Bernardini}, {Bi}, {Cai}, {Chang}, {Chen}, {Chen}, {Chen}, {Chen}, {Cui}, {Cui}, {Dai}, {D'Amone}, {De Benedittis}, {De Mitri}, {Di Santo}, {Ding}, {Dong}, {Dong}, {Dong}, {Donvito}, {Droz}, {Duan}, {Duan}, {D'Urso}, {Fan}, {Fan}, {Fang}, {Feng}, {Feng}, {Fusco}, {Gallo}, {Gan}, {Gao}, {Gargano}, {Gong}, {Gong}, {Guo}, {Guo}, {Guo}, {Han}, {Hu}, {Huang}, {Huang}, {Huang}, {Ionica}, {Jiang}, {Jin}, {Kong}, {Lei}, {Li}, {Li}, {Li}, {Li}, {Li}, {Liang}, {Liang}, {Liao}, {Liu}, {Liu}, {Liu}, {Liu}, {Liu}, {Liu}, {Loparco}, {Luo}, {Ma}, {Ma}, {Ma}, {Ma}, {Ma}, {Marsella}, {Mazziotta}, {Mo}, {Niu}, {Pan}, {Peng}, {Peng}, {Qiao}, {Rao}, {Salinas}, {Shang}, {Shen}, {Shen}, {Shen}, {Song}, {Su}, {Su}, {Sun}, {Surdo}, {Teng}, {Tykhonov}, {Vitillo}, {Wang}, {Wang}, {Wang}, {Wang}, {Wang}, {Wang}, {Wang}, {Wang}, {Wang}, {Wang}, {Wang}, {Wang}, {Wang}, {Wei}, {Wei}, {Wei}, {Wen}, {Wu}, {Wu}, {Wu}, {Wu}, {Wu}, {Xi}, {Xia}, {Xu}, {Xu}, {Xu}, {Xu}, {Xue}, {Yang}, {Yang}, {Yang}, {Yang}, {Yao}, {Yu}, {Yuan}, {Yue}, {Zang}, {Zhang}, {Zhang}, {Zhang}, {Zhang}, {Zhang}, {Zhang}, {Zhang}, {Zhang}, {Zhang}, {Zhang}, {Zhang}, {Zhang}, {Zhang}, {Zhao}, {Zhao}, {Zhao}, {Zhou}, {Zhou}, {Zhu}, {Zhu}, \& {Zimmer}}]{DAMPE2019b}
{An}, Q., {Asfandiyarov}, R., {Azzarello}, P., {et~al.} 2019, \bibinfo{title}{{Measurement of the cosmic ray proton spectrum from 40 GeV to 100 TeV with the DAMPE satellite},} Science Advances, 5, eaax3793, \dodoi{10.1126/sciadv.aax3793}

\bibitem[{S. {Ando} {et~al.}(2020){Ando}, {Geringer-Sameth}, {Hiroshima}, {Hoof}, {Trotta}, \& {Walker}}]{2020PhRvD.102f1302A}
{Ando}, S., {Geringer-Sameth}, A., {Hiroshima}, N., {et~al.} 2020, \bibinfo{title}{{Structure formation models weaken limits on WIMP dark matter from dwarf spheroidal galaxies},} \prd, 102, 061302, \dodoi{10.1103/PhysRevD.102.061302}

\bibitem[{J. {Ballet} {et~al.}(2023){Ballet}, {Bruel}, {Burnett}, \& {Lott}}]{Ballet2023_4FGLDR4}
{Ballet}, J., {Bruel}, P., {Burnett}, T.~H., \& {Lott}, B. 2023, \bibinfo{title}{{Fermi Large Area Telescope Fourth Source Catalog Data Release 4 (4FGL-DR4)},} ArXiv e-prints, arXiv:2307.12546.
\newblock \doarXiv{2307.12546}

\bibitem[{R. {Bartels} {et~al.}(2018){Bartels}, {Storm}, {Weniger}, \& {Calore}}]{Bartels2018}
{Bartels}, R., {Storm}, E., {Weniger}, C., \& {Calore}, F. 2018, \bibinfo{title}{{The Fermi-LAT GeV excess as a tracer of stellar mass in the Galactic bulge},} Nature Astronomy, 2, 819, \dodoi{10.1038/s41550-018-0531-z}

\bibitem[{E.~M. {Berkhuijsen}(1971){Berkhuijsen}}]{Berkhuijsen1971}
{Berkhuijsen}, E.~M. 1971, \bibinfo{title}{{A Survey of the Continuum Radiation at 820 MHz between Declinations -7{\textdegree} and +85{\textdegree}. A Study of the Galactic Radiation and the Degree of Polarization with Special Reference to the Loops and Spurs},} \aap, 14, 359

\bibitem[{G. {Bertone} \& D. {Hooper}(2018){Bertone} \& {Hooper}}]{Bertone2018}
{Bertone}, G., \& {Hooper}, D. 2018, \bibinfo{title}{{History of dark matter},} Reviews of Modern Physics, 90, 045002, \dodoi{10.1103/RevModPhys.90.045002}

\bibitem[{G. {Bertone} {et~al.}(2005){Bertone}, {Hooper}, \& {Silk}}]{Bertone2005}
{Bertone}, G., {Hooper}, D., \& {Silk}, J. 2005, \bibinfo{title}{{Particle dark matter: evidence, candidates and constraints},} \physrep, 405, 279, \dodoi{10.1016/j.physrep.2004.08.031}

\bibitem[{T.~D. {Brandt} \& B. {Kocsis}(2015){Brandt} \& {Kocsis}}]{Brandt2015}
{Brandt}, T.~D., \& {Kocsis}, B. 2015, \bibinfo{title}{{Disrupted Globular Clusters Can Explain the Galactic Center Gamma-Ray Excess},} \apj, 812, 15, \dodoi{10.1088/0004-637X/812/1/15}

\bibitem[{M. Buschmann {et~al.}(2020)Buschmann, Rodd, Safdi, Chang, Mishra-Sharma, Lisanti, \& Macias}]{Buschmann:2020adf}
Buschmann, M., Rodd, N.~L., Safdi, B.~R., {et~al.} 2020, \bibinfo{title}{{Foreground Mismodeling and the Point Source Explanation of the Fermi Galactic Center Excess},} Phys. Rev. D, 102, 023023, \dodoi{10.1103/PhysRevD.102.023023}

\bibitem[{F. {Calore} {et~al.}(2015{\natexlab{a}}){Calore}, {Cholis}, {McCabe}, \& {Weniger}}]{Calore2015}
{Calore}, F., {Cholis}, I., {McCabe}, C., \& {Weniger}, C. 2015{\natexlab{a}}, \bibinfo{title}{{A tale of tails: Dark matter interpretations of the Fermi GeV excess in light of background model systematics},} \prd, 91, 063003, \dodoi{10.1103/PhysRevD.91.063003}

\bibitem[{F. {Calore} {et~al.}(2015{\natexlab{b}}){Calore}, {Cholis}, \& {Weniger}}]{Calore2015b}
{Calore}, F., {Cholis}, I., \& {Weniger}, C. 2015{\natexlab{b}}, \bibinfo{title}{{Background model systematics for the Fermi GeV excess},} \jcap, 2015, 038, \dodoi{10.1088/1475-7516/2015/03/038}

\bibitem[{J.-M. {Casandjian}(2015){Casandjian}}]{Casandjian2015}
{Casandjian}, J.-M. 2015, \bibinfo{title}{{Local H i Emissivity Measured with Fermi-LAT and Implications for Cosmic-Ray Spectra},} \apj, 806, 240, \dodoi{10.1088/0004-637X/806/2/240}

\bibitem[{J.-M. {Casandjian} \& I. {Grenier}(2009){Casandjian} \& {Grenier}}]{Casandjian2009}
{Casandjian}, J.-M., \& {Grenier}, I. 2009, \bibinfo{title}{{High Energy Gamma-Ray Emission from the Loop I region},} ArXiv e-prints, arXiv:0912.3478.
\newblock \doarXiv{0912.3478}

\bibitem[{G.~L. {Case} \& D. {Bhattacharya}(1998){Case} \& {Bhattacharya}}]{Case1998}
{Case}, G.~L., \& {Bhattacharya}, D. 1998, \bibinfo{title}{{A New {\ensuremath{\Sigma}}-D Relation and Its Application to the Galactic Supernova Remnant Distribution},} \apj, 504, 761, \dodoi{10.1086/306089}

\bibitem[{R. {Catena} \& P. {Ullio}(2010){Catena} \& {Ullio}}]{Catena2010}
{Catena}, R., \& {Ullio}, P. 2010, \bibinfo{title}{{A novel determination of the local dark matter density},} \jcap, 2010, 004, \dodoi{10.1088/1475-7516/2010/08/004}

\bibitem[{M. {Cautun} {et~al.}(2020){Cautun}, {Ben{\'\i}tez-Llambay}, {Deason}, {Frenk}, {Fattahi}, {G{\'o}mez}, {Grand}, {Oman}, {Navarro}, \& {Simpson}}]{Cantun2020}
{Cautun}, M., {Ben{\'\i}tez-Llambay}, A., {Deason}, A.~J., {et~al.} 2020, \bibinfo{title}{{The milky way total mass profile as inferred from Gaia DR2},} \mnras, 494, 4291, \dodoi{10.1093/mnras/staa1017}

\bibitem[{J. {Chang} {et~al.}(2017){Chang}, {Ambrosi}, {An}, {et~al.}}]{Chang2017}
{Chang}, J., {Ambrosi}, G., {An}, Q., {et~al.} 2017, \bibinfo{title}{{The DArk Matter Particle Explorer mission},} Astropart. Phys., 95, 6, \dodoi{10.1016/j.astropartphys.2017.08.005}

\bibitem[{E. {Charles} {et~al.}(2016){Charles}, {S{\'a}nchez-Conde}, {Anderson}, {Caputo}, {Cuoco}, {Di Mauro}, {Drlica-Wagner}, {Gomez-Vargas}, {Meyer}, {Tibaldo}, {Wood}, {Zaharijas}, {Zimmer}, {Ajello}, {Albert}, {Baldini}, {Bechtol}, {Bloom}, {Ceraudo}, {Cohen-Tanugi}, {Digel}, {Gaskins}, {Gustafsson}, {Mirabal}, \& {Razzano}}]{Charles2016}
{Charles}, E., {S{\'a}nchez-Conde}, M., {Anderson}, B., {et~al.} 2016, \bibinfo{title}{{Sensitivity projections for dark matter searches with the Fermi large area telescope},} \physrep, 636, 1, \dodoi{10.1016/j.physrep.2016.05.001}

\bibitem[{J.-G. Cheng {et~al.}(2023)Cheng, Liang, \& Liang}]{Cheng:2023chi}
Cheng, J.-G., Liang, Y.-F., \& Liang, E.-W. 2023, \bibinfo{title}{{Search for the gamma-ray spectral lines with the DAMPE and the Fermi-LAT observations},} Phys. Rev. D, 108, 063015, \dodoi{10.1103/PhysRevD.108.063015}

\bibitem[{K.-S. {Cheng} {et~al.}(2011){Cheng}, {Chernyshov}, {Dogiel}, {Ko}, \& {Ip}}]{Cheng2011}
{Cheng}, K.-S., {Chernyshov}, D.~O., {Dogiel}, V.~A., {Ko}, C.-M., \& {Ip}, W.-H. 2011, \bibinfo{title}{{Origin of the Fermi Bubble},} \apjl, 731, L17, \dodoi{10.1088/2041-8205/731/1/L17}

\bibitem[{M. {Chernyakova} {et~al.}(2011){Chernyakova}, {Malyshev}, {Aharonian}, {Crocker}, \& {Jones}}]{Chernyakova2011}
{Chernyakova}, M., {Malyshev}, D., {Aharonian}, F.~A., {Crocker}, R.~M., \& {Jones}, D.~I. 2011, \bibinfo{title}{{The High-energy, Arcminute-scale Galactic Center Gamma-ray Source},} \apj, 726, 60, \dodoi{10.1088/0004-637X/726/2/60}

\bibitem[{I. {Cholis} {et~al.}(2021){Cholis}, {Zhong}, {McDermott}, \& {Surdutovich}}]{Cholis2022_params}
{Cholis}, I., {Zhong}, Y.-M., {McDermott}, S.~D., \& {Surdutovich}, J.~P. 2021, The Return of the Templates: Revisiting the Galactic Center Excess with Multi-Messenger Observations, Zenodo, \dodoi{10.5281/ZENODO.6423495}

\bibitem[{I. {Cholis} {et~al.}(2022){Cholis}, {Zhong}, {McDermott}, {et~al.}}]{Cholis2022}
{Cholis}, I., {Zhong}, Y.-M., {McDermott}, S.~D., {et~al.} 2022, \bibinfo{title}{{Return of the templates: Revisiting the Galactic Center excess with multimessenger observations},} \prd, 105, 103023, \dodoi{10.1103/PhysRevD.105.103023}

\bibitem[{M. {Cirelli} {et~al.}(2011){Cirelli}, {Corcella}, {Hektor}, {et~al.}}]{PPPC2011}
{Cirelli}, M., {Corcella}, G., {Hektor}, A., {et~al.} 2011, \bibinfo{title}{{PPPC 4 DM ID: a poor particle physicist cookbook for dark matter indirect detection},} \jcap, 2011, 051, \dodoi{10.1088/1475-7516/2011/03/051}

\bibitem[{B. {Coleman} {et~al.}(2020){Coleman}, {Paterson}, {Gordon}, {Macias}, \& {Ploeg}}]{Coleman2020}
{Coleman}, B., {Paterson}, D., {Gordon}, C., {Macias}, O., \& {Ploeg}, H. 2020, \bibinfo{title}{{Maximum entropy estimation of the Galactic bulge morphology via the VVV Red Clump},} \mnras, 495, 3350, \dodoi{10.1093/mnras/staa1281}

\bibitem[{R.~M. {Crocker} \& F. {Aharonian}(2011){Crocker} \& {Aharonian}}]{Crocker2011}
{Crocker}, R.~M., \& {Aharonian}, F. 2011, \bibinfo{title}{{Fermi Bubbles: Giant, Multibillion-Year-Old Reservoirs of Galactic Center Cosmic Rays},} \prl, 106, 101102, \dodoi{10.1103/PhysRevLett.106.101102}

\bibitem[{M.-Y. {Cui} {et~al.}(2017){Cui}, {Yuan}, {Tsai}, \& {Fan}}]{Cui2017}
{Cui}, M.-Y., {Yuan}, Q., {Tsai}, Y.-L.~S., \& {Fan}, Y.-Z. 2017, \bibinfo{title}{{Possible Dark Matter Annihilation Signal in the AMS-02 Antiproton Data},} \prl, 118, 191101, \dodoi{10.1103/PhysRevLett.118.191101}

\bibitem[{Y.-X. {Cui} {et~al.}(2023){Cui}, {Ma}, {Yuan}, {et~al.}}]{Cui2023}
{Cui}, Y.-X., {Ma}, P.-X., {Yuan}, G.-W., {et~al.} 2023, \bibinfo{title}{{Study of the global alignment for the DAMPE detector},} Nucl. Inst. Methods A, 1046, 167670, \dodoi{10.1016/j.nima.2022.167670}

\bibitem[{A. {Cuoco} {et~al.}(2017){Cuoco}, {Kr{\"a}mer}, \& {Korsmeier}}]{Cuoco2017}
{Cuoco}, A., {Kr{\"a}mer}, M., \& {Korsmeier}, M. 2017, \bibinfo{title}{{Novel Dark Matter Constraints from Antiprotons in Light of AMS-02},} \prl, 118, 191102, \dodoi{10.1103/PhysRevLett.118.191102}

\bibitem[{T. {Daylan} {et~al.}(2016){Daylan}, {Finkbeiner}, {Hooper}, {et~al.}}]{Daylan2016}
{Daylan}, T., {Finkbeiner}, D.~P., {Hooper}, D., {et~al.} 2016, \bibinfo{title}{{The characterization of the gamma-ray signal from the central Milky Way: A case for annihilating dark matter},} Phys. Dark Univ., 12, 1, \dodoi{10.1016/j.dark.2015.12.005}

\bibitem[{P.~F. {de Salas} \& A. {Widmark}(2021){de Salas} \& {Widmark}}]{deSalas2021}
{de Salas}, P.~F., \& {Widmark}, A. 2021, \bibinfo{title}{{Dark matter local density determination: recent observations and future prospects},} Reports on Progress in Physics, 84, 104901, \dodoi{10.1088/1361-6633/ac24e7}

\bibitem[{H. Dembinski {et~al.}(2020)Dembinski, Ongmongkolkul, {et~al.}}]{iminuit}
Dembinski, H., Ongmongkolkul, P., {et~al.} 2020, \bibinfo{title}{scikit-hep/iminuit,} Zenodo, \dodoi{10.5281/zenodo.3949207}

\bibitem[{M. {Di Mauro}(2021){Di Mauro}}]{DiMauro2021}
{Di Mauro}, M. 2021, \bibinfo{title}{{Characteristics of the Galactic Center excess measured with 11 years of F e r m i -LAT data},} \prd, 103, 063029, \dodoi{10.1103/PhysRevD.103.063029}

\bibitem[{T. {Do} {et~al.}(2019){Do}, {Hees}, {Ghez}, {Martinez}, {et~al.}}]{Keck2019}
{Do}, T., {Hees}, A., {Ghez}, A., {Martinez}, G.~D., {et~al.} 2019, \bibinfo{title}{{Relativistic redshift of the star S0-2 orbiting the Galactic Center supermassive black hole},} Science, 365, 664, \dodoi{10.1126/science.aav8137}

\bibitem[{K.-K. {Duan} {et~al.}(2019){Duan}, {Jiang}, {Liang}, {et~al.}}]{Duan2019}
{Duan}, K.-K., {Jiang}, W., {Liang}, Y.-F., {et~al.} 2019, \bibinfo{title}{{DmpIRFs and DmpST: DAMPE instrument response functions and science tools for gamma-ray data analysis},} \raa, 19, 132, \dodoi{10.1088/1674-4527/19/9/132}

\bibitem[{K.-K. {Duan} {et~al.}(2025{\natexlab{a}}){Duan}, {Shen}, {Xu}, {Jiang}, \& {Li}}]{Duan2025}
{Duan}, K.-K., {Shen}, Z.-Q., {Xu}, Z.-L., {Jiang}, W., \& {Li}, X. 2025{\natexlab{a}}, \bibinfo{title}{{PSF calibration of DAMPE for gamma-ray observations},} Astroparticle Physics, 165, 103058, \dodoi{10.1016/j.astropartphys.2024.103058}

\bibitem[{K.-K. {Duan} {et~al.}(2025{\natexlab{b}}){Duan}, {Wang}, {Li}, {Xu}, {Sming Tsai}, \& {Fan}}]{Duan2025b}
{Duan}, K.-K., {Wang}, X., {Li}, W.-H., {et~al.} 2025{\natexlab{b}}, \bibinfo{title}{{Scrutinizing the impact of the solar modulation on AMS-02 antiproton excess},} \jcap, 2025, 049, \dodoi{10.1088/1475-7516/2025/10/049}

\bibitem[{R.~J. {Egger} \& B. {Aschenbach}(1995){Egger} \& {Aschenbach}}]{Egger1995}
{Egger}, R.~J., \& {Aschenbach}, B. 1995, \bibinfo{title}{{Interaction of the Loop I supershell with the Local Hot Bubble.},} \aap, 294, L25, \dodoi{10.48550/arXiv.astro-ph/9412086}

\bibitem[{Y.-Z. {Fan} {et~al.}(2022){Fan}, {Tang}, {Tsai}, \& {Wu}}]{Fan2022a}
{Fan}, Y.-Z., {Tang}, T.-P., {Tsai}, Y.-L.~S., \& {Wu}, L. 2022, \bibinfo{title}{{Inert Higgs Dark Matter for CDF II W -Boson Mass and Detection Prospects},} \prl, 129, 091802, \dodoi{10.1103/PhysRevLett.129.091802}

\bibitem[{Y.~Z. {Fan} {et~al.}(2022){Fan}, {Chang}, {Guo}, {Yuan}, {Hu}, {Li}, {Yue}, {Huang}, {Liu}, {Feng}, {Zhang}, {Wei}, {Sun}, {Yu}, {Kong}, {Zhao}, {Zang}, {Jiang}, {Pan}, {Wei}, {Wang}, {Duan}, {Shen}, {Xia}, {Xu}, {Feng}, {Huang}, {TSAI}, {Wei}, {Zeng}, {He}, {Li}, {Yang}, {Yan}, {Zhang}, {Wu}, \& {Wei}}]{Fan2022b}
{Fan}, Y.~Z., {Chang}, J., {Guo}, J.~H., {et~al.} 2022, \bibinfo{title}{{Very Large Area Gamma-ray Space Telescope (VLAST)},} Acta Astronomica Sinica, 63, 27

\bibitem[{Y.-Z. Fan {et~al.}(2024)Fan, Shen, Liang, Li, Duan, Xia, Huang, Feng, \& Yuan}]{Fan:2024rcr}
Fan, Y.-Z., Shen, Z.-Q., Liang, Y.-F., {et~al.} 2024, \bibinfo{title}{{A $\sim 43$ GeV $\gamma$-ray line signature in the directions of a group of nearby massive galaxy clusters},} \doarXiv{2407.11737}

\bibitem[{ {Fermi-LAT Collaboration}(2009){Fermi-LAT Collaboration}}]{gll_iem_v02}
{Fermi-LAT Collaboration}. 2009, {Description and Caveats for the LAT Team Model of Diffuse Gamma-Ray Emission Version: gll iem v02.fit}, \url{https://fermi.gsfc.nasa.gov/ssc/data/access/lat/ring_for_FSSC_final4.pdf}

\bibitem[{H.~T. {Freudenreich}(1998){Freudenreich}}]{Freudenreich1998}
{Freudenreich}, H.~T. 1998, \bibinfo{title}{{A COBE Model of the Galactic Bar and Disk},} \apj, 492, 495, \dodoi{10.1086/305065}

\bibitem[{A. {Gautam} {et~al.}(2022){Gautam}, {Crocker}, {Ferrario}, {et~al.}}]{Gautam2022}
{Gautam}, A., {Crocker}, R.~M., {Ferrario}, L., {et~al.} 2022, \bibinfo{title}{{Millisecond pulsars from accretion-induced collapse as the origin of the Galactic Centre gamma-ray excess signal},} Nature Astronomy, 6, 703, \dodoi{10.1038/s41550-022-01658-3}

\bibitem[{A.~M. {Ghez} {et~al.}(2008){Ghez}, {Salim}, {Weinberg}, {Lu}, {Do}, {Dunn}, {Matthews}, {Morris}, {Yelda}, {Becklin}, {Kremenek}, {Milosavljevic}, \& {Naiman}}]{Ghez2008}
{Ghez}, A.~M., {Salim}, S., {Weinberg}, N.~N., {et~al.} 2008, \bibinfo{title}{{Measuring Distance and Properties of the Milky Way's Central Supermassive Black Hole with Stellar Orbits},} \apj, 689, 1044, \dodoi{10.1086/592738}

\bibitem[{L. {Goodenough} \& D. {Hooper}(2009){Goodenough} \& {Hooper}}]{Goodenough2009}
{Goodenough}, L., \& {Hooper}, D. 2009, \bibinfo{title}{{Possible Evidence For Dark Matter Annihilation In The Inner Milky Way From The Fermi Gamma Ray Space Telescope},} ArXiv e-prints, arXiv:0910.2998.
\newblock \doarXiv{0910.2998}

\bibitem[{C. {Gordon} \& O. {Mac{\'\i}as}(2013){Gordon} \& {Mac{\'\i}as}}]{Gordon2013}
{Gordon}, C., \& {Mac{\'\i}as}, O. 2013, \bibinfo{title}{{Dark matter and pulsar model constraints from Galactic Center Fermi-LAT gamma-ray observations},} \prd, 88, 083521, \dodoi{10.1103/PhysRevD.88.083521}

\bibitem[{K.~M. {G{\'o}rski} {et~al.}(2005){G{\'o}rski}, {Hivon}, {Banday}, {Wandelt}, {Hansen}, {Reinecke}, \& {Bartelmann}}]{Healpix2005}
{G{\'o}rski}, K.~M., {Hivon}, E., {Banday}, A.~J., {et~al.} 2005, \bibinfo{title}{{HEALPix: A Framework for High-Resolution Discretization and Fast Analysis of Data Distributed on the Sphere},} \apj, 622, 759, \dodoi{10.1086/427976}

\bibitem[{F. {Guo} \& W.~G. {Mathews}(2012){Guo} \& {Mathews}}]{Guo2012}
{Guo}, F., \& {Mathews}, W.~G. 2012, \bibinfo{title}{{The Fermi Bubbles. I. Possible Evidence for Recent AGN Jet Activity in the Galaxy},} \apj, 756, 181, \dodoi{10.1088/0004-637X/756/2/181}

\bibitem[{C.~R. {Harris} {et~al.}(2020){Harris} {et~al.}}]{numpy2020}
{Harris}, C.~R., {et~al.} 2020, \bibinfo{title}{{Array Programming with NumPy},} Nature, 585, 357, \dodoi{10.1038/s41586-020-2649-2}

\bibitem[{C.~G.~T. {Haslam} {et~al.}(1982){Haslam}, {Salter}, {Stoffel}, \& {Wilson}}]{Haslam1982}
{Haslam}, C.~G.~T., {Salter}, C.~J., {Stoffel}, H., \& {Wilson}, W.~E. 1982, \bibinfo{title}{{A 408-MHZ All-Sky Continuum Survey. II. The Atlas of Contour Maps},} \aaps, 47, 1

\bibitem[{D. {Hooper} \& L. {Goodenough}(2011){Hooper} \& {Goodenough}}]{Hooper2011}
{Hooper}, D., \& {Goodenough}, L. 2011, \bibinfo{title}{{Dark matter annihilation in the Galactic Center as seen by the Fermi Gamma Ray Space Telescope},} Phys. Lett. B, 697, 412, \dodoi{10.1016/j.physletb.2011.02.029}

\bibitem[{D. {Hooper} \& T. {Linden}(2011){Hooper} \& {Linden}}]{2011PhRvD..84l3005H}
{Hooper}, D., \& {Linden}, T. 2011, \bibinfo{title}{{Origin of the gamma rays from the Galactic Center},} \prd, 84, 123005, \dodoi{10.1103/PhysRevD.84.123005}

\bibitem[{D. {Hooper} \& T.~R. {Slatyer}(2013){Hooper} \& {Slatyer}}]{Hooper2013}
{Hooper}, D., \& {Slatyer}, T.~R. 2013, \bibinfo{title}{{Two emission mechanisms in the Fermi Bubbles: A possible signal of annihilating dark matter},} Phys. Dark Univ., 2, 118, \dodoi{10.1016/j.dark.2013.06.003}

\bibitem[{X. {Huang} {et~al.}(2016){Huang}, {En{\ss}lin}, \& {Selig}}]{Huang2016}
{Huang}, X., {En{\ss}lin}, T., \& {Selig}, M. 2016, \bibinfo{title}{{Galactic dark matter search via phenomenological astrophysics modeling},} \jcap, 2016, 030, \dodoi{10.1088/1475-7516/2016/04/030}

\bibitem[{X. {Huang} {et~al.}(2021){Huang}, {Yuan}, \& {Fan}}]{Huang2021}
{Huang}, X., {Yuan}, Q., \& {Fan}, Y.-Z. 2021, \bibinfo{title}{{A GeV-TeV particle component and the barrier of cosmic-ray sea in the Central Molecular Zone},} Nature Communications, 12, 6169, \dodoi{10.1038/s41467-021-26436-z}

\bibitem[{J.~D. Hunter(2007)Hunter}]{matplotlib2007}
Hunter, J.~D. 2007, \bibinfo{title}{Matplotlib: A 2D graphics environment,} Comput. Sci. Eng., 9, 90, \dodoi{10.1109/MCSE.2007.55}

\bibitem[{F. {James} \& M. {Roos}(1975){James} \& {Roos}}]{MINUIT1975}
{James}, F., \& {Roos}, M. 1975, \bibinfo{title}{{Minuit - a system for function minimization and analysis of the parameter errors and correlations},} Comput. Phys. Commun., 10, 343, \dodoi{10.1016/0010-4655(75)90039-9}

\bibitem[{W. {Jiang} {et~al.}(2020){Jiang}, {Li}, {Duan}, {et~al.}}]{Jiang2020}
{Jiang}, W., {Li}, X., {Duan}, K.-K., {et~al.} 2020, \bibinfo{title}{The boresight alignment of the {DArk} Matter Particle Explorer,} \raa, 20, 092, \dodoi{10.1088/1674-4527/20/6/92}

\bibitem[{G. {J{\'o}hannesson} {et~al.}(2018){J{\'o}hannesson}, {Porter}, \& {Moskalenko}}]{Johannesson2018}
{J{\'o}hannesson}, G., {Porter}, T.~A., \& {Moskalenko}, I.~V. 2018, \bibinfo{title}{{The Three-dimensional Spatial Distribution of Interstellar Gas in the Milky Way: Implications for Cosmic Rays and High-energy Gamma-ray Emissions},} \apj, 856, 45, \dodoi{10.3847/1538-4357/aab26e}

\bibitem[{C. {Karwin} {et~al.}(2017){Karwin}, {Murgia}, {Tait}, {et~al.}}]{Karwin2017}
{Karwin}, C., {Murgia}, S., {Tait}, T. M.~P., {et~al.} 2017, \bibinfo{title}{{Dark matter interpretation of the Fermi-LAT observation toward the Galactic Center},} \prd, 95, 103005, \dodoi{10.1103/PhysRevD.95.103005}

\bibitem[{U. {Keshet} \& I. {Gurwich}(2017){Keshet} \& {Gurwich}}]{Keshet2017}
{Keshet}, U., \& {Gurwich}, I. 2017, \bibinfo{title}{{Fermi Bubble Edges: Spectrum and Diffusion Function},} \apj, 840, 7, \dodoi{10.3847/1538-4357/aa6936}

\bibitem[{R.~K. {Leane} \& T.~R. {Slatyer}(2019){Leane} \& {Slatyer}}]{Leane2019}
{Leane}, R.~K., \& {Slatyer}, T.~R. 2019, \bibinfo{title}{{Revival of the Dark Matter Hypothesis for the Galactic Center Gamma-Ray Excess},} \prl, 123, 241101, \dodoi{10.1103/PhysRevLett.123.241101}

\bibitem[{S. {Li}(2026){Li}}]{2026JHEAp..5100536L}
{Li}, S. 2026, \bibinfo{title}{{Search for {\ensuremath{\gamma}}-ray emission from dwarf spheroidal galaxies with Fermi-LAT data},} Journal of High Energy Astrophysics, 51, 100536, \dodoi{10.1016/j.jheap.2025.100536}

\bibitem[{D.~R. {Lorimer} {et~al.}(2006){Lorimer}, {Faulkner}, {Lyne}, {Manchester}, {Kramer}, {McLaughlin}, {Hobbs}, {Possenti}, {Stairs}, {Camilo}, {Burgay}, {D'Amico}, {Corongiu}, \& {Crawford}}]{Lorimer2006}
{Lorimer}, D.~R., {Faulkner}, A.~J., {Lyne}, A.~G., {et~al.} 2006, \bibinfo{title}{{The Parkes Multibeam Pulsar Survey - VI. Discovery and timing of 142 pulsars and a Galactic population analysis},} \mnras, 372, 777, \dodoi{10.1111/j.1365-2966.2006.10887.x}

\bibitem[{O. {Macias} {et~al.}(2018){Macias}, {Gordon}, {Crocker}, {Coleman}, {Paterson}, {Horiuchi}, \& {Pohl}}]{Macias2018}
{Macias}, O., {Gordon}, C., {Crocker}, R.~M., {et~al.} 2018, \bibinfo{title}{{Galactic bulge preferred over dark matter for the Galactic centre gamma-ray excess},} Nature Astronomy, 2, 387, \dodoi{10.1038/s41550-018-0414-3}

\bibitem[{O. {Macias} {et~al.}(2019){Macias}, {Horiuchi}, {et~al.}}]{Macias2019}
{Macias}, O., {Horiuchi}, S., {et~al.} 2019, \bibinfo{title}{{Strong evidence that the galactic bulge is shining in gamma rays},} \jcap, 2019, 042, \dodoi{10.1088/1475-7516/2019/09/042}

\bibitem[{J.~R. {Mattox} {et~al.}(1996){Mattox}, {Bertsch}, {Chiang}, {Dingus}, {Digel}, {Esposito}, {Fierro}, {Hartman}, {Hunter}, {Kanbach}, {Kniffen}, {Lin}, {Macomb}, {Mayer-Hasselwander}, {Michelson}, {von Montigny}, {Mukherjee}, {Nolan}, {Ramanamurthy}, {Schneid}, {Sreekumar}, {Thompson}, \& {Willis}}]{Mattox1996}
{Mattox}, J.~R., {Bertsch}, D.~L., {Chiang}, J., {et~al.} 1996, \bibinfo{title}{{The Likelihood Analysis of EGRET Data},} \apj, 461, 396, \dodoi{10.1086/177068}

\bibitem[{A. {McDaniel} {et~al.}(2024){McDaniel}, {Ajello}, {Karwin}, {Di Mauro}, {Drlica-Wagner}, \& {S{\'a}nchez-Conde}}]{McDaniel2024}
{McDaniel}, A., {Ajello}, M., {Karwin}, C.~M., {et~al.} 2024, \bibinfo{title}{{Legacy analysis of dark matter annihilation from the Milky Way dwarf spheroidal galaxies with 14 years of Fermi -LAT data},} \prd, 109, 063024, \dodoi{10.1103/PhysRevD.109.063024}

\bibitem[{S.~D. {McDermott} {et~al.}(2023){McDermott}, {Zhong}, \& {Cholis}}]{McDermott2023}
{McDermott}, S.~D., {Zhong}, Y.-M., \& {Cholis}, I. 2023, \bibinfo{title}{{On the morphology of the gamma-ray galactic centre excess},} \mnras, 522, L21, \dodoi{10.1093/mnrasl/slad035}

\bibitem[{P.~J. {McMillan}(2017){McMillan}}]{McMillan2017}
{McMillan}, P.~J. 2017, \bibinfo{title}{{The mass distribution and gravitational potential of the Milky Way},} \mnras, 465, 76, \dodoi{10.1093/mnras/stw2759}

\bibitem[{N. {Mirabal}(2013){Mirabal}}]{Mirabal2013}
{Mirabal}, N. 2013, \bibinfo{title}{{Dark matter versus pulsars: catching the impostor},} \mnras, 436, 2461, \dodoi{10.1093/mnras/stt1740}

\bibitem[{G. {Mou} {et~al.}(2014){Mou}, {Yuan}, {Bu}, {Sun}, \& {Su}}]{Mou2014}
{Mou}, G., {Yuan}, F., {Bu}, D., {Sun}, M., \& {Su}, M. 2014, \bibinfo{title}{{Fermi Bubbles Inflated by Winds Launched from the Hot Accretion Flow in Sgr A*},} \apj, 790, 109, \dodoi{10.1088/0004-637X/790/2/109}

\bibitem[{S. {Murgia}(2020){Murgia}}]{Murgia2020}
{Murgia}, S. 2020, \bibinfo{title}{{The Fermi{\textendash}LAT Galactic Center Excess: Evidence of Annihilating Dark Matter?},} Annu. Rev. Nucl. Part. Sci., 70, 455, \dodoi{10.1146/annurev-nucl-101916-123029}

\bibitem[{M.~M. {Muru} {et~al.}(2025){Muru}, {Silk}, {Libeskind}, {Gottl{\"o}ber}, \& {Hoffman}}]{Muru2025}
{Muru}, M.~M., {Silk}, J., {Libeskind}, N.~I., {Gottl{\"o}ber}, S., \& {Hoffman}, Y. 2025, \bibinfo{title}{{Fermi-LAT Galactic Center Excess Morphology of Dark Matter in Simulations of the Milky Way Galaxy},} \prl, 135, 161005, \dodoi{10.1103/g9qz-h8wd}

\bibitem[{S.~A. {Narayanan} \& T.~R. {Slatyer}(2017){Narayanan} \& {Slatyer}}]{Narayanan2017}
{Narayanan}, S.~A., \& {Slatyer}, T.~R. 2017, \bibinfo{title}{{A latitude-dependent analysis of the leptonic hypothesis for the Fermi Bubbles},} \mnras, 468, 3051, \dodoi{10.1093/mnras/stx577}

\bibitem[{J.~F. {Navarro} {et~al.}(1996){Navarro}, {Frenk}, \& {White}}]{Navarro1996}
{Navarro}, J.~F., {Frenk}, C.~S., \& {White}, S. D.~M. 1996, \bibinfo{title}{{The Structure of Cold Dark Matter Halos},} \apj, 462, 563, \dodoi{10.1086/177173}

\bibitem[{X. {Pan} {et~al.}(2024){Pan}, {Jiang}, {Yue}, {Lei}, {Cui}, \& {Yuan}}]{PanX2024}
{Pan}, X., {Jiang}, W., {Yue}, C., {et~al.} 2024, \bibinfo{title}{{Simulation study of the performance of the Very Large Area gamma-ray Space Telescope},} Nuclear Science and Techniques, 35, 149, \dodoi{10.1007/s41365-024-01499-x}

\bibitem[{M. {Pohl} {et~al.}(2022{\natexlab{a}}){Pohl}, {Macias}, {Coleman}, \& {Gordon}}]{Pohl2022}
{Pohl}, M., {Macias}, O., {Coleman}, P., \& {Gordon}, C. 2022{\natexlab{a}}, \bibinfo{title}{{Assessing the Impact of Hydrogen Absorption on the Characteristics of the Galactic Center Excess},} \apj, 929, 136, \dodoi{10.3847/1538-4357/ac6032}

\bibitem[{M. {Pohl} {et~al.}(2022{\natexlab{b}}){Pohl}, {Macias}, \& {Gordon}}]{Pohl2022_data}
{Pohl}, M., {Macias}, O., \& {Gordon}, C. 2022{\natexlab{b}}, Analysis Templates for Pohl, Macias, Coleman, and Gordon (2022), Zenodo, \dodoi{10.5281/ZENODO.6276721}

\bibitem[{T.~A. {Porter} {et~al.}(2017){Porter}, {J{\'o}hannesson}, \& {Moskalenko}}]{Porter2017}
{Porter}, T.~A., {J{\'o}hannesson}, G., \& {Moskalenko}, I.~V. 2017, \bibinfo{title}{{High-energy Gamma Rays from the Milky Way: Three-dimensional Spatial Models for the Cosmic-Ray and Radiation Field Densities in the Interstellar Medium},} \apj, 846, 67, \dodoi{10.3847/1538-4357/aa844d}

\bibitem[{T.~A. {Porter} {et~al.}(2008){Porter}, {Moskalenko}, {Strong}, {Orlando}, \& {Bouchet}}]{Porter:2008ve}
{Porter}, T.~A., {Moskalenko}, I.~V., {Strong}, A.~W., {Orlando}, E., \& {Bouchet}, L. 2008, \bibinfo{title}{{Inverse Compton Origin of the Hard X-Ray and Soft Gamma-Ray Emission from the Galactic Ridge},} \apj, 682, 400, \dodoi{10.1086/589615}

\bibitem[{P. {Predehl} {et~al.}(2020){Predehl}, {Sunyaev}, {Becker}, {Brunner}, {Burenin}, {Bykov}, {Cherepashchuk}, {Chugai}, {Churazov}, {Doroshenko}, {Eismont}, {Freyberg}, {Gilfanov}, {Haberl}, {Khabibullin}, {Krivonos}, {Maitra}, {Medvedev}, {Merloni}, {Nandra}, {Nazarov}, {Pavlinsky}, {Ponti}, {Sanders}, {Sasaki}, {Sazonov}, {Strong}, \& {Wilms}}]{Predehl2020}
{Predehl}, P., {Sunyaev}, R.~A., {Becker}, W., {et~al.} 2020, \bibinfo{title}{{Detection of large-scale X-ray bubbles in the Milky Way halo},} \nat, 588, 227, \dodoi{10.1038/s41586-020-2979-0}

\bibitem[{A.~M. {Price-Whelan} {et~al.}(2022){Price-Whelan}, {Lim}, {et~al.}}]{astropy2022}
{Price-Whelan}, A.~M., {Lim}, P.~L., {et~al.} 2022, \bibinfo{title}{{The Astropy Project: Sustaining and Growing a Community-oriented Open-source Project and the Latest Major Release (v5.0) of the Core Package},} \apj, 935, 167, \dodoi{10.3847/1538-4357/ac7c74}

\bibitem[{E.~D. {Ramirez} {et~al.}(2025){Ramirez}, {Sun}, {Buckley}, {Mishra-Sharma}, \& {Slatyer}}]{Ramirez2025}
{Ramirez}, E.~D., {Sun}, Y., {Buckley}, M.~R., {Mishra-Sharma}, S., \& {Slatyer}, T.~R. 2025, \bibinfo{title}{{Inferring the morphology of the Galactic Center excess with Gaussian processes},} \prd, 111, 063065, \dodoi{10.1103/PhysRevD.111.063065}

\bibitem[{M. {Remazeilles} {et~al.}(2015){Remazeilles}, {Dickinson}, {Banday}, {Bigot-Sazy}, \& {Ghosh}}]{Remazeilles2015}
{Remazeilles}, M., {Dickinson}, C., {Banday}, A.~J., {Bigot-Sazy}, M.-A., \& {Ghosh}, T. 2015, \bibinfo{title}{{An improved source-subtracted and destriped 408-MHz all-sky map},} \mnras, 451, 4311, \dodoi{10.1093/mnras/stv1274}

\bibitem[{C.~J. {Salter}(1983){Salter}}]{Salter1983}
{Salter}, C.~J. 1983, \bibinfo{title}{{Loop-I the North Polar Spur - a Major Feature of the Local Interstellar Environment},} Bulletin of the Astronomical Society of India, 11, 1

\bibitem[{K.~C. Sarkar(2024)Sarkar}]{Sarkar:2024mjm}
Sarkar, K.~C. 2024, \bibinfo{title}{{The Fermi/eROSITA bubbles: a look into the nuclear outflow from the Milky Way},} Astron. Astrophys. Rev., 32, 1, \dodoi{10.1007/s00159-024-00152-1}

\bibitem[{Z.-Q. {Shen} {et~al.}(2023){Shen}, {Duan}, {Jiang}, {Xu}, \& {Li}}]{Shen2023ICRC2}
{Shen}, Z.-Q., {Duan}, K.-K., {Jiang}, W., {Xu}, Z.-L., \& {Li}, X. 2023, \bibinfo{title}{{Recent progresses on the $\gamma$-ray observations of DAMPE},} in ICRC2023, Vol. 444, Proc. Sci., 670, \dodoi{10.22323/1.444.0670}

\bibitem[{Z.-Q. {Shen} {et~al.}(2024){Shen}, {Li}, {Duan}, {Jiang}, {Xu}, {Yue}, \& {Li}}]{Shen2024}
{Shen}, Z.-Q., {Li}, W.-H., {Duan}, K.-K., {et~al.} 2024, \bibinfo{title}{{The Calibrations of the DAMPE {\ensuremath{\gamma}}-Ray Effective Area},} \apj, 976, 53, \dodoi{10.3847/1538-4357/ad834b}

\bibitem[{Y. {Sofue}(2015){Sofue}}]{Sofue2015}
{Sofue}, Y. 2015, \bibinfo{title}{{The North Polar Spur and Aquila Rift},} \mnras, 447, 3824, \dodoi{10.1093/mnras/stu2661}

\bibitem[{D. {Song} {et~al.}(2024){Song}, {Eckner}, {Gordon}, {Calore}, {Macias}, {Abazajian}, {Horiuchi}, {Kaplinghat}, \& {Pohl}}]{Song2024}
{Song}, D., {Eckner}, C., {Gordon}, C., {et~al.} 2024, \bibinfo{title}{{Robust inference of the Galactic Centre gamma-ray excess spatial properties},} \mnras, 530, 4395, \dodoi{10.1093/mnras/stae923}

\bibitem[{A.~W. {Strong} \& I.~V. {Moskalenko}(1998){Strong} \& {Moskalenko}}]{Strong:1998pw}
{Strong}, A.~W., \& {Moskalenko}, I.~V. 1998, \bibinfo{title}{{Propagation of Cosmic-Ray Nucleons in the Galaxy},} \apj, 509, 212, \dodoi{10.1086/306470}

\bibitem[{A.~W. {Strong} {et~al.}(2000){Strong}, {Moskalenko}, \& {Reimer}}]{Strong:1998fr}
{Strong}, A.~W., {Moskalenko}, I.~V., \& {Reimer}, O. 2000, \bibinfo{title}{{Diffuse Continuum Gamma Rays from the Galaxy},} \apj, 537, 763, \dodoi{10.1086/309038}

\bibitem[{M. {Su} \& D.~P. {Finkbeiner}(2012){Su} \& {Finkbeiner}}]{Su2012}
{Su}, M., \& {Finkbeiner}, D.~P. 2012, \bibinfo{title}{{Evidence for Gamma-Ray Jets in the Milky Way},} \apj, 753, 61, \dodoi{10.1088/0004-637X/753/1/61}

\bibitem[{M. {Su} {et~al.}(2010){Su}, {Slatyer}, \& {Finkbeiner}}]{Su2010a}
{Su}, M., {Slatyer}, T.~R., \& {Finkbeiner}, D.~P. 2010, \bibinfo{title}{{Giant Gamma-ray Bubbles from Fermi-LAT: Active Galactic Nucleus Activity or Bipolar Galactic Wind?},} \apj, 724, 1044, \dodoi{10.1088/0004-637X/724/2/1044}

\bibitem[{P. {Virtanen} {et~al.}(2020){Virtanen} {et~al.}}]{scipy2020}
{Virtanen}, P., {et~al.} 2020, \bibinfo{title}{SciPy 1.0: Fundamental Algorithms for Scientific Computing in Python,} Nature Methods, 17, 261, \dodoi{10.1038/s41592-019-0686-2}

\bibitem[{V. {Vitale} \& A. {Morselli}(2009){Vitale} \& {Morselli}}]{2009arXiv0912.3828V}
{Vitale}, V., \& {Morselli}, A. 2009, \bibinfo{title}{{Indirect Search for Dark Matter from the center of the Milky Way with the Fermi-Large Area Telescope},} arXiv e-prints, arXiv:0912.3828, \dodoi{10.48550/arXiv.0912.3828}

\bibitem[{Q. {Wan} {et~al.}(2023){Wan}, {Guo}, {Xu}, {Wang}, {Zhang}, {Hu}, {Zhang}, {Pan}, {Li}, {Yue}, {Jiang}, {Cui}, \& {Chen}}]{WangQ2023}
{Wan}, Q., {Guo}, J.-H., {Xu}, X., {et~al.} 2023, \bibinfo{title}{{Design of a high-dynamic-range prototype readout system for VLAST calorimeter},} Nuclear Science and Techniques, 34, 149, \dodoi{10.1007/s41365-023-01291-3}

\bibitem[{S.~S. Wilks(1938)Wilks}]{Wilks1938}
Wilks, S.~S. 1938, \bibinfo{title}{The Large-Sample Distribution of the Likelihood Ratio for Testing Composite Hypotheses,} Ann. Math. Statist., 9, 60, \dodoi{10.1214/aoms/1177732360}

\bibitem[{M. {Wolleben}(2007){Wolleben}}]{Wolleben2007}
{Wolleben}, M. 2007, \bibinfo{title}{{A New Model for the Loop I (North Polar Spur) Region},} \apj, 664, 349, \dodoi{10.1086/518711}

\bibitem[{Z.-L. {Xu} {et~al.}(2022){Xu}, {Duan}, {Jiang}, {et~al.}}]{Xu2022}
{Xu}, Z.-L., {Duan}, K.-K., {Jiang}, W., {et~al.} 2022, \bibinfo{title}{{Optimal gamma-ray selections for monochromatic line searches with DAMPE},} Front. Phys., 17, 34501, \dodoi{10.1007/s11467-021-1121-6}

\bibitem[{Z.-L. {Xu} {et~al.}(2018){Xu}, {Duan}, {Shen}, {et~al.}}]{Xu2018}
{Xu}, Z.-L., {Duan}, K.-K., {Shen}, Z.-Q., {et~al.} 2018, \bibinfo{title}{{An algorithm to resolve {\ensuremath{\gamma}}-rays from charged cosmic rays with DAMPE},} \raa, 18, 027, \dodoi{10.1088/1674-4527/18/3/27}

\bibitem[{H.-Y.~K. {Yang} {et~al.}(2022){Yang}, {Ruszkowski}, \& {Zweibel}}]{Yang2022}
{Yang}, H.-Y.~K., {Ruszkowski}, M., \& {Zweibel}, E.~G. 2022, \bibinfo{title}{{Fermi and eROSITA bubbles as relics of the past activity of the Galaxy's central black hole},} Nature Astronomy, 6, 584, \dodoi{10.1038/s41550-022-01618-x}

\bibitem[{R.-Z. {Yang} {et~al.}(2014){Yang}, {Aharonian}, \& {Crocker}}]{Yang2014}
{Yang}, R.-Z., {Aharonian}, F., \& {Crocker}, R. 2014, \bibinfo{title}{{The Fermi bubbles revisited},} \aap, 567, A19, \dodoi{10.1051/0004-6361/201423562}

\bibitem[{Z. {Yang} {et~al.}(2026){Yang}, {Yan}, {She}, {An}, {Fang}, {Zhang}, {Yu}, {Liu}, {Wei}, {An}, {Guo}, {Wan}, {Zhang}, {Sun}, \& {Kong}}]{YangZ2026}
{Yang}, Z., {Yan}, J., {She}, Q., {et~al.} 2026, \bibinfo{title}{{Front-end readout electronics design for the anti-coincidence detector and charge detector in VLAST},} Nuclear Instruments and Methods in Physics Research A, 1082, 171015, \dodoi{10.1016/j.nima.2025.171015}

\bibitem[{Q. {Yuan} \& K. {Ioka}(2015){Yuan} \& {Ioka}}]{Yuan2015}
{Yuan}, Q., \& {Ioka}, K. 2015, \bibinfo{title}{{Testing the Millisecond Pulsar Scenario of the Galactic Center Gamma-Ray Excess With Very High Energy Gamma-Rays},} \apj, 802, 124, \dodoi{10.1088/0004-637X/802/2/124}

\bibitem[{Q. Yuan \& B. Zhang(2014)Yuan \& Zhang}]{Yuan:2014rca}
Yuan, Q., \& Zhang, B. 2014, \bibinfo{title}{{Millisecond pulsar interpretation of the Galactic center gamma-ray excess},} JHEAp, 3-4, 1, \dodoi{10.1016/j.jheap.2014.06.001}

\bibitem[{Y. {Zhang} {et~al.}(2025){Zhang}, {Chen}, {Chen}, {Liu}, {Hu}, {Zhang}, {Wei}, {Shen}, {Feng}, {Guo}, {Liu}, {Huang}, {Wang}, \& {Xu}}]{ZhangY2025}
{Zhang}, Y., {Chen}, Q., {Chen}, D., {et~al.} 2025, \bibinfo{title}{{The development of a high granular crystal calorimeter prototype of VLAST},} arXiv e-prints, arXiv:2509.24851, \dodoi{10.48550/arXiv.2509.24851}

\bibitem[{Y.-M. {Zhong} \& I. {Cholis}(2024){Zhong} \& {Cholis}}]{Zhong2024}
{Zhong}, Y.-M., \& {Cholis}, I. 2024, \bibinfo{title}{{Robustness of the Galactic Center excess morphology against masking},} \prd, 109, 123017, \dodoi{10.1103/PhysRevD.109.123017}

\bibitem[{Y.-M. {Zhong} {et~al.}(2020){Zhong}, {McDermott}, {Cholis}, {et~al.}}]{Zhong2020}
{Zhong}, Y.-M., {McDermott}, S.~D., {Cholis}, I., {et~al.} 2020, \bibinfo{title}{{Testing the Sensitivity of the Galactic Center Excess to the Point Source Mask},} \prl, 124, 231103, \dodoi{10.1103/PhysRevLett.124.231103}

\bibitem[{B. {Zhou} {et~al.}(2015){Zhou}, {Liang}, {Huang}, {et~al.}}]{Zhou2015}
{Zhou}, B., {Liang}, Y.-F., {Huang}, X., {et~al.} 2015, \bibinfo{title}{{GeV excess in the Milky Way: The role of diffuse galactic gamma-ray emission templates},} \prd, 91, 123010, \dodoi{10.1103/PhysRevD.91.123010}

\bibitem[{C.-R. {Zhu} {et~al.}(2022){Zhu}, {Cui}, {Xia}, {Yu}, {Huang}, {Yuan}, \& {Fan}}]{Zhu2022}
{Zhu}, C.-R., {Cui}, M.-Y., {Xia}, Z.-Q., {et~al.} 2022, \bibinfo{title}{{Explaining the GeV Antiproton Excess, GeV {\ensuremath{\gamma}} -Ray Excess, and W -Boson Mass Anomaly in an Inert Two Higgs Doublet Model},} \prl, 129, 231101, \dodoi{10.1103/PhysRevLett.129.231101}

\bibitem[{A. Zonca {et~al.}(2019)Zonca, Singer, Lenz, Reinecke, Rosset, Hivon, \& Gorski}]{healpy2019}
Zonca, A., Singer, L., Lenz, D., {et~al.} 2019, \bibinfo{title}{healpy: equal area pixelization and spherical harmonics transforms for data on the sphere in Python,} Journal of Open Source Software, 4, 1298, \dodoi{10.21105/joss.01298}

\bibitem[{K. {Zubovas} \& S. {Nayakshin}(2012){Zubovas} \& {Nayakshin}}]{Zubovas2012}
{Zubovas}, K., \& {Nayakshin}, S. 2012, \bibinfo{title}{{Fermi bubbles in the Milky Way: the closest AGN feedback laboratory courtesy of Sgr A*?},} \mnras, 424, 666, \dodoi{10.1111/j.1365-2966.2012.21250.x}

\end{thebibliography}
\bibliographystyle{aasjournalv7}



\end{document}